\documentclass[11pt]{article}
\usepackage{geometry, multirow, amsmath, amsfonts, amsthm, amssymb, graphicx, enumerate, natbib, bibentry, float, caption, subcaption, longtable,xcolor,bm}
\usepackage[breaklinks=true]{hyperref} 
\usepackage[nameinlink]{cleveref} 
\usepackage{xcolor} 
\hypersetup{
	bookmarksnumbered=true,
	colorlinks=true,
	anchorcolor={red!50!black},
	linkcolor={red!50!black},
	citecolor={blue!50!black},
	urlcolor={blue!50!black}
}

\usepackage{txfonts}
\usepackage[title]{appendix}
\usepackage{setspace}

\newenvironment{keywords}{
       \list{}{\advance\topsep by0.35cm\relax\small
       \leftmargin=1cm
       \labelwidth=0.35cm
       \listparindent=0.35cm
       \itemindent\listparindent
       \rightmargin\leftmargin}\item[\hskip\labelsep
                                     \bfseries Keywords:]}
     {\endlist}

\usepackage{tikz}
\usepackage{pgfplots}
\usepackage{natbib}
\usepackage{bookmark}
\bookmarksetup{
	numbered,
	addtohook={%
		\ifnum\bookmarkget{level}>1 %
		\bookmarksetup{numbered=false}%
		\fi
	},
}
\usepackage{latexsym,amsmath,url,epsfig}
\usepackage{textcomp}
\usepackage{amsfonts,euscript}

\usepackage{graphicx}
\usepackage{amsmath}
\usepackage{amssymb}
\usepackage{amsfonts}
\usepackage{amsthm}
\usepackage{mathrsfs}
\usepackage[all]{xy}
\usepackage{epstopdf}
\usepackage{rotating}
\usepackage[title]{appendix}
\usepackage{multirow,rotating}
\usepackage{url}
\usepackage{algorithm}
\usepackage{algorithmicx}
\usepackage{algpseudocode}
\usepackage{setspace}
\usepackage{listings}
\usepackage{color}
\usepackage{tikz}
\usetikzlibrary{calc,positioning}
\usepackage{graphics}
\usepackage[english]{babel}
\usepackage{datetime}
\usepackage{booktabs}
\usepackage{rotfloat}
\usepackage{framed}
\usepackage{natbib}
\usepackage{graphicx}

\usepackage{tabularx} 
\usepackage{array}

\usepackage{makecell}
\usepackage{todonotes}
\usetikzlibrary{chains,shapes,decorations,arrows,calc,arrows.meta,fit,positioning}
\tikzset{
    -Latex,auto,node distance =1 cm and 1 cm,semithick,
    state/.style ={ellipse, draw, minimum width = 0.7 cm},
    point/.style = {circle, draw, inner sep=0.04cm,fill,node contents={}},
    bidirected/.style={Latex-Latex,dashed},
    el/.style = {inner sep=2pt, align=left, sloped}
}
\usepackage{subcaption}
\usepackage{pgfplots}

\definecolor{cVS}{RGB}{20,20,20}
\definecolor{cVA}{RGB}{31,119,180}
\definecolor{cVP}{RGB}{200,40,40}
\pgfplotsset{
  payoffaxis/.style={
    width=\linewidth, height=6.6cm, axis on top,
    xlabel={$\theta_1$}, tick align=outside, tick pos=left,
    xtick distance=0.4, label style={font=\small},
    tick label style={font=\footnotesize}, enlarge x limits=false, clip=true,
    scaled y ticks=false, scaled x ticks=false,
    yticklabel style={/pgf/number format/.cd, fixed, precision=3},
  },
}
\tikzset{
  vsline/.style={cVS, thick},
  valine/.style={cVA, thick, dashed},
  vpline/.style={cVP, very thick},
  cutline/.style={gray, densely dotted},
  vlabel/.style={font=\scriptsize, gray, anchor=north, inner sep=1pt},
}

\newtheorem{theorem}{Theorem}

\newtheorem{definition}{Definition}
\newtheorem{lemma}{Lemma}
\newtheorem{proposition}{Proposition}

\theoremstyle{definition}

\algdef{SE}[DOWHILE]{Do}{doWhile}{\algorithmicdo}[1]{\algorithmicwhile\ #1}

\allowdisplaybreaks

\let\epsilon\varepsilon

\usepackage[normalem]{ulem}

\begin{document}
\setstretch{1.5}

\begin{titlepage}
\title{\textbf{False Positives and Innovation Incentives: Contracting for Noisy Experimentation}}
\author{Chang Liu\thanks{\scriptsize{School of Economics, UNSW Business School; \href{mailto:chang.liu36@unsw.edu.au}{\nolinkurl{chang.liu36@unsw.edu.au}}.}} \and Haoyu Liu\thanks{\scriptsize{Department of Industrial Engineering and Operations Research, University of California, Berkeley; \href{mailto:haoyuliu@berkeley.edu}{\nolinkurl{haoyuliu@berkeley.edu}}.}} \and Nian Si\thanks{\scriptsize{Department of Industrial Engineering and Decision Analytics, Hong Kong University of Science and Technology; \href{mailto:niansi@ust.hk}{\nolinkurl{niansi@ust.hk}}.}} \and Zeyu Zheng\thanks{\scriptsize{Department of Industrial Engineering and Operations Research, University of California, Berkeley; \href{mailto:zyzheng@berkeley.edu}{\nolinkurl{zyzheng@berkeley.edu}}.}}}
\date{\begin{tabular}{ rl } 
This version:&November 25, 2026
\end{tabular}}
\maketitle
\thispagestyle{empty}
\begin{abstract}
In many data-driven organizations, employees propose product changes that are launched only after passing noisy online experiments, and their compensation is often tied to successful launches. Because even low-value features may appear successful by chance, such launch-based incentives can encourage agents to conduct experiments that reduce total surplus. We study this problem in a principal-agent model in which an agent with private knowledge about feature quality chooses whether to conduct a costly experiment and is compensated through an affine contract consisting of a launch-contingent bonus and a long-run value-sharing component, subject to limited liability. We evaluate each contract by the largest fraction of the principal's full-information payoff that it guarantees uniformly across the agent's private knowledge. Our analysis reveals a fundamental tension between experimental noise and limited liability in incentive design. Launch bonuses create incentives for false-positive experimentation by rewarding features that succeed by chance, whereas limited liability can render pure value sharing insufficient to implement socially efficient experimentation. Therefore, the optimal contract balances launch-contingent rewards and long-run value sharing according to the default feature quality, the degree of distributional uncertainty, and the severity of the limited-liability constraint.
\end{abstract}
\end{titlepage}

\setstretch{1.5}\clearpage
\newpage\setcounter{page}{2}

\section{Introduction}

Controlled experimentation has become a central tool for product innovation in data-driven firms. New features are typically proposed by engineers, evaluated offline, tested through online A/B experiments, and launched only if the experiment reports a statistically significant positive effect. This workflow has made experimentation scalable: firms can evaluate many product changes quickly, relying on real user behavior rather than managerial judgment alone. As emphasized by \citet{thomke2003experimentation,thomke2020experimentation}, modern experimentation technologies have changed not only how firms learn, but also how innovation itself is organized.

\begin{figure}[ht]
    \centering
    \resizebox{\linewidth}{!}{%
    \begin{tikzpicture}[
        node distance=0.35cm,
        font=\scriptsize,
        recstate/.style={
            rectangle, draw, text width=2.0cm,
            minimum height=0.65cm, align=center,
            inner sep=1.5pt
        },
        roundstate/.style={
            ellipse, draw, text width=2.0cm,
            minimum height=0.65cm, align=center,
            inner sep=1.5pt
        }
    ]
        \node[recstate] (1) {Propose a feature};
        \node[recstate, right=of 1] (2) {Implement and test offline};
        \node[recstate, right=of 2] (3) {Test via online A/B tests};
        \node[recstate, right=of 3] (4) {Launch and roll out};
        \node[roundstate, right=of 4] (5) {Performance evaluation};

        \path (1) edge (2);
        \path (2) edge (3);
        \path (3) edge (4);
        \path (4) edge (5);
    \end{tikzpicture}%
    }
    \caption{A standard workflow in a data-driven company.}
    \label{fig:scheme}
\end{figure}
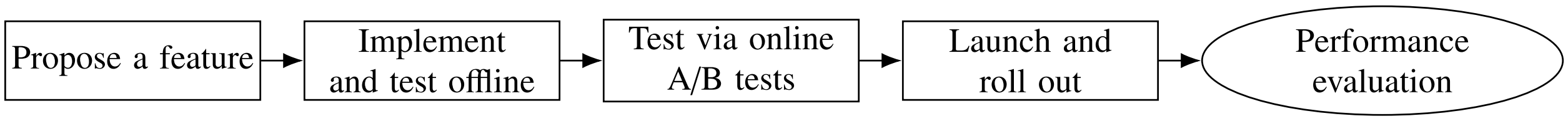

The standard workflow of an engineer in a data-driven company is as follows (as illustrated in Figure \ref{fig:scheme}):
\begin{enumerate}
    \item The engineer proposes an innovation and develops the feature in an offline environment.
    \item The engineer evaluates the feature on offline datasets and compares its performance against a predefined benchmark.
    \item If the offline evaluations are positive, the feature is implemented in an online environment for controlled experiments with real user traffic. 
    \item If the primary metric shows a statistically significant improvement and other critical metrics remain within expected norms, the innovation is officially launched and rolled out.
    \item Within certain periods, the engineer's performance is assessed based on the number of features successfully launched and their sustained performance over time.
\end{enumerate}

While this workflow is effective and has helped many companies succeed, the inherent randomness of controlled experiments creates a misalignment between the company's and the engineer's interests.  Even a feature with no true effect can pass a statistical test by chance, and a feature with negative long-run value can occasionally appear successful in the experiment. When an engineer's evaluation or compensation depends on whether proposed features are launched, this randomness becomes an incentive instrument. An engineer may then prefer to propose and test weak features, not because they are likely to create value, but because each noisy test carries some probability of a successful launch. Such behavior is costly for the firm: implementation, experimentation, and rollout all consume resources, and falsely launched features may reduce long-run product value.

The resulting problem can be formulated as a principal-agent model with an experimental measurement technology. The principal, interpreted as the firm or platform, seeks to incentivize the agent, interpreted as an engineer or product team, to conduct experiments whenever experimentation generates positive total surplus. Experimentation may reveal, for instance, whether a proposed feature improves long-run system performance, user engagement, or platform revenue. The agent, however, has private knowledge about the intrinsic quality of the proposed feature, such as its expected user impact, implementation risk, or likelihood of causing downstream harm. The principal cannot directly observe this latent quality ex ante; instead, she observes only the outcome of the experiment, filtered through a noisy measurement process in the form of short-run A/B test metrics. Compensation can therefore be conditioned on observable outcomes and long-run value: whether the feature is ultimately launched, and how successful the experimental feature appears afterwards.

The key friction is the combination of noisy launch decisions and limited liability. If experiments perfectly revealed feature quality, incentive alignment would be relatively straightforward: the principal could reward the agent for launching features that are known to create positive long-run value, and a simple linear sharing rule based on long-run value could make the agent internalize the relevant share of the experimentation cost. Because experiments are noisy, however, harmful features are sometimes launched, and limited liability prevents the principal from fully charging the agent ex post for such launches. As a result, long-run value sharing leaves the agent less than his share of the losses from harmful launches, and it can make an experiment with negative total surplus privately profitable. This motivates the use of a launch-contingent payment, or launch bonus, defined as a fixed transfer paid to the agent whenever the feature is launched, regardless of its realized long-run value. The launch bonus creates a subtle contract-design tradeoff. On one hand, it can encourage false-positive experimentation by making noisy successful launches attractive to the agent. On the other hand, it may still be necessary when the liability floor binds, because the principal has limited ability to use negative transfers or ex post penalties to discipline bad launches. The optimal contract must therefore choose the launch bonus carefully: large enough to support socially valuable experimentation, but not so large that it makes experimentation on low-value features privately attractive.

To model this tradeoff, we propose a baseline model in Section~\ref{sec:baseline-model}. We assume that the true value of the new feature can take one of two values: a default value known to both parties, or an alternative value known only to the agent. Both parties share the same belief over which value is realized. The principal offers an \textit{affine contract} consisting of a \textit{launch bonus} and a long-run \textit{value share}. The launch bonus is a fixed payment made conditional on launching the feature. The value-sharing payment is proportional to the realized true value of the feature, but is subject to an exogenous lower bound reflecting limited liability: when the feature turns out to be harmful, the principal cannot charge the agent beyond a fixed maximum amount.

We restrict attention to this affine contract class for three reasons. First, it is highly interpretable: each contract is governed by two transparent parameters, a fixed launch bonus and a linear share of long-run value, making the resulting incentives easy for both managers and engineers to understand. Second, it is readily implementable in practice, as it mirrors how data-driven organizations tie performance evaluations and compensation to successful launches and sustained product impact without relying on complex nonlinear schedules that are difficult to administer or audit. Third, restricting attention to affine contracts entails little loss in performance: as our numerical experiments in Appendix~\ref{app:affine-vs-unrestricted} show, they perform nearly as well as unrestricted monotone contracts.

We summarize our findings as follows:

First, we apply a prior-free, scale-free robustness criterion based on a \textit{uniform profit guarantee} to contract design in experiment-driven firms. Taking the principal's full-information payoff as the benchmark, we evaluate each affine contract by the largest fraction of this benchmark that it guarantees uniformly across the agent's private knowledge. We show that this ratio criterion endogenously unifies two fundamental incentive requirements: providing any guarantee (even zero) is equivalent to \textit{principal protection}, meaning the principal earns a nonnegative expected payoff whenever the agent experiments, whereas providing a strictly positive guarantee additionally requires \textit{socially efficient experimentation}, meaning the agent experiments if and only if expected total surplus is nonnegative.  

Second, we characterize optimal affine contracts and their uniform profit guarantees in the baseline model, showing how the optimal mix of launch bonuses and value sharing depends on the default feature value and the tightness of limited liability. When the default value is pessimistic, a strictly positive guarantee is always attainable regardless of the liability cap, and an optimal contract uses a single instrument, either a pure launch bonus or proportional value sharing. When the default value is optimistic, limited liability directly governs the contract structure: no contract provides a positive guarantee under a tight cap, and under any other cap a unique contract implements efficient experimentation, combining value sharing with a launch bonus under an intermediate cap and using proportional sharing under a loose cap. Whenever a positive guarantee is attainable, this contract is uniquely optimal.

Third, we study profit guarantees above a \textit{profitability threshold}, requiring the profit-ratio guarantee only for experiments whose expected total surplus is at least a given level while maintaining principal protection everywhere. In the optimistic-default setting, whenever a positive guarantee is attainable, every optimal contract uses a strictly positive launch bonus: pure value sharing is never optimal because replacing part of the value share with a launch bonus avoids overpaying the agent on high-value launches. Moreover, when the profitability threshold is sufficiently high, the optimal contract is unique, independent of the liability cap, and reduces to a pure launch bonus.

Fourth, we extend the analysis to \textit{distributional uncertainty}, where the agent privately knows a general quality distribution subject only to a finite mean and a lower bound on the probability of nonnegative quality. Across all such distributions, no affine contract provides a strictly positive unrestricted guarantee. However, above a sufficiently high profitability threshold, a strictly positive guarantee becomes attainable, and every optimal contract uses both instruments.

Taken together, our results bridge two literatures that have developed largely in isolation: limited-liability contracting and strategic hypothesis testing. Classical limited-liability models study how payment floors shape incentives for effort or output. Here, the contractible event is instead the outcome of a noisy launch rule applied to an experiment that takes place only if the agent chooses to test. Because the agent privately knows feature quality, limited liability distorts not only how profits are split, but also whether the agent chooses to experiment in the first place.

The central insight is that limited liability can undermine pure value sharing as a tool for disciplining noisy experimentation. Without a liability cap, proportional value sharing aligns the agent's payoff with total surplus and guarantees the principal a fixed share of first-best profit. With limited liability, however, low-quality features can still pass the noisy launch rule by chance while the agent's downside is capped, allowing pure value sharing to reward precisely the inefficient experiments it is meant to deter. Consequently, a launch bonus can be necessary: under an optimistic default with an intermediate cap, every contract that provides a positive guarantee uses one. A launch bonus can also be optimal for a reason that does not involve limited liability. With a zero default value, which the evidence in \citet{berman2022false} supports in many practical settings, the pure launch bonus is uniquely optimal for every cap, because its expected payment rises with the alternative value only through the launch probability and therefore leaves the agent a smaller share of total surplus on barely efficient experiments than any contract that shares value. Both results challenge the conventional view that launch bonuses inevitably invite false-positive gaming.

The rest of the paper is organized as follows. Section~\ref{sec:literature} reviews the related literature. Section~\ref{sec:baseline-model} introduces the baseline model, the full-information benchmark, and the uniform profit guarantee. Section~\ref{sec:robust-affine-contracts} characterizes optimal contracts in the baseline model, both across all alternative values and above a profitability threshold. Section~\ref{sec:distributional-uncertainty} extends the analysis to general distributional uncertainty. Section~\ref{sec:numerical-illustration} provides numerical illustrations of our main findings. Section~\ref{sec:conclusion} concludes, and all proofs are deferred to Appendix~\ref{app:proofs}.

\section{Related Literature}
\label{sec:literature}
\subsection{Contract Theory}
Our paper builds on the classical theory of incentive contracts under moral hazard \citep{holmstrom1979}. A recurring theme of this literature is that the form of the optimal contract is shaped by the constraints on the payments the principal can impose on the agent. When the agent is protected by limited liability, optimal contracts typically take a simple option-like or affine form that rewards the agent only in sufficiently favorable states \citep{sappington1983, innes1990}. Our affine contract, which combines a launch-contingent bonus with a capped value-sharing payment, inherits this structure: limited liability prevents the principal from charging the agent for harmful launches, and this is precisely the force that prevents pure value sharing from disciplining inefficient experimentation. A related strand studies incentive provision when the available performance measure is noisy or distorted, so that rewarding measured success also rewards luck and can induce gaming \citep{holmstrom1991, baker1992}. In our setting the launch decision is exactly such a measure, because launches are triggered by noisy experiments and a launch-contingent reward therefore compensates the agent for outcomes that are partly the product of experimental noise.

Our paper also relates to two additional strands of contract theory. First, the literature on contracting for innovation and experimentation studies how to motivate an agent to acquire information or learn over time \citep{bergemann2005, manso2011, horner2013, halac2016}. That literature largely focuses on dynamic effort and the option value of continued learning. We instead study the static decision of whether to run a single costly experiment whose outcome is a noisy launch signal, and we characterize which affine contracts maximize the principal's uniform profit guarantee under limited liability. Second, our prior-free design objective connects to the literature on robust contracting and mechanism design \citep{carroll2015robustness, carroll2019robustness}, which evaluates contracts by their worst-case performance without assuming a prior over the agent's environment. Because feature values in our setting can vary substantially in magnitude with what the agent privately knows, we follow \citet{hartline2025scale,hartline2026robustness} and adopt a max-min profit-ratio criterion that evaluates performance relative to the full-information benchmark case by case, thereby giving equal consideration to cases of different scales.

\subsection{A/B Tests}
With the rise of the digital economy, the study of A/B testing on platforms has received significant attention over the past decade. Most existing work focuses on experimental design methodologies that enable accurate estimation of treatment effects for a single experiment. Examples include two-sided randomization designs \citep{bajari2021multiple,johari2022experimental}, bipartite experiments \citep{eckles2017design,pouget2019variance,harshaw2023design}, and adaptive experimental designs \citep{qin2022adaptivity,simchi2022multi,kuang2023weak,cook2023semiparametric}. More recently, \citet{johari2024does} and \citet{ni2025decision} have examined A/B testing from a decision-making perspective. Rather than focusing solely on unbiased treatment effect estimation, they study how experimental data can be used to make better decisions, even when the experiment itself is biased. While this literature focuses on the design and interpretation of experiments for a single experimenter, it largely abstracts away from the organizational incentive problem that arises when experiments are proposed and conducted by self-interested agents.

Closest to our incentive-design perspective is the recent literature on strategic hypothesis testing, which studies how statistical protocols or contract menus should be designed when self-interested, risk-sensitive, or heterogeneous agents strategically choose whether to generate data or run costly trials \citep{bates2022principal,shi2024sharp,shi2025instance}. In contrast, our contract is not a statistical testing protocol; rather, it is an internal compensation scheme for delegated product experimentation under noisy launch decisions and limited liability.

Another closely related line of literature studies how firms should use A/B tests to select and implement new ideas under scarce experimental resources. \citet{azevedo2020b}  show that the optimal experimentation strategy depends critically on the tail behavior of the prior distribution of innovation quality: when this distribution is fat-tailed, firms should adopt a lean experimentation strategy that tests many ideas with relatively smaller sample sizes, rather than concentrating resources on a few high-powered experiments. Building on this framework, \citet{azevedo2023b} study the case of Gaussian priors and derive closed-form characterizations of the optimal implementation rule, the value of experimentation, and the optimal experiment size. While our paper shares this decision-theoretic perspective, we study a fundamentally different incentive-design problem in which experimentation is delegated to an informed agent who chooses whether to run a costly test and is compensated based on noisy launch outcomes.

\section{Model and Assumptions}
\label{sec:baseline-model}

\subsection{Baseline Model}

The baseline model is a moral hazard problem between a principal (the company) and an agent (the engineer or product team). The agent develops a new feature whose true quality is denoted by $\theta$. With probability $1-p$, the feature has a default quality $\theta_0$; with probability $p\in(0,1)$, it has an alternative quality $\theta_1$. The default value $\theta_0$ and the probability $p$ are common knowledge, while the alternative value $\theta_1$ is known only to the agent. Neither party initially observes which quality is realized, and both are risk neutral.

After developing the feature, the agent decides whether to conduct a costly experiment that generates information about $\theta$. Conducting the experiment costs the agent $c>0$ and the principal $\kappa>0$. We interpret $c$ as the agent's residual effort or opportunity cost and $\kappa$ as the organization's resource cost. The experiment produces a noisy signal $\hat{\theta} = \theta + \varepsilon$, where $\varepsilon \sim N(0, s^2)$ is independent of $\theta$ and $s>0$ is common knowledge. The feature is launched if and only if $\hat{\theta} > \Delta$, where $\Delta$ is a predetermined launch threshold set by the principal. In practice, $\Delta$ is typically set to the 97.5th or 99.5th percentile of the distribution $N(0,s^2)$, the critical values of two-sided tests at the 5 and 1 percent levels, and both the statistical significance threshold $\Delta$ and the noise level $s$ (governed by traffic allocation and experimentation latency constraints) are standardized at the platform-infrastructure level rather than tailored to individual projects as contracting instruments. Throughout our analysis, we therefore take $\Delta$ and $s$ as fixed.

To provide incentives, the principal offers the agent a contract before the agent develops the feature and decides whether to conduct the experiment. The contract specifies the agent's compensation conditional on experimentation and launch. We focus on affine contracts with limited liability, characterized by two nonnegative parameters, $\alpha$ and $\beta$. The launch bonus $\beta$ is a fixed payment made if and only if the new feature is launched, and the value share $\alpha$ gives the agent a stake in the feature's long-run value. Conditional on launch, the agent receives $\beta + \max\{\alpha\theta, -r\}$, where $r > 0$ is a liability cap: even if the feature eventually proves harmful, the principal cannot charge the agent more than $r$ through the value-sharing term. We say that the liability floor binds at quality $\theta$ if $\alpha\theta<-r$, so that the floor raises the payment, and that it does not bind if $\alpha\theta\geq-r$. No performance-related transfer is made when the feature is not launched. 

The timing of the game is as follows.
\begin{enumerate}
    \item The principal offers a contract $w=(\alpha,\beta)$, where $\alpha,\beta \ge 0$.
    \item After observing the contract, the agent develops the feature, privately observes its alternative value $\theta_1$, and decides whether to conduct the experiment. If he does not experiment, the game ends and both parties receive a payoff of zero.
    \item If the agent conducts the experiment, he incurs cost $c$, the principal incurs cost $\kappa$, and the signal $\hat{\theta} = \theta + \varepsilon$ is realized. The feature is launched if and only if $\hat{\theta} > \Delta$.
    \item Payoffs are then realized. Conditional on experimentation, the principal's expected payoff is
$$
    V_P(\theta_1;w)
    = \mathbb{E}\!\left[
        \left(\theta - \beta - \max\{\alpha \theta,-r\}\right)
        \mathbb{I}\{\hat{\theta} > \Delta\}
        - \kappa
    \right],
$$
    and the agent's expected payoff is
$$
    V_A(\theta_1;w)
    = \mathbb{E}\!\left[
        \left(\beta + \max\{\alpha \theta,-r\}\right)
        \mathbb{I}\{\hat{\theta} > \Delta\}
        - c
    \right],
$$
    where the expectation is taken with respect to the randomness in $\theta$ and $\hat{\theta}$.
\end{enumerate}

When there is no risk of confusion, we suppress the arguments $(\theta_1;w)$. Given a contract $w$ and an alternative value $\theta_1$, the agent experiments if and only if $V_A(\theta_1;w) \ge 0$. Throughout, we resolve indifference in favor of experimentation.\footnote{The boundary case $r=0$ prohibits negative value-sharing payments and can create additional intervals of alternative values on which the agent is exactly indifferent. Under this tie-breaking rule, the agent's decisions at $r=0$ then need not be the limits of his decisions as $r\downarrow0$. We assume $r>0$ to rule out these boundary effects.} Let
$$U_P(\theta_1;w) = V_P(\theta_1;w)\mathbb{I}\{V_A(\theta_1;w)\geq0\}$$
denote the principal's expected payoff under this decision rule. That is, $V_P$ is the principal's expected payoff conditional on experimentation, whereas $U_P$ also accounts for the agent's decision whether to experiment. These expected payoffs are averaged over realized quality and experimental noise for each fixed alternative value, but not over alternative values. Section~\ref{subsec:contracting-objective} defines the principal's objective by comparing $U_P$ with the full-information benchmark uniformly across alternative values.

Conditional on experimentation, total surplus is
\begin{align}
     V_S(\theta_1) &= V_P(\theta_1;w) + V_A(\theta_1;w) = \mathbb{E} \left[ \theta \cdot \mathbb{I}\left\{\hat{\theta} > \Delta\right\} -c - \kappa \right] \nonumber \\
   &=
    p\,\theta_1\Phi\!\left(\frac{\theta_1-\Delta}{s}\right)
    +(1-p)\,\theta_0\Phi\!\left(\frac{\theta_0-\Delta}{s}\right)
    -c-\kappa.
     \label{eq:expected-total-surplus}
\end{align}
Here $\Phi$ denotes the standard normal CDF. Total surplus does not depend on the contract, so we write $V_S(\theta_1)$ without $w$. We adopt a utilitarian welfare criterion and define experimentation to be \textit{socially efficient} if and only if $V_S \ge 0$. Total surplus matters for the principal's objective because it determines the full-information benchmark: Section~\ref{subsec:discussion-of-assumptions} shows that the benchmark equals $\max\{V_S,0\}$. Section~\ref{subsec:socially-efficient-experimentation} then characterizes the alternative values $\theta_1$ at which experimentation is socially efficient. Section~\ref{subsec:contracting-objective} shows that a contract can guarantee the principal a strictly positive fraction of the benchmark uniformly across alternative values only if it induces the agent to experiment exactly when doing so is socially efficient.

\subsection{Full-Information Benchmark and the Role of the Assumptions}
\label{subsec:discussion-of-assumptions}

We first establish the benchmark used to evaluate contracts. Suppose the principal observes the alternative value $\theta_1$ before contracting, while the experimentation costs, noisy launch rule, liability constraint, and available compensation instruments remain unchanged. Define
$$\overline{U}_P(\theta_1) = \max_{w=(\alpha,\beta):\,\alpha,\beta\geq0} U_P(\theta_1;w),$$
where the maximization is performed separately at each alternative value.

\begin{lemma}[Full-information benchmark]\label{lem:full-information-benchmark}
For every $\theta_1$, $\overline{U}_P(\theta_1) = \max\{ V_S(\theta_1),0\}$.
\end{lemma}

The benchmark therefore equals first-best expected total surplus under the maintained experimentation technology and launch rule, so comparing $U_P(\cdot;w)$ with $\overline{U}_P(\cdot)$ measures the fraction of this surplus that the principal retains. Full information here concerns the alternative value, not realized quality or experimental noise. Under full information, she can choose a different contract for each alternative value. In the baseline, she must instead use one contract across all alternative values. The question is how large a fraction of the full-information benchmark a single contract can guarantee. The following comparisons show how removing either the liability floor or experimental noise permits proportional sharing, defined below, to guarantee a positive fraction even when $\theta_1$ remains privately known.

\paragraph{Unlimited liability.}
Keep the experimental noise but remove the liability floor (formally, $r\to\infty$), and consider proportional sharing:
$$\beta=0,\quad\alpha=\frac{c}{c+\kappa}.$$
Since the expected launched value equals $V_S+c+\kappa$ by \eqref{eq:expected-total-surplus}, the agent's and principal's expected payoffs are fixed fractions of total surplus:
$$V_A=\frac{c}{c+\kappa}V_S,\quad V_P=\frac{\kappa}{c+\kappa}V_S.$$
The agent therefore experiments exactly when $V_S\geq0$, and the principal receives zero otherwise. Consequently, proportional sharing guarantees the fraction $\kappa/(c+\kappa)$ of the full-information benchmark at every alternative value:
$$U_P(\theta_1;w)=\frac{\kappa}{c+\kappa}\overline{U}_P(\theta_1)\quad\text{for}\quad w=\left(\frac{c}{c+\kappa},0\right).$$
This contract thus implements efficient experimentation with a positive guarantee. 

\paragraph{Noiseless experiment.}
Now retain the liability floor, remove the experimental noise so that $\hat\theta=\theta$, and hold the launch threshold fixed. The feature is then launched if and only if $\theta>\Delta$. For $\Delta\geq0$, every launched feature therefore has positive long-run value, so $\alpha\theta\geq0$ and the liability floor never binds. Proportional sharing again makes each party's expected payoff a fixed fraction of total surplus, so it implements efficient experimentation and guarantees the principal the fraction $\kappa/(c+\kappa)$ of the corresponding full-information benchmark. The absence of false-positive launches removes the distortion caused by the liability floor, despite the principal's uncertainty about $\theta_1$.

\paragraph{Why the combination matters.}
With noisy launch decisions and a finite liability cap, proportional sharing no longer makes the agent's payoff a fixed fraction of total surplus: the liability floor raises his payoff after sufficiently harmful launches. Because $\theta_1$ is privately known, the principal cannot recalibrate the contract separately at each alternative value. A common contract must therefore address the experimentation decision, not only the rents left to the agent. These two considerations enter a single profit-guarantee criterion. Since $\overline{U}_P(\theta_1)=\max\{V_S(\theta_1),0\}$, the benchmark is positive precisely where experimentation generates strictly positive surplus. As Section~\ref{subsec:contracting-objective} establishes, a strictly positive uniform guarantee requires efficient experimentation, while the size of the guarantee depends on the rents left to the agent. We therefore characterize the efficient experimentation region next, before formalizing the contracting objective.

\subsection{Socially Efficient Experimentation}
\label{subsec:socially-efficient-experimentation}

Lemma~\ref{lem:full-information-benchmark} links the full-information benchmark to total surplus. We now characterize the alternative values for which experimentation is socially efficient and hence the region where the benchmark is positive, together with the zero-surplus cutoffs that bound it. Throughout this section, $\theta_0$, $p$, $\Delta$, $s$, $c$, and $\kappa$ are held fixed. Recall that experimentation is socially efficient at $\theta_1$ if and only if $V_S(\theta_1)\geq0$, where, from \eqref{eq:expected-total-surplus},
$$
V_S(\theta_1)
=
p\,\theta_1\Phi\!\left(\frac{\theta_1-\Delta}{s}\right)
+
(1-p)\,\theta_0\Phi\!\left(\frac{\theta_0-\Delta}{s}\right)
-c-\kappa .
$$

The only term in $V_S(\theta_1)$ that varies with $\theta_1$ is $p\,\theta_1\Phi\!\left(\frac{\theta_1-\Delta}{s}\right),$ which has a single trough. That is, there is a unique $\underline \theta_1<0$ such that $V_S(\theta_1)$ is decreasing for $\theta_1<\underline \theta_1$ and increasing for $\theta_1>\underline \theta_1$.\footnote{With $\phi$ the standard normal density, the derivative in $\theta_1$ is $p\left(\Phi(\frac{\theta_1-\Delta}{s})+\frac{\theta_1}{s}\phi(\frac{\theta_1-\Delta}{s})\right)$, which tends to $0^-$ as $\theta_1\to-\infty$, equals $p\Phi(-\Delta/s)>0$ at $\theta_1=0$, and tends to $p$ as $\theta_1\to+\infty$. It therefore crosses zero at some $\underline\theta_1<0$, and Lemma~\ref{lem:surplus-shape} shows that this crossing is unique.} The intuition is simple. If $\theta_1$ is very negative, the alternative state almost never clears the launch threshold, so it has little effect on expected surplus. As $\theta_1$ becomes less negative, the alternative state is still harmful but is launched more often, which lowers expected total surplus. Once $\theta_1$ is high enough, both the value conditional on launch and the launch probability increase, so expected surplus rises.

Throughout the baseline analysis, we focus on the nondegenerate case in which the trough is strictly negative, $V_S(\underline \theta_1)<0.$ If instead $V_S(\underline \theta_1)\geq0$, then experimentation is socially efficient for every value of $\theta_1\in \mathbb{R}$, and the implementation problem is degenerate.

Furthermore, we note that $\lim_{\theta_1\to-\infty} p\theta_1\Phi\!\left(\frac{\theta_1-\Delta}{s}\right) =0,$ which implies this term  takes the same value  at $\theta_1=0$ as in the limit $\theta_1\to-\infty$. Therefore
\begin{equation}
    \label{eq:limit-equals-zero}
    \lim_{\theta_1\to-\infty} V_S(\theta_1)=
    (1-p)\,\theta_0\Phi\!\left(\frac{\theta_0-\Delta}{s}\right)
    -c-\kappa=V_S(0).
\end{equation}
Based on \eqref{eq:limit-equals-zero}, we analyze $V_S(0)$ below, since its sign determines whether the surplus curve crosses zero once or twice. To proceed, we first define $\overline\theta_0$ as the solution to
$$
(1-p)\,\overline\theta_0
\Phi\!\left(\frac{\overline\theta_0-\Delta}{s}\right)
=
c+\kappa.
$$
We observe that $\overline\theta_0$ is positive and unique, since $\theta_0\Phi((\theta_0-\Delta)/s)$ is strictly increasing for $\theta_0\ge 0$. Therefore, $V_S(0)$ is nonnegative if and only if $\theta_0\ge \overline\theta_0$.

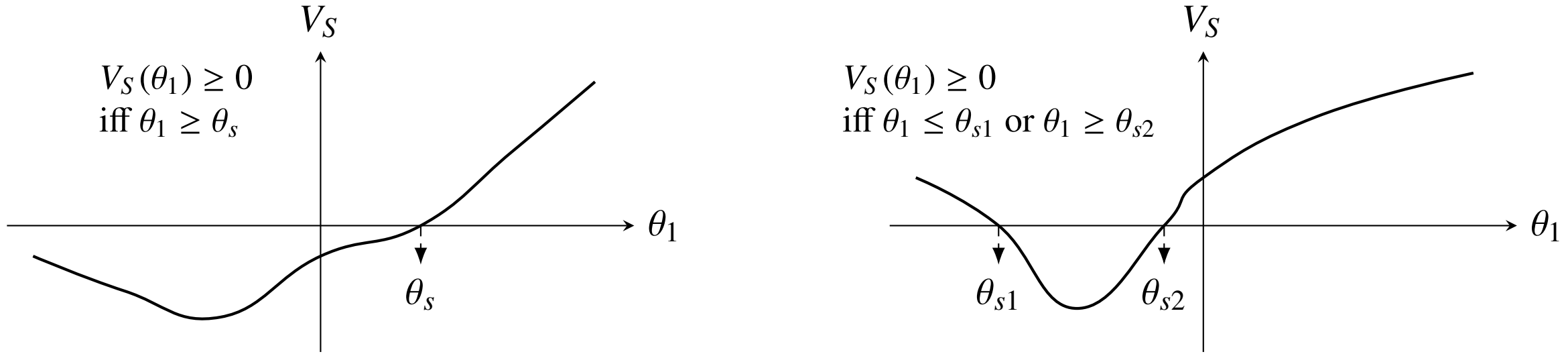
\begin{figure}[!htbp]
    \centering
    \begin{subfigure}[t]{0.49\textwidth}
        \centering
        \begin{tikzpicture}[scale=0.86]
            \draw[->,>=stealth] (-3.6,0) -- (3.6,0) node[right] {$\theta_1$};
            \draw[->,>=stealth] (0,-1.45) -- (0,2.0) node[above] {$V_S$};
            \draw[thick, smooth, tension=0.8,-]
                plot coordinates {(-3.3,-0.35) (-2.25,-0.75) (-1.15,-1.05) (0,-0.35) (1.15,0) (2.2,0.85) (3.15,1.65)};
            \draw[densely dashed] (1.15,0) -- (1.15,-0.45) node[below] {$\theta_s$};
            \node[anchor=north west, align=left, font=\small] at (-2.7,2)
                {$V_S(\theta_1)\ge 0$\\ iff $\theta_1\ge\theta_s$};
        \end{tikzpicture}
        \caption{Pessimistic default: one cutoff}
        \label{fig:surplus-one-cutoff}
    \end{subfigure}\hfill
    \begin{subfigure}[t]{0.49\textwidth}
        \centering
        \begin{tikzpicture}[scale=0.86]
            \draw[->,>=stealth] (-3.6,0) -- (3.6,0) node[right] {$\theta_1$};
            \draw[->,>=stealth] (0,-1.45) -- (0,2.0) node[above] {$V_S$};
            \draw[thick, smooth, tension=0.8,-]
                plot coordinates {(-3.3,0.55) (-2.35,0) (-1.45,-0.95) (-0.45,0) (0,0.55) (1.3,1.25) (3.1,1.75)};
            \draw[densely dashed] (-2.35,0) -- (-2.35,-0.45) node[below] {$\theta_{s1}$};
            \draw[densely dashed] (-0.45,0) -- (-0.45,-0.45) node[below] {$\theta_{s2}$};
            \node[anchor=north west, align=left, font=\small] at (-4.3,2)
                {$V_S(\theta_1)\ge 0$\\ iff $\theta_1\le\theta_{s1}$ or $\theta_1\ge\theta_{s2}$};
        \end{tikzpicture}
        \caption{Optimistic default: two cutoffs}
        \label{fig:surplus-two-cutoffs}
    \end{subfigure}
    \caption{The shape of expected total surplus. With a pessimistic default, the curve crosses zero once, so experimentation is socially efficient only for sufficiently high alternative values. With an optimistic default, the curve crosses zero twice, so experimentation is socially efficient both for very low and for sufficiently high alternative values.}
\end{figure}

\paragraph{Pessimistic default ($\theta_0\le\overline\theta_0$).}
The total surplus at $\theta_1=0$, $V_S(0)$, is nonpositive. Since $V_S$ has a single trough and $V_S(0)\le0$, $V_S(\cdot)$ is nonpositive on $(-\infty,0]$. For $\theta_1>0$,  $V_S(\theta_1)$ is strictly increasing and eventually diverges to $+\infty$. Hence, there is a single cutoff $\theta_s\ge0$ satisfying $V_S(\theta_s)=0$ such that experimentation is socially efficient if and only if $\theta_1\ge\theta_s .$ In this case, the socially efficient region is the connected interval $[\theta_s,\infty).$ The shape of expected total surplus as a function of $\theta_1$ is visualized in Figure~\ref{fig:surplus-one-cutoff}.

\paragraph{Optimistic default ($\theta_0>\overline\theta_0$).}
The total surplus at $\theta_1=0$, $V_S(0)$, is positive. Since the trough is strictly negative, the curve $V_S(\cdot)$ starts above zero as $\theta_1\to -\infty$, decreases below zero, and then rises above zero again before reaching $\theta_1=0$. Consequently, it crosses the horizontal axis twice, yielding two negative cutoff values,
$$
\theta_{s1}<\underline \theta_1<\theta_{s2}<0.
$$
Experimentation is socially efficient if and only if $\theta_1\in (-\infty,\theta_{s1}] \cup [\theta_{s2},\infty),$ which consists of two disconnected regions. The upper region corresponds to the familiar high-alternative-quality regime. The lower region represents a screening regime: when $\theta_1$ is sufficiently low, the experiment almost always rules out the alternative state while still allowing the optimistic default state to be launched. Consequently, experimentation can be socially efficient even when $\theta_1$ is very low. Figure~\ref{fig:surplus-two-cutoffs} visualizes the optimistic-default case.

The efficient cutoffs separate the regions where the full-information benchmark is positive from those where it is zero. Section~\ref{subsec:contracting-objective} next formalizes the uniform profit guarantee and establishes how this geometry restricts the contracts that can provide a positive guarantee.

\subsection{Uniform Profit Guarantees}
\label{subsec:contracting-objective}
The principal chooses a contract without observing the alternative value. She evaluates a contract by the fraction of the full-information benchmark that it guarantees uniformly across alternative values. By Lemma~\ref{lem:full-information-benchmark}, the principal then retains at least this fraction of first-best expected total surplus at every alternative value.

\begin{definition}[Uniform profit guarantee]
\label{def:uniform-profit-guarantee}
An affine contract $w=(\alpha,\beta)$ with $\alpha,\beta\geq0$ provides a uniform profit guarantee $\lambda\in[0,1]$ if
$$U_P(\theta_1;w)\geq\lambda\,\overline{U}_P(\theta_1)\quad\text{for every }\theta_1\in\mathbb{R}.$$
The optimal guarantee is
$$\lambda^*=\sup\left\{\lambda\in[0,1]:\ \text{some affine contract provides the guarantee }\lambda\right\},$$
and a contract is optimal if it provides $\lambda^*$.
\end{definition}

Two properties of a contract connect the guarantee criterion to experimentation incentives. A contract satisfies one of these rules if its condition holds at every $\theta_1\in\mathbb{R}$.

\textbf{Rule 1: Principal protection.} Whenever the agent experiments, the principal's expected payoff is nonnegative:
$$
V_A(\theta_1;w)\geq 0 \quad\implies\quad V_P(\theta_1;w)\geq 0 .
$$

\textbf{Rule 2: Efficient experimentation.} The agent experiments if and only if experimentation is socially efficient:
$$
V_A(\theta_1;w)\geq 0 \quad\iff\quad V_S(\theta_1)\geq 0 .
$$

We call a contract \emph{protected} if it satisfies Rule 1 and \emph{implementing} if it satisfies Rule 2. The following lemma relates these properties to the uniform profit guarantee.

\begin{lemma}[Profit guarantees and implementation]
\label{lem:guarantee-implementation}
A contract provides the guarantee $\lambda=0$ if and only if it satisfies Rule 1. Under the maintained baseline assumptions, every contract that provides a strictly positive uniform guarantee also satisfies Rule 2.
\end{lemma}

Providing any guarantee, even $\lambda=0$, therefore requires principal protection: a contract that provides a guarantee gives the principal a nonnegative expected payoff at every alternative value, including those at which the full-information benchmark is zero. Stating the guarantee as an inequality rather than as a ratio imposes this requirement at those values without dividing by a zero benchmark. A strictly positive guarantee additionally requires efficient experimentation.

For a protected contract $w$, the largest guarantee it provides is
$$\lambda(w)=\inf_{\theta_1:\,\overline{U}_P(\theta_1)>0}\frac{U_P(\theta_1;w)}{\overline{U}_P(\theta_1)}.$$
We define $\lambda(w)$ only for protected contracts. An unprotected contract fails the guarantee inequality for every $\lambda\in[0,1]$, and a ratio evaluated only where the benchmark is positive would not impose protection where it is zero. By Lemma~\ref{lem:guarantee-implementation}, $\lambda^*$ is the supremum of $\lambda(w)$ over protected contracts. If $w$ is also implementing, the agent experiments wherever $V_S>0$, so
$$\lambda(w)=1-\sup_{\theta_1:\,V_S(\theta_1)>0}\frac{V_A(\theta_1;w)}{V_S(\theta_1)}.$$
Among protected implementing contracts, maximizing the guarantee is therefore equivalent to minimizing the worst-case share of total surplus left to the agent.

A protected contract can have guarantee zero for two reasons. It may fail to induce experimentation at some alternative value with strictly positive surplus. Alternatively, it may implement efficient experimentation but leave the principal a retained fraction whose infimum is zero over the positive-surplus region. In either case, a zero guarantee can coexist with strictly positive profits at many alternative values. When $\lambda^*=0$, every protected contract is optimal under Definition~\ref{def:uniform-profit-guarantee}, even though these contracts may differ in the profitable experiments they induce and the profits they generate. We return to these differences after characterizing the optimal guarantee.

\section{Robust Optimal Contracts}
\label{sec:robust-affine-contracts}
We now characterize the contracts that maximize the uniform profit guarantee. By Lemma~\ref{lem:guarantee-implementation}, a strictly positive guarantee requires efficient experimentation, so the efficient cutoffs of Section~\ref{subsec:socially-efficient-experimentation} restrict the contracts that can provide one. Under a pessimistic default, the single efficient cutoff leaves a family of implementing contracts that differ in the guarantees they provide. Under an optimistic default, at most one contract can provide a positive guarantee, and the two efficient cutoffs identify it. We first solve these unrestricted problems. When the optimal guarantee is zero, we use pointwise profit comparisons to distinguish some of the remaining contracts. We then examine how the optimal guarantee changes when the principal requires coverage only for experiments whose expected total surplus is at least a specified profitability threshold.

\subsection{Pessimistic Default: One-Cutoff Implementation}
\label{subsec:pessimistic-default}

Consider first the case $\theta_0\leq\overline\theta_0$. Experimentation is socially efficient if and only if $\theta_1\geq\theta_s$, where $\theta_s\geq0$ is the unique efficient cutoff. Define
$$\Phi_s=\Phi\left(\frac{\theta_s-\Delta}{s}\right),\quad
\Phi_0=\Phi\left(\frac{\theta_0-\Delta}{s}\right),\quad
\phi_s=\frac{1}{s}\phi\left(\frac{\theta_s-\Delta}{s}\right),$$
where $\phi$ is the standard normal density.

Any contract providing a strictly positive uniform guarantee must induce experimentation exactly when $\theta_1\geq\theta_s$ and hence, by continuity, make the agent indifferent at $\theta_s$. Since $\theta_s\geq0$, the liability floor does not bind in the alternative state there, and the indifference condition is
\begin{equation}\label{eq:pessimistic-rule2}
V_A(\theta_s;w) = p\Phi_s\cdot(\beta + \alpha\theta_s)  + (1-p)\Phi_0\cdot\left(\beta + \max\{\alpha\theta_0, -r\}\right) - c =0.
\end{equation}
The following proposition reduces the uniform guarantee requirement to this cutoff condition and a single inequality in the contract coefficients.

\begin{proposition}[Uniform profit guarantees under a pessimistic default]
\label{prop:rule1}
Suppose $\theta_0\leq\overline\theta_0$. For any $\lambda\in(0,1]$, an affine contract $w=(\alpha,\beta)$ provides the uniform profit guarantee $\lambda$ if and only if it satisfies~\eqref{eq:pessimistic-rule2} and
\begin{equation}
\label{eq:pessimistic-positive-guarantee}
\alpha+\frac{\beta\phi_s}{\Phi_s+\theta_s\phi_s}\leq1-\lambda.
\end{equation}
For every contract satisfying~\eqref{eq:pessimistic-rule2}, the ratio $V_P(\theta_1;w)/V_S(\theta_1)$ is nondecreasing on $(\theta_s,\infty)$, strictly increasing if $\beta>0$, and
$$
\inf_{\theta_1>\theta_s}\frac{V_P(\theta_1;w)}{V_S(\theta_1)}=\frac{\left.\frac{\partial V_P(\theta_1;w)}{\partial\theta_1}\right|_{\theta_1=\theta_s}}{\left.\frac{\partial V_S(\theta_1)}{\partial\theta_1}\right|_{\theta_1=\theta_s}}=1-\alpha-\frac{\beta\phi_s}{\Phi_s+\theta_s\phi_s}.
$$
\end{proposition}

The principal's least favorable relative payoff is approached as the alternative value tends to the efficient cutoff from above. Since the surplus slope at the cutoff does not depend on the contract, maximizing the guaranteed fraction reduces to minimizing the agent's payoff slope subject to~\eqref{eq:pessimistic-rule2}:
$$
\left.\frac{\partial V_A(\theta_1;w)}{\partial\theta_1}\right|_{\theta_1=\theta_s}
=p\phi_s(\beta+\alpha\theta_s)+p\alpha\Phi_s.
$$
Theorem~\ref{thm:pessimistic-default} establishes that the resulting guarantee is strictly positive, so by Proposition~\ref{prop:rule1} this reduction solves the original optimization over the entire affine class.

An optimum is attained by at least one of two single-instrument contracts: the pure launch bonus
$$
(\alpha_B,\beta_B)=\left(0,\frac{c}{p\Phi_s+(1-p)\Phi_0}\right),
$$
or proportional sharing, introduced in Section~\ref{subsec:discussion-of-assumptions},
$$
(\alpha_V,\beta_V)=\left(\frac{c}{c+\kappa},0\right).
$$
Their marginal agent payoffs at the efficient cutoff are
$$
M_B = p\phi_s\beta_B, \quad M_V = p(\phi_s\theta_s + \Phi_s)\alpha_V.
$$

\begin{theorem}[Optimal profit guarantees under a pessimistic default]
\label{thm:pessimistic-default}
Suppose $\theta_0\leq\overline\theta_0$. A strictly positive uniform guarantee is attainable, and the optimal contracts are independent of $r$.

If $\theta_s\geq\theta_0$, the pure launch bonus $(\alpha_B,\beta_B)$ is uniquely optimal. If $\theta_s<\theta_0$, the unique optimum is $(\alpha_B,\beta_B)$ when $M_B<M_V$ and $(\alpha_V,\beta_V)$ when $M_B>M_V$; when $M_B=M_V$, the optimal contracts are exactly those on the line segment joining $(\alpha_B,\beta_B)$ and $(\alpha_V,\beta_V)$. The optimal guarantee is $\lambda^*=1-\beta_B\phi_s/\left(\Phi_s+\theta_s\phi_s\right)$ when the pure launch bonus is optimal, and $\lambda^*=\kappa/(c+\kappa)$ when proportional sharing is optimal.
\end{theorem}

The two instruments differ in how the agent's rents grow once participation is secured at the efficient cutoff. As the alternative value rises, a launch bonus raises expected compensation only through the increasing probability of launch. Value sharing also raises compensation conditional on an alternative-state launch. The theorem compares these effects with the growth of total surplus.

The case $\theta_s\geq\theta_0$ includes $\theta_0=0$, since $\theta_s\geq0$. That default is reasonable in many practical settings, as supported by the empirical evidence in \cite{berman2022false}. With a zero default, the pure launch bonus is uniquely optimal, in contrast to the conventional wisdom that value-sharing contracts are the preferred incentive instrument.

When $\theta_s<\theta_0$, the floor binds at neither $\theta_s$ nor $\theta_0$, and the contracts satisfying~\eqref{eq:pessimistic-rule2} form the line segment joining the pure launch bonus and proportional sharing. The comparison between $M_B$ and $M_V$ determines which endpoint provides the larger guarantee.

When $\theta_s<\theta_0$ and $M_B=M_V$, every contract on the segment provides the same optimal guarantee, so the guarantee criterion does not distinguish among them. A comparison of their profits at each alternative value nonetheless yields a sharper conclusion.

\begin{definition}[Pointwise dominance]
A contract $\tilde w$ \emph{dominates} a contract $w$ if
$$
U_P(\theta_1;\tilde w)\geq U_P(\theta_1;w)
\quad\text{for every }\theta_1\in\mathbb R,
$$
with strict inequality at some alternative value. A contract is \emph{dominated} if some contract dominates it, and \emph{undominated} otherwise.
\end{definition}

Any contract that dominates an optimal contract also provides the optimal guarantee, so pointwise dominance sharpens the conclusion only within the optimal set.\footnote{See \citet{borgers2025undominated} for the use of undominated mechanisms to distinguish mechanisms that are optimal under other robust design criteria.} In this equality case, the pure launch bonus gives the principal strictly greater expected profit at every $\theta_1>\theta_s$ than any other contract on the optimal segment, while all contracts on the segment yield zero profit at $\theta_1\leq\theta_s$. The pure launch bonus is therefore the unique undominated optimal contract (Proposition~\ref{prop:pessimistic-equality-selection}).

The liability cap does not affect these conclusions. Under the pure launch bonus, $\alpha_B=0$, so the floor never enters compensation. Whenever proportional sharing or a mixture is optimal, $\theta_0>\theta_s\geq0$, and the floor binds neither at the default value nor anywhere in the efficient experimentation region. The optimal coefficients and guaranteed fractions therefore do not depend on $r$.

\subsection{Optimistic Default: Two-Cutoff Implementation}
\label{subsec:optimistic-default}

Now suppose $\theta_0>\overline\theta_0$. Experimentation is socially efficient if and only if
$$
\theta_1\in(-\infty,\theta_{s1}]\cup[\theta_{s2},\infty),
\quad\theta_{s1}<\theta_{s2}<0.
$$
A strictly positive uniform guarantee requires the contract to induce experimentation on both regions and to exclude the intervening alternative values, where surplus is negative. Continuity therefore requires indifference at both efficient cutoffs. Define
$$\Phi_{si}=\Phi\left(\frac{\theta_{si}-\Delta}{s}\right),\quad\phi_{si}=\frac{1}{s}\phi\left(\frac{\theta_{si}-\Delta}{s}\right),\quad i=1,2.
$$
Since $\theta_0>\overline\theta_0>0$, the floor does not bind in the default state, and the two necessary conditions are
\begin{align}
&p\Phi_{s1}\left(\beta+\max\{\alpha\theta_{s1},-r\}\right)
 +(1-p)\Phi_0(\beta+\alpha\theta_0)=c,
\label{eq:optimistic-lower-cutoff}\\
&p\Phi_{s2}\left(\beta+\max\{\alpha\theta_{s2},-r\}\right)
 +(1-p)\Phi_0(\beta+\alpha\theta_0)=c.
\label{eq:optimistic-upper-cutoff}
\end{align}
In contrast to the pessimistic case, the two cutoff conditions determine a unique candidate rather than a family. Whether that candidate delivers a positive guarantee depends on the liability cap and on the rents it leaves to the agent.

Define the two cap thresholds
$$
\underline r=-\frac{c\theta_{s2}}{(1-p)(\theta_0-\theta_{s2})\Phi_0},\quad\overline r=-\frac{c\theta_{s1}}{c+\kappa},
$$
which satisfy $0<\underline r<\overline r$.

\begin{theorem}[Optimal profit guarantees under an optimistic default]
\label{thm:optimistic-default}
Suppose $\theta_0>\overline\theta_0$. The cutoff equations~\eqref{eq:optimistic-lower-cutoff} and~\eqref{eq:optimistic-upper-cutoff} have a unique admissible solution, that is, a unique solution $w^*=(\alpha^*,\beta^*)$ with $\alpha^*,\beta^*\geq0$.

\begin{itemize}
\item \emph{Tight cap} ($r\leq\underline r$): no contract implements efficient experimentation, and $\lambda^*=0$.

\item \emph{Intermediate cap} ($\underline r<r<\overline r$): $w^*$ is the unique implementing contract and uses both instruments, with $\alpha^*>0$ and $0<\beta^*<r$.\footnote{In this regime, the cutoff equations reduce to the linear system \eqref{eq:optimistic-intermediate}, whose unique solution is characterized in Appendix~\ref{app:optimistic-default}.}

\item \emph{Loose cap} ($r\geq\overline r$): proportional sharing $w^*=(\alpha_V,0)$ is the unique implementing contract.
\end{itemize}

For $r>\underline r$, the optimal guarantee is
$$\lambda^*=\max\left\{0,\,1-\sup_{\theta_1:\,V_S(\theta_1)>0}\frac{V_A(\theta_1;w^*)}{V_S(\theta_1)}\right\}.$$
If $\lambda^*>0$, $w^*$ is uniquely optimal. If $\lambda^*=0$, the optimal contracts are exactly the protected contracts.
\end{theorem}

The positive-guarantee condition has an explicit expression in each implementing regime (Proposition~\ref{prop:optimistic-explicit-conditions}). Under an intermediate cap, $\lambda^*>0$ if and only if
\begin{equation}
\label{eq:optimistic-intermediate-guarantee}
\max\left\{
\frac{r-\beta^*}{-\theta_{s1}-{\Phi_{s1}}/{\phi_{s1}}},\,\alpha^*+\frac{\beta^*}{\theta_{s2}+{\Phi_{s2}}/{\phi_{s2}}}\right\}<1.
\end{equation}
Under a loose cap, $\lambda^*>0$ if and only if
\begin{equation}
\label{eq:optimistic-loose-guarantee}
\sup_{\theta_1<-r/\alpha_V}\frac{p\Phi\left(\frac{\theta_1-\Delta}{s}\right)(-r-\alpha_V\theta_1)}{V_S(\theta_1)}<\frac{\kappa}{c+\kappa}.
\end{equation}

Under a tight cap, matching the two efficient cutoffs forces $\beta^*=r$. The net payment following an alternative-state launch is then zero wherever the floor binds. Because $\alpha^*\theta_{s2}\leq-r$, the floor binds on the whole inefficient interval $(\theta_{s1},\theta_{s2})$. The cutoff equations therefore leave the agent indifferent there, and the tie-breaking rule induces experimentation. Since every implementing contract must solve the cutoff equations, no contract implements efficient experimentation, and none provides a strictly positive uniform guarantee, even though the cutoff equations have a solution.

Under an intermediate cap, the floor binds at the lower efficient cutoff but not at the upper one. Relative to proportional sharing, $w^*$ lowers the value share and introduces a positive launch bonus, and together the two instruments match both cutoffs. The two terms in~\eqref{eq:optimistic-intermediate-guarantee} are the agent's worst-case shares of total surplus over the entire lower and upper efficient regions, respectively. Because the agent's share falls as the alternative value moves away from either cutoff into the efficient region, each worst case is approached at the corresponding cutoff. A strictly positive uniform guarantee therefore requires both terms to be strictly less than one. Since every contract providing a strictly positive guarantee must equal $w^*$, and $\beta^*>0$, the launch bonus is indispensable whenever such a guarantee is attainable in this regime.

Under a loose cap, proportional sharing matches both efficient cutoffs. The floor can nevertheless bind at more adverse alternative values within the lower efficient region. The supremum in~\eqref{eq:optimistic-loose-guarantee} measures the additional rent created where the floor binds, relative to total surplus. Because it is taken over the floor-binding region, the agent's worst-case share can occur inside the lower efficient region rather than at its boundary. A cap that is loose for implementation is thus not necessarily irrelevant for profit retention. As $r$ increases, the supremum falls toward zero, so the condition holds for all sufficiently large caps and the optimal guarantee approaches $\kappa/(c+\kappa)$.

If the relevant condition, \eqref{eq:optimistic-intermediate-guarantee} or \eqref{eq:optimistic-loose-guarantee}, fails, the principal's least favorable relative payoff under $w^*$ is nonpositive. When it is negative, $w^*$ gives the principal a negative payoff at some efficient alternative value and provides no guarantee at all. In either case, $\lambda^*=0$, and the optimal contracts are exactly the protected contracts. The counterexamples in Appendix~\ref{app:rule1-counterexample} illustrate both an upper-region loss caused by the required launch bonus and a lower-region loss caused by the liability floor. The null contract $w=(0,0)$, under which the agent never experiments, is optimal whenever $\lambda^*=0$, but, as we show below, it is dominated.

\paragraph{Small agent costs.}
The intermediate and loose cap regimes do not by themselves ensure a strictly positive uniform guarantee. The next theorem shows that the relevant condition, \eqref{eq:optimistic-intermediate-guarantee} or \eqref{eq:optimistic-loose-guarantee}, holds for every cap above $\underline r$ when the agent bears a sufficiently small share of the total experimentation cost, and that the optimal guarantee tends to one, uniformly over these caps, as that share shrinks.

\begin{theorem}[Near-benchmark guarantees when the agent's cost is small]
\label{thm:small-agent-cost}
Suppose $\theta_0>\overline\theta_0$. Fix $\Delta$, $p$, $\theta_0$, $s$, and total experimentation cost $c+\kappa$, and suppose that the minimum of $V_S$ is strictly negative. For every $\lambda\in(0,1)$, there exists $\overline c_\lambda=\overline c_\lambda(\Delta,p,\theta_0,s,c+\kappa)>0$, independent of $r$, such that whenever $0<c<\overline c_\lambda$ and $r>\underline r$, the contract $w^*$ of Theorem~\ref{thm:optimistic-default} provides the uniform profit guarantee $\lambda$. In particular, $w^*$ is then uniquely optimal.
\end{theorem}

When the organization bears most of the total experimentation cost and the agent bears only a small residual effort or opportunity cost, the optimal contract thus guarantees the principal nearly all of first-best expected total surplus. Because $\overline c_\lambda$ does not depend on $r$, this bound holds uniformly across all caps above $\underline r$, which admit efficient experimentation.

Small agent costs do not by themselves eliminate the launch bonus. Both cap thresholds are proportional to $c$, so whether the cap is intermediate or loose depends on the cap relative to the agent's cost, $r/c$, rather than on $c$ alone. Under an intermediate cap, $w^*$ uses a positive launch bonus, and this bonus is necessary for any strictly positive uniform guarantee, however close to one that guarantee is.

\paragraph{Pointwise comparisons when the optimal guarantee is zero.}
When $\lambda^*=0$, pointwise comparisons can identify dominated optimal contracts, but they do not in general select a unique one. The following family makes the remaining multiplicity explicit.

\begin{proposition}[Undominated zero-guarantee launch bonuses]
\label{prop:zero-guarantee-bonus-family}
Suppose $\theta_0>\overline\theta_0$. For every $\theta_a\geq\theta_0$, define
$$
\beta_a=\frac{c}{p\Phi\left(\frac{\theta_a-\Delta}{s}\right)+(1-p)\Phi_0},
\quad
w_a=(0,\beta_a).
$$
Each $w_a$ is protected, has $\lambda(w_a)=0$, and is undominated. Consequently, if $\lambda^*=0$, there is a continuum of undominated optimal contracts.
\end{proposition}

Each $w_a$ induces experimentation exactly on $[\theta_a,\infty)$, gives the principal strictly positive profit there, and therefore dominates the null contract. For two cutoffs $\theta_0\leq\theta_a<\theta_b$, the contract $w_a$ gives the principal higher profit on $[\theta_a,\theta_b)$, where $w_b$ induces no experimentation, whereas $w_b$ gives higher profit on $[\theta_b,\infty)$, because inducing experimentation from a lower cutoff requires a larger launch bonus. Both give zero below $\theta_a$. We do not characterize the full set of undominated contracts.

Choosing among the contracts in this family amounts to choosing the cutoff $\theta_a$ above which the agent experiments. When $\lambda^*=0$, neither the guarantee criterion nor pointwise dominance pins down this cutoff. We therefore turn to a version of the guarantee that applies only at alternative values where expected total surplus is at least a given threshold but still requires principal protection at every alternative value.

\subsection{Profit Guarantees Above a Profitability Threshold}
\label{subsec:profitability-threshold}

The unrestricted guarantee evaluates the principal's profit fraction at every alternative value with positive expected total surplus, including values at which that surplus is arbitrarily close to zero. A platform may care less about the fraction it retains on such marginal features than about the fraction it retains on features that generate substantial surplus, while still requiring that it never incur an expected loss on an experiment the agent chooses to run. We therefore ask how optimal incentives change when the profit-ratio guarantee is required only for experiments whose expected total surplus is at least a specified threshold $\eta > 0$. This restriction can distinguish contracts when the unrestricted guarantee is zero, and it can also change the optimal contract when that guarantee is positive. We maintain the optimistic-default setting of Section~\ref{subsec:optimistic-default}, including the strictly negative minimum of $V_S$.

For $\eta>0$, let
$$
\Theta_\eta=\{\theta_1:V_S(\theta_1)\geq\eta\}
$$
denote the set of alternative values whose expected total surplus is at least $\eta$. Let $\mathcal W_P$ denote the set of protected contracts, and for $w\in\mathcal W_P$ define
$$
\lambda_\eta(w)=\inf_{\theta_1\in\Theta_\eta}\frac{U_P(\theta_1;w)}{V_S(\theta_1)},
\quad
\lambda_\eta^*=\sup_{w\in\mathcal W_P}\lambda_\eta(w).
$$
The set $\Theta_\eta$ is nonempty, because $V_S(\theta_1)\to\infty$ as $\theta_1\to\infty$. Protection remains required at every alternative value; the threshold restricts only where the profit fraction must be guaranteed. Each threshold defines a separate optimization over the same class of protected contracts. Because a higher threshold shrinks $\Theta_\eta$, $\lambda_\eta^*$ is nondecreasing in $\eta$. As $\eta\downarrow0$, $\Theta_\eta$ increases to the set of alternative values with positive surplus, so $\lambda_\eta(w)$ decreases to the unrestricted guarantee $\lambda(w)$ for each protected contract $w$. Proposition~\ref{prop:threshold-guarantee-frontier} shows that, for caps above $\underline r$, the optimal guarantees converge as well.

\paragraph{The value of a launch bonus.}
Our first result applies at every threshold for which a positive guarantee is attainable.

\begin{proposition}[Strict value of launch bonuses]
\label{prop:eta-strict-bonus}
Fix $\eta>0$ and suppose $\lambda_\eta^*>0$. For every protected pure-sharing contract $w=(\alpha,0)$, there is a protected contract $\tilde w=(\tilde\alpha,\tilde\beta)$ with $\tilde\alpha,\tilde\beta>0$ such that $\lambda_\eta(\tilde w)>\lambda_\eta(w)$. Consequently, every contract attaining $\lambda_\eta^*$ has a strictly positive launch bonus.
\end{proposition}

A small launch bonus can replace part of the value share while preserving experimentation throughout $\Theta_\eta$, and the lower share reduces the rents the agent earns on highly valuable experiments. Appendix~\ref{app:profitability-threshold} shows that this substitution strictly lowers the supremum of the agent's share of surplus over $\Theta_\eta$, accounting for the liability floor and for protection outside $\Theta_\eta$. Pure sharing can still provide a positive guarantee, but whenever $\lambda_\eta^*>0$ it cannot be optimal.

\paragraph{Connection with the unrestricted guarantee.}
The premise $\lambda_\eta^*>0$ of Proposition~\ref{prop:eta-strict-bonus} is not automatic. For caps that admit efficient experimentation, whether an arbitrarily small threshold yields a positive guarantee depends on the protection status of the implementing contract in the unrestricted problem.

\begin{proposition}[Convergence and positivity of threshold guarantees]
\label{prop:threshold-guarantee-frontier}
Suppose $r>\underline r$. For every $\eta>0$, some contract attains $\lambda_\eta^*$. Moreover, $\lim_{\eta\downarrow0}\lambda_\eta^*=\lambda^*$, and $\lambda_\eta^*>0$ for every $\eta>0$ if and only if the implementing contract $w^*$ in Theorem~\ref{thm:optimistic-default} is protected.
\end{proposition}

When the unrestricted optimal guarantee is zero, the proposition distinguishes two reasons for it. If $w^*$ is protected, every threshold permits a positive guarantee, although $\lambda_\eta^*$ tends to zero as $\eta\downarrow0$. At each threshold, a small reduction in compensation from $w^*$ can preserve experimentation throughout $\Theta_\eta$ while improving profit retention. If $w^*$ is unprotected, there exists $\eta_0>0$ such that $\lambda_\eta^*=0$ for every $0<\eta<\eta_0$. Excluding only experiments with arbitrarily small positive surplus then does not suffice to yield a positive guarantee.

\paragraph{Mixed contracts and pure launch bonuses.}
Proposition~\ref{prop:eta-strict-bonus} rules out pure sharing whenever the optimal threshold guarantee is positive. The remaining distinction depends on whether $\Theta_\eta$ includes part of the lower efficient region. If it does, any contract providing a positive guarantee must also use a positive value share, so every optimum is mixed. If it does not, a pure launch bonus can instead be optimal. The next proposition characterizes this case.

\begin{proposition}[Optimal incentives in the upper region]
\label{prop:bonus-threshold-guarantee}
Suppose $\Theta_\eta=[\theta_\eta,\infty)$ for some $\theta_\eta\geq0$. Then $\lambda_\eta^*>0$, and exactly one contract $w_\eta^*$ attains it. Both the contract and its guarantee are independent of $r$. If $\theta_\eta\geq\theta_0$, this contract is the pure launch bonus
$$w_\eta^*=\left(0,\,\frac{c}{p\Phi\left(\frac{\theta_\eta-\Delta}{s}\right)+(1-p)\Phi_0}\right).$$
\end{proposition}

The pure-bonus conclusion connects directly to Proposition~\ref{prop:zero-guarantee-bonus-family}. For every $\theta_a\geq\theta_0$, setting $\eta=V_S(\theta_a)$ gives $\Theta_\eta=[\theta_a,\infty)$, and the contract $w_a$ in that proposition is the unique optimum. Thus every member of the undominated pure-bonus family is uniquely optimal for a corresponding profitability threshold.

For a given threshold with $\theta_\eta\geq\theta_0$, the selected member has $\theta_a=\theta_\eta$. When $\lambda^*=0$, the threshold criterion therefore selects a unique contract from the continuum of undominated optimal contracts identified at the end of Section~\ref{subsec:optimistic-default}. This selection comes from the threshold; under pointwise dominance, the contracts in the family remain pairwise incomparable.

The condition $\theta_\eta\geq\theta_0$ is sufficient for pure-bonus optimality. In this case, every alternative value in $\Theta_\eta$ is at least the known default value, and among contracts that make the agent indifferent at $\theta_\eta$, the pure launch bonus minimizes the expected payment at each of these values. When $\theta_\eta<\theta_0$, the unique optimum may instead use both instruments. The optimal contract is fully characterized throughout the upper-region case (Proposition~\ref{prop:upper-region-classification}).

\section{Profit Guarantees Under Distributional Uncertainty}
\label{sec:distributional-uncertainty}

The baseline model assumes that the principal knows the default value $\theta_0$ and the probability $p$ but not the alternative value $\theta_1$. She therefore offers a single affine contract for all alternative values and evaluates it by the fraction of the full-information benchmark that it guarantees uniformly across them. We now allow the agent's private knowledge to be a general distribution of feature quality rather than a single alternative value. Realized quality is still denoted by $\theta$, and its privately known distribution by $\mu$. The principal has no prior over $\mu$. She knows only that quality has a finite mean and is nonnegative with probability at least $1-p$, where $p\in(0,1)$ is the baseline parameter. Let $\mathcal M$ be the set of Borel probability measures $\mu$ on $\mathbb R$ satisfying
$$
\int_{\mathbb R}|\theta|\,\mu(d\theta)<\infty,\quad\mu([0,\infty))\geq1-p.
$$
We impose no common bound on the supports or means of distributions in $\mathcal M$.

For any fixed $\theta_0\geq0$, $\mathcal M$ contains the baseline distribution $(1-p)\delta_{\theta_0}+p\delta_{\theta_1}$ for every $\theta_1\in\mathbb R$, where $\delta_x$ is the point mass at $x$. As in the baseline with a nonnegative default value, the feature is harmful with probability at most $p$, yet a harmful feature may pass the noisy launch rule by chance. After developing the feature, the agent observes $\mu$ and then decides whether to experiment; neither party initially observes realized quality.

The rest of the model, including the experimentation technology, launch rule, costs, and contract class with its finite liability cap $r>0$, is unchanged. Conditional on experimentation, total surplus and the agent's expected payoff are
\begin{align*}
V_S(\mu)&=\int_{\mathbb R}\theta\Phi\left(\frac{\theta-\Delta}{s}\right)\,\mu(d\theta)-c-\kappa,\\
V_A(\mu;w)&=\int_{\mathbb R}\Phi\left(\frac{\theta-\Delta}{s}\right)
 \left(\beta+\max\{\alpha\theta,-r\}\right)\,\mu(d\theta)-c.
\end{align*}
As before, $V_P=V_S-V_A$, the agent experiments if and only if $V_A\geq0$, and $U_P=V_P\mathbb I\{V_A\geq0\}$. The full-information benchmark remains $\overline U_P(\mu)=\max\{V_S(\mu),0\}$. Let
$$
\mathcal W_P=\left\{w=(\alpha,\beta)\in[0,\infty)^2:U_P(\mu;w)\geq0\text{ for every }\mu\in\mathcal M\right\}
$$
now denote the set of protected contracts.

\paragraph{Unrestricted impossibility.}
We extend Definition~\ref{def:uniform-profit-guarantee} to the distributional environment. The principal again chooses one contract without access to the agent's private knowledge, but now requires the guarantee to hold across all distributions in $\mathcal M$.

\begingroup
\renewcommand{\thedefinition}{1$'$}
\begin{definition}[Uniform profit guarantee under distributional uncertainty]
An affine contract $w=(\alpha,\beta)$ with $\alpha,\beta\geq0$ provides a uniform profit guarantee $\lambda\in[0,1]$ if $$U_P(\mu;w)\geq\lambda\,\overline U_P(\mu)\quad\text{for every }\mu\in\mathcal M.$$ The optimal guarantee is $$\lambda^*=\sup\left\{\lambda\in[0,1]:\ \text{some affine contract provides the guarantee }\lambda\right\},$$ and a contract is optimal if it provides $\lambda^*$.
\end{definition}
\endgroup
\addtocounter{definition}{-1}

As in the baseline, the guarantee applies even where the full-information benchmark is zero. A zero guarantee is therefore equivalent to principal protection, while a positive guarantee additionally requires experimentation at every distribution with positive total surplus. For a protected contract $w\in\mathcal W_P$, let $\lambda(w)$ denote the largest fraction it guarantees:
$$
\lambda(w)=\inf_{\substack{\mu\in\mathcal M\\V_S(\mu)>0}}\frac{U_P(\mu;w)}{V_S(\mu)},
$$
so $\lambda^*=\sup_{w\in\mathcal W_P}\lambda(w)$. The null contract, under which the agent never experiments, provides a zero guarantee. The following result shows that no affine contract provides a positive one.

\begin{proposition}[No positive unrestricted guarantee]
\label{prop:distributional-unrestricted-impossibility}
No affine contract implements efficient experimentation throughout $\mathcal M$, and $\lambda^*=0$.
\end{proposition}

A contract with a positive guarantee must induce every positive-surplus experiment, while protection rules out every negative-surplus one. Continuity therefore requires the agent to be indifferent at every break-even distribution. For distributions supported on nonnegative qualities, the liability floor never binds. Such distributions can have the same expected launched value but different launch probabilities. The value share $\alpha$ then yields the same expected payment, whereas a positive launch bonus $\beta$ does not. Indifference at every break-even distribution thus forces $\beta=0$ and, since expected launched value at break-even equals $c+\kappa$, $\alpha=c/(c+\kappa)$.

Without a liability cap, this proportional-sharing contract would guarantee the principal the fraction $\kappa/(c+\kappa)$ throughout $\mathcal M$. A finite cap breaks this proportionality: the agent retains his share of gains but bears less than his share of sufficiently large losses. Break-even distributions with such losses then give him a positive rent, leaving the principal with a loss. Adjusting compensation cannot remove this rent while preserving indifference at all other break-even distributions. Every protected contract must therefore forgo some positive-surplus experiments, making its unrestricted guarantee zero. Since efficient experimentation imposes the same indifference requirement, no affine contract implements it.\footnote{Neither conclusion of Proposition~\ref{prop:distributional-unrestricted-impossibility} requires the full set $\mathcal M$. Both already hold when $\mathcal M$ is restricted to two-point distributions of the baseline form with fixed, known probabilities. The default value $\theta_0$ can take any nonnegative value and the alternative value $\theta_1$ any nonpositive value; only these two support values are unknown to the principal.}

\paragraph{Positive mixed guarantees above a profitability threshold.}
Because the impossibility is driven by experiments with total surplus near zero, we follow Section~\ref{subsec:profitability-threshold} and impose the ratio requirement only on distributions whose total surplus is at least $\eta>0$. Let $\mathcal M_\eta=\{\mu\in\mathcal M:V_S(\mu)\geq\eta\}$, and for $w\in\mathcal W_P$ define
$$
\lambda_\eta(w)=\inf_{\mu\in\mathcal M_\eta} \frac{U_P(\mu;w)}{V_S(\mu)},
\quad
\lambda_\eta^*=\sup_{w\in\mathcal W_P}\lambda_\eta(w).
$$
Protection remains required at every distribution in $\mathcal M$; the threshold restricts only where the profit fraction must be guaranteed. The following result shows that a positive guarantee becomes attainable once the profitability threshold is sufficiently high, and that maximizing the guarantee then requires both instruments.

\begin{theorem}[Mixed contracts under distributional uncertainty]
\label{thm:distributional-mixed-guarantees}\ 
\begin{enumerate}
\item For all $\eta>0$, every protected pure launch bonus $w=(0,\beta)$ satisfies $\lambda_\eta(w)=0$.
\item There exists $\overline{\eta}>0$ such that for all $\eta>\overline{\eta}$, the optimal guarantee $\lambda_\eta^*$ is positive and attained, and every contract $w=(\alpha,\beta)$ attaining $\lambda_\eta^*$ has $\alpha>0$ and $\beta>0$.
\end{enumerate}
\end{theorem}

Under a pure launch bonus, the agent's expected payment depends only on how often the feature launches. A feature that is sometimes severely harmful and only rarely highly valuable can generate large expected surplus yet launch less often than a feature of zero quality, because the launch rule almost always rejects its harmful outcomes. Since protection requires the bonus to deter experimentation on the zero-quality feature, it also deters some highly valuable experiments. Under distributional uncertainty, value sharing is therefore necessary for any positive threshold guarantee, whereas in the baseline a pure launch bonus can be uniquely optimal.

The launch bonus nevertheless remains valuable in combination with value sharing. Above the threshold $\overline{\eta}$, replacing part of the value share with a launch bonus raises the guarantee and preserves both protection and participation. The bonus keeps the agent experimenting, while the smaller share lets the principal retain more of the gains from highly valuable outcomes. Value sharing is thus indispensable, but above this threshold it is never optimal on its own.\footnote{Neither part of Theorem~\ref{thm:distributional-mixed-guarantees} requires the full set $\mathcal M$. Both already hold when $\mathcal M$ is restricted to a three-outcome family. In the default state, the feature has no effect; in the alternative state, it either causes harm or produces an improvement, with fixed, known conditional probabilities under which improvements are sufficiently rare. Only the magnitudes of harm and improvement are unknown to the principal; each can take any nonnegative value.}

\section{Numerical Illustration}
\label{sec:numerical-illustration}
In this section, we illustrate our main findings from Section~\ref{sec:robust-affine-contracts} under a unified parameter setup. Throughout this section, we fix $p = 0.6$, total experimentation cost $c + \kappa = 0.75$, and $\Delta=1.96s$, so that a feature is launched if and only if the experiment reports a statistically significant positive effect in a two-sided test at the 5 percent level. Section~\ref{subsec:num-pessimistic} varies the default quality $\theta_0$ and noise level $s$ to map the pessimistic-default contract regions. Sections~\ref{subsec:num-optimistic} and~\ref{subsec:num-threshold} further fix $s = 7/8$, $\Delta = 1.715$, and $\theta_0 = 2.4$ in the optimistic-default regime, using an agent cost share $c/(c+\kappa) = 0.15$ for the contract-coefficient plot.

\subsection{Pessimistic Default}
\label{subsec:num-pessimistic}
\begin{figure}[!htbp]
    \centering
    \includegraphics[width=0.5\textwidth]{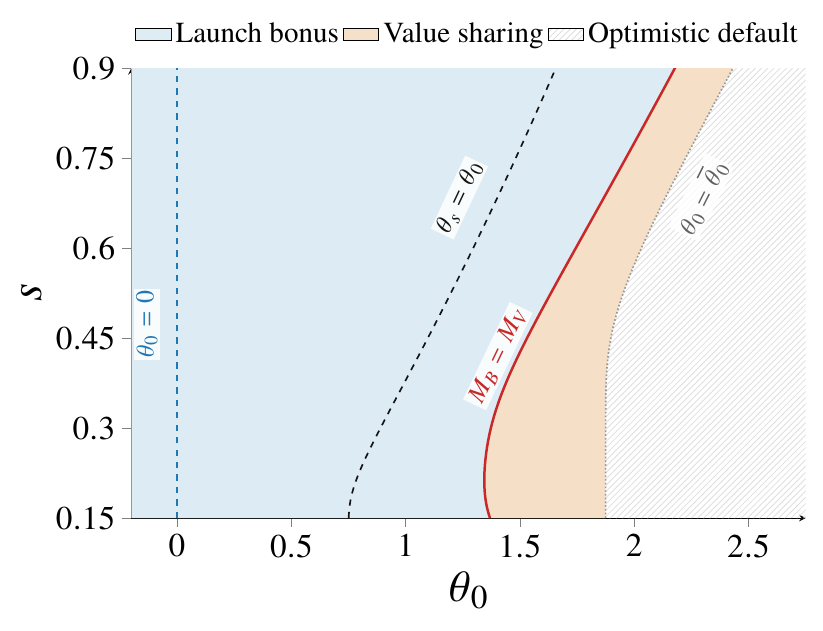}
    \caption{Optimal contract regions under a pessimistic default.}
    \label{fig:num-pessimistic-phase}
\end{figure}

Figure~\ref{fig:num-pessimistic-phase} illustrates Theorem~\ref{thm:pessimistic-default} across combinations of default quality $\theta_0$ and experimental noise $s$. Throughout the noise range, the zero default $\theta_0 = 0$ lies strictly inside the pure launch-bonus region. Pure value sharing is optimal only in a narrow band of high default values just below the optimistic-default boundary $\theta_0 = \overline{\theta}_0$.

\subsection{Liability Caps and Cost Shares Under an Optimistic Default}
\label{subsec:num-optimistic}
\begin{figure}[!htbp]
    \centering
    \begin{subfigure}[t]{0.49\textwidth}
        \centering
        \includegraphics[width=\linewidth]{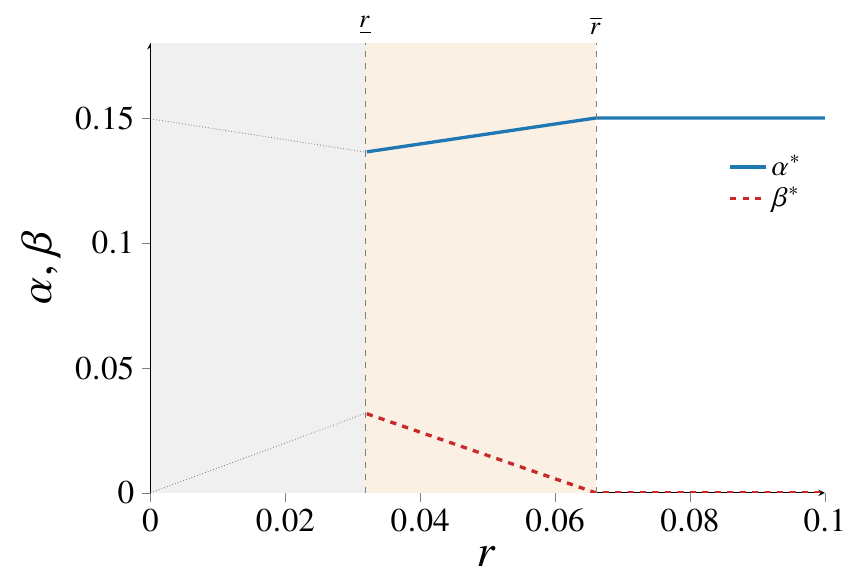}
        \caption{Contract coefficients}
    \end{subfigure}\hfill
    \begin{subfigure}[t]{0.49\textwidth}
        \centering
        \includegraphics[width=\linewidth]{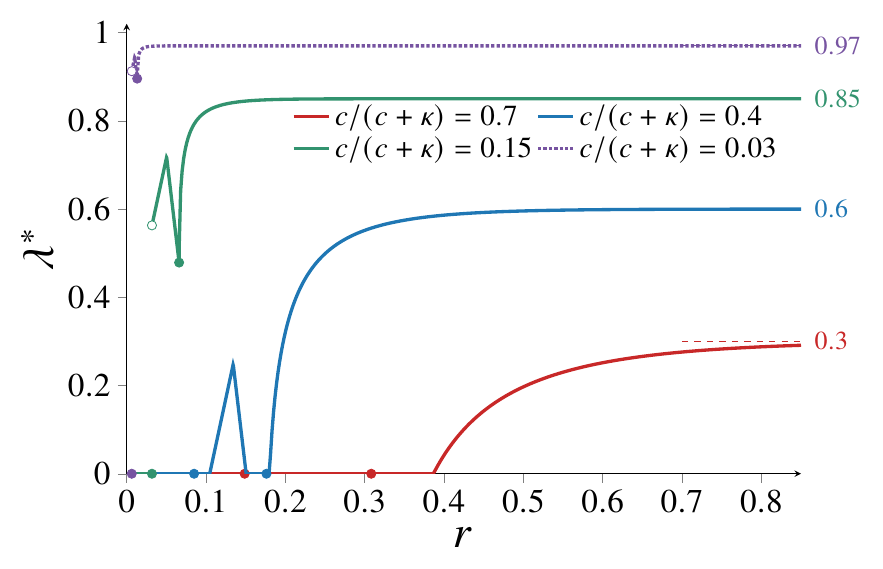}
        \caption{Profit guarantee across agent cost shares}
    \end{subfigure}
    \caption{Optimistic-default contracts and uniform profit guarantees as functions of the liability cap $r$. In panel~a, the gray curves in the tight-cap region show the solution of the cutoff equations, which does not implement efficient experimentation; for $r>\underline r$, the curves show the unique optimal contract.}
    \label{fig:num-optimistic-cap}
\end{figure}

Figure~\ref{fig:num-optimistic-cap} illustrates Theorems~\ref{thm:optimistic-default} and~\ref{thm:small-agent-cost} as the liability cap $r$ varies. In panel~a, no contract implements efficient experimentation under a tight cap, the intermediate-cap region smoothly substitutes value sharing for the launch bonus as $r$ relaxes, and the loose-cap region locks the contract at proportional sharing. Panel~b compares the optimal guarantee $\lambda^*$ across four agent cost shares. Raising $r$ into the intermediate or loose regime does not ensure a positive $\lambda^*$ when the agent bears a large share of costs, and $\lambda^*$ can rise and then fall within the intermediate cap. As the agent's cost share shrinks, however, $\lambda^*$ rises sharply across all caps above $\underline{r}$ and approaches one, confirming Theorem~\ref{thm:small-agent-cost}.

\subsection{Profit Guarantees Above a Profitability Threshold}
\label{subsec:num-threshold}
\begin{figure}[!htbp]
    \centering
    \begin{subfigure}[t]{0.49\textwidth}
        \centering
        \includegraphics[width=\linewidth]{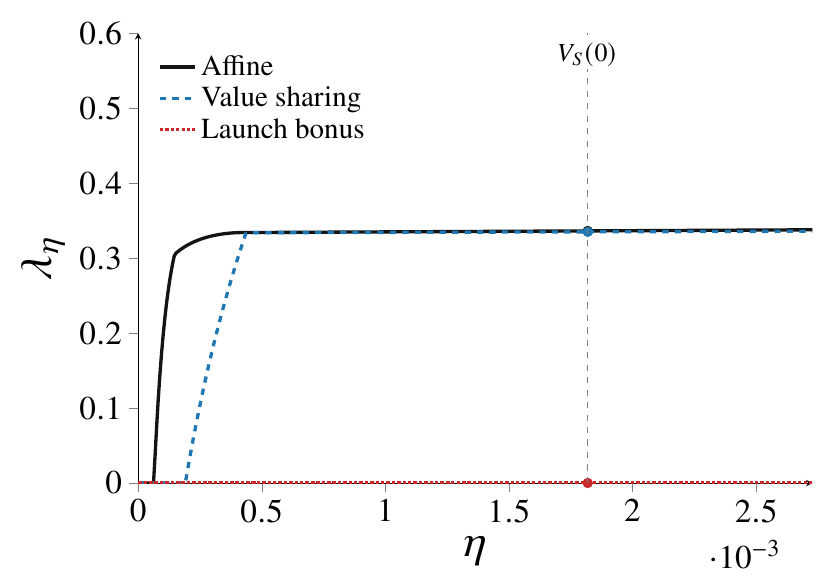}
        \caption{Low-threshold detail}
    \end{subfigure}\hfill
    \begin{subfigure}[t]{0.49\textwidth}
        \centering
        \includegraphics[width=\linewidth]{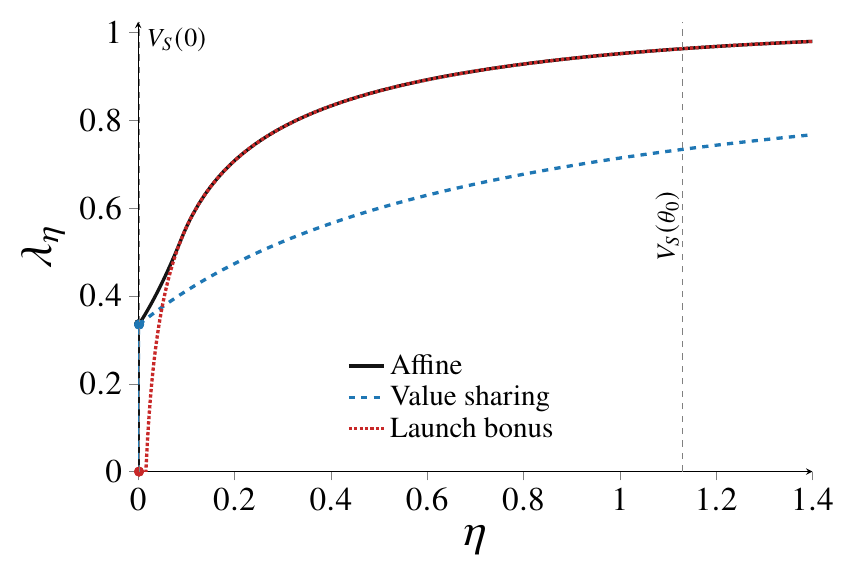}
        \caption{Full threshold range}
    \end{subfigure}
    \caption{Threshold profit guarantees $\lambda_\eta$ for optimal affine, pure value-sharing, and pure launch-bonus contracts.}
    \label{fig:num-threshold}
\end{figure}

We now illustrate Section~\ref{subsec:profitability-threshold} by setting $c = 1/2$, $\kappa = 1/4$, and $r = 0.3$. Under these parameters, the liability cap lies in the loose regime, yet the implementing contract fails principal protection and the unrestricted guarantee is $\lambda^* = 0$. Figure~\ref{fig:num-threshold} tracks how the profit guarantee $\lambda_\eta$ recovers as the principal restricts the ratio requirement to experiments with expected total surplus at least $\eta > 0$, comparing optimal affine, pure value-sharing, and pure launch-bonus contracts.

Panel~a zooms in on small thresholds around $\eta = V_S(0)$. For very small $\eta$, all three contract classes remain at zero, confirming Proposition~\ref{prop:threshold-guarantee-frontier}. Once $\eta$ rises slightly further, a mixed affine contract quickly restores a substantial guarantee and strictly outperforms pure value sharing, illustrating Proposition~\ref{prop:eta-strict-bonus}. In contrast, pure launch bonuses yield a zero guarantee across this entire low-threshold range. Panel~b shows the full threshold range. As $\eta$ increases well above $V_S(0)$, the guarantee is required only for experiments with high surplus, so a smaller launch bonus suffices without threatening principal protection. Pure launch bonuses then turn positive, overtake pure value sharing, and coincide with the optimal affine contract.

\section{Conclusion}
\label{sec:conclusion}
This paper shows that noisy experimentation and limited liability create a central tension in innovation incentives. Launch bonuses can motivate experimentation, but they also reward false positives; value sharing can discipline bad launches, but its effectiveness is limited by liability constraints. Whenever a positive uniform guarantee is attainable, every optimal affine contract induces socially efficient experimentation while protecting the principal from losses. In the baseline model, which instruments the optimal contract uses depends on the default value and the liability cap, as Table~\ref{tab:optimal-contract-summary} summarizes.

Two further results refine this picture. First, when the principal requires the profit-ratio guarantee only for experiments whose expected total surplus is at least a profitability threshold $\eta>0$, every optimal contract under an optimistic default uses a positive launch bonus whenever a positive guarantee is attainable, because replacing part of the value share with a launch bonus limits payouts on high-value launches; for sufficiently high thresholds, the unique optimal contract is a pure launch bonus. Second, when the agent privately knows a general quality distribution, no affine contract provides a positive unrestricted guarantee, but above a sufficiently high profitability threshold a positive guarantee is attainable, and neither a pure launch bonus nor pure value sharing is then optimal on its own.

Our results offer three managerial insights for firms that tie engineers' compensation to experimental launches.
\begin{itemize}
\item First, launch bonuses do not simply reward luck. With a zero default value ($\theta_0=0$), which the evidence in \citet{berman2022false} supports in many practical settings, Theorem~\ref{thm:pessimistic-default} shows that a pure launch bonus is uniquely optimal. Because stronger features are more likely to clear the launch threshold, a fixed bonus still rewards better ideas through a higher launch probability without scaling up the payout itself. By contrast, value sharing raises both the probability and the size of the payout, leaving the firm with a smaller share of the gains.
\item Second, when the default feature is already valuable enough to justify testing on its own, the main challenge shifts from encouraging experimentation to preventing over-experimentation. In this case, an engineer may test features that carry a moderate risk of harm simply because they can occasionally pass the launch rule by chance. Deterring such experiments requires exposing the engineer to part of the downside from harmful launches, making limited liability decisive, as indicated by Theorem~\ref{thm:optimistic-default}. 
\item Third, lowering the personal effort required to run experiments directly reduces incentive costs. Theorem~\ref{thm:small-agent-cost} shows that when the organization bears most of the experimentation cost and the engineer bears only a small share, the optimal contract lets the firm retain nearly all of first-best surplus. Consequently, investments that lower engineers' private cost of testing, such as self-serve experimentation platforms, not only streamline testing but also reduce the payouts required to motivate efficient experimentation.
\end{itemize}

\begin{table}[ht]
\centering
\footnotesize
\begin{tabular}{lccccc}
\toprule
\multirow{3}{*}{Setting}
& \multicolumn{2}{c}{Pessimistic default}
& \multicolumn{3}{c}{Optimistic default} \\
\cmidrule(lr){2-3}\cmidrule(lr){4-6}
& Low default
& High default
& Tight cap
& Intermediate cap
& Loose cap \\
& $\theta_s\ge \theta_0$
& $\theta_s<\theta_0$
& $r\le \underline{r}$
& $\underline{r}<r<\overline{r}$
& $r\ge \overline{r}$ \\
\midrule

\makecell[l]{Optimal \\ contract}

&
\makecell{
Pure launch bonus\\
$(0,\beta_B)$
}
&
\makecell{
$(0,\beta_B)$ if $M_B<M_V$\\
$(\alpha_V,0)$ if $M_B>M_V$
}
&
\makecell{
No positive\\
guarantee
}
&
\makecell{
Mixed contract\\
$(\alpha^*,\beta^*)$
}
&
\makecell{
Pure value sharing\\
$(\alpha_V,0)$
} \\
\bottomrule
\end{tabular}

\caption{Summary of optimal contracts across default and liability-cap regimes in the baseline model. Under a pessimistic default, the listed contract is uniquely optimal, except that every contract on the segment joining $(0,\beta_B)$ and $(\alpha_V,0)$ is optimal when $M_B=M_V$. Under an optimistic default with an intermediate or loose cap, the listed contract is uniquely optimal whenever $\lambda^* > 0$, and every protected contract is optimal when $\lambda^*=0$.}
\label{tab:optimal-contract-summary}
\end{table}

\bibliographystyle{myref}
\bibliography{mybib}

\newpage
\begin{appendices}
\setstretch{1}

\renewcommand{\theequation}{\thesection.\arabic{equation}}
\setcounter{equation}{0}
\renewcommand{\thelemma}{\thesection.\arabic{lemma}}
\setcounter{lemma}{0}
\renewcommand{\theproposition}{\thesection.\arabic{proposition}}
\setcounter{proposition}{0}
\renewcommand{\thefigure}{\thesection.\arabic{figure}}
\setcounter{figure}{0}
\renewcommand{\thetable}{\thesection.\arabic{table}}
\setcounter{table}{0}

\section{Proofs}\label{app:proofs}

\subsection{Proofs for Section~\ref{sec:baseline-model}}

\begin{proof}[Proof of Lemma~\ref{lem:full-information-benchmark}]
Fix $\theta_1\in\mathbb R$. Under any contract $w$, if the agent experiments, then $V_A(\theta_1;w)\geq0$ and hence $U_P(\theta_1;w)=V_S(\theta_1)-V_A(\theta_1;w) \leq V_S(\theta_1)$. If the agent does not experiment, the principal receives zero. Therefore, $U_P(\theta_1;w)\leq\max\{V_S(\theta_1),0\}$ under every contract.

Suppose first that $V_S(\theta_1)\geq0$. The pure launch bonus
$$
w=\left(0,\,\frac{c}{p\Phi\left(\frac{\theta_1-\Delta}{s}\right)+(1-p)\Phi\left(\frac{\theta_0-\Delta}{s}\right)}\right)
$$
makes the agent indifferent between experimenting and not experimenting. The denominator is the launch probability conditional on experimentation, which is strictly positive, so the expected bonus equals $c$ and $V_A(\theta_1;w)=0$. Under the maintained tie-breaking rule, the agent experiments, so the principal receives $U_P(\theta_1;w)=V_S(\theta_1)$.

If instead $V_S(\theta_1)<0$, the null contract $w=(0,0)$ gives the agent payoff $-c<0$ from experimentation. The agent therefore does not experiment, and the principal receives zero.

Thus the upper bound is attained at every alternative value, so $\overline U_P(\theta_1)=\max\{V_S(\theta_1),0\}$. This completes the proof.
\end{proof}

We next establish three lemmas on the shapes of total surplus and the agent's and principal's expected payoffs as the alternative value $\theta_1$ varies. These results will be used to characterize experimentation incentives and profit guarantees in Section~\ref{sec:robust-affine-contracts}.

\begin{lemma}[Shape of total surplus]\label{lem:surplus-shape}
The marginal total surplus
$$\frac{\partial}{\partial\theta_1}V_S(\theta_1)=p\left(\Phi\left(\frac{\theta_1-\Delta}{s}\right)+\frac{\theta_1}{s}\,\phi\left(\frac{\theta_1-\Delta}{s}\right)\right)$$
equals zero at a unique point $\underline\theta_1<0$. Consequently, $V_S(\,\cdot\,)$ is strictly decreasing on $(-\infty,\underline\theta_1)$ and strictly increasing on $(\underline\theta_1,\infty)$. Moreover,
$$\frac{\partial}{\partial\theta_1}V_S(\theta_1)=\frac{p}{s}\phi\left(\frac{\theta_1-\Delta}{s}\right)H(\theta_1),\quad H(t)=t+\frac{s\Phi\left(\frac{t-\Delta}{s}\right)}{\phi\left(\frac{t-\Delta}{s}\right)},$$
and $H$ is strictly increasing.
\end{lemma}

\begin{proof}
Differentiating once more,
$$\frac{\partial^2}{\partial\theta_1^2}V_S(\theta_1)=\frac{p}{s}\phi\left(\frac{\theta_1-\Delta}{s}\right)\left(-\frac{1}{s^2}\theta_1^2+\frac{\Delta}{s^2}\theta_1+2\right).$$
The bracket is a concave quadratic in $\theta_1$ with positive constant term, so its two roots have opposite signs, $\theta_l<0<\theta_u$. Hence $\partial V_S/\partial\theta_1$ is strictly decreasing on $(-\infty,\theta_l)$ and on $(\theta_u,\infty)$, and strictly increasing on $(\theta_l,\theta_u)$. Since
$$\lim_{\theta_1\to-\infty}\frac{\partial V_S}{\partial\theta_1}=0,
   \quad
   \lim_{\theta_1\to+\infty}\frac{\partial V_S}{\partial\theta_1}=p>0,$$
this forces $\partial V_S/\partial\theta_1$ to fall below zero on $(-\infty,\theta_l)$, rise through zero exactly once, and stay positive thereafter. Let $\underline\theta_1$ be that unique zero, so $\partial V_S/\partial\theta_1<0$ for $\theta_1<\underline\theta_1$ and $>0$ for $\theta_1>\underline\theta_1$. Evaluating at the origin,
$$\left.\frac{\partial V_S}{\partial\theta_1}\right|_{\theta_1=0}
   =p\Phi\left(-\frac{\Delta}{s}\right)>0,$$
so the marginal is already positive at $\theta_1=0$; the crossing therefore lies strictly to its left, $\underline\theta_1<0$, and the monotonicity of $V_S$ follows. 

Finally, the factorization of the marginal total surplus is immediate. The standard normal CDF is strictly log-concave, so $\Phi(z)/\phi(z)$ is strictly increasing. It follows that $H$ is strictly increasing. This completes the proof.
\end{proof}

\begin{lemma}[Shape of the agent's payoff]\label{lem:agent-shape}
Fix $w=(\alpha,\beta)$ with $\alpha>0$. The agent's payoff $V_A(\,\cdot\,;w)$ has a kink at $\theta_1=-r/\alpha$, and away from it the marginal equals
$$\frac{\partial}{\partial\theta_1}V_A(\theta_1;w)=
\left\{\begin{aligned}
   &\frac{p}{s}\phi\left(\frac{\theta_1-\Delta}{s}\right)(\beta+\alpha\theta_1)
   +p\alpha\Phi\left(\frac{\theta_1-\Delta}{s}\right), && \theta_1>-r/\alpha,\\
   &\frac{p}{s}\phi\left(\frac{\theta_1-\Delta}{s}\right)(\beta-r), && \theta_1<-r/\alpha.
 \end{aligned}\right.$$
If $\beta\ge r$, then $\partial V_A/\partial\theta_1\ge 0$ on both branches, so $V_A(\,\cdot\,;w)$ is nondecreasing. If $\beta<r$, then the upper-branch expression (regarded as a function on $\mathbb{R}$) has a unique zero $\theta_A^*$, and $\partial V_A/\partial\theta_1<0$ for $\theta_1<\overline\theta_A$ and $\partial V_A/\partial\theta_1>0$ for $\theta_1>\overline\theta_A$, where $\overline\theta_A=\max\{\theta_A^*,-r/\alpha\}$.
\end{lemma}

\begin{proof}
The default term $(1-p)\Phi_0\left(\beta+\max\{\alpha\theta_0,-r\}\right)$ does not involve $\theta_1$. Differentiating the alternative-state term $p\Phi((\theta_1-\Delta)/s)(\beta+\max\{\alpha\theta_1,-r\})$ on each side of the kink gives the two branches, which we denote $g_1$ (upper) and $g_2$ (lower).

If $\beta\ge r$, then $g_2\ge0$ on the lower branch; on the upper branch $\beta+\alpha\theta_1>\beta-r\ge0$ and $p\alpha\Phi\ge0$ give $g_1\ge0$. Hence $\partial V_A/\partial\theta_1\ge0$ throughout and $V_A(\,\cdot\,;w)$ is nondecreasing.

Now suppose $\beta<r$. On the lower branch $g_2=(p/s)\phi((\theta_1-\Delta)/s)(\beta-r)<0$. For the upper-branch expression, $\partial g_1/\partial\theta_1$ equals $(p/s)\phi((\theta_1-\Delta)/s)$ times a concave quadratic in $\theta_1$ (its leading coefficient is $-\alpha/s^2<0$), and $\lim_{\theta_1\to-\infty}g_1=0$, $\lim_{\theta_1\to+\infty}g_1=p\alpha>0$. These limits rule out $g_1$ being nonincreasing on $\mathbb{R}$, so the quadratic is positive somewhere and therefore has two distinct real roots. The argument of Lemma~\ref{lem:surplus-shape} then gives a unique zero $\theta_A^*\in\mathbb{R}$ of the upper-branch expression, with $g_1<0$ for $\theta_1<\theta_A^*$ and $g_1>0$ for $\theta_1>\theta_A^*$.

Set $\overline\theta_A=\max\{\theta_A^*,-r/\alpha\}$. If $\theta_A^*\ge-r/\alpha$, the zero lies on the upper branch: $\partial V_A/\partial\theta_1=g_2<0$ for $\theta_1<-r/\alpha$, and $=g_1$ on $\theta_1>-r/\alpha$, which is negative up to $\theta_A^*$ and positive beyond. If instead $\theta_A^*<-r/\alpha$, then $g_1>0$ already at the kink, so $\partial V_A/\partial\theta_1=g_2<0$ for $\theta_1<-r/\alpha$ and $=g_1>0$ for $\theta_1>-r/\alpha$. In both cases $\partial V_A/\partial\theta_1<0$ for $\theta_1<\overline\theta_A$ and $>0$ for $\theta_1>\overline\theta_A$. This completes the proof.
\end{proof}

\begin{lemma}[Shape of the principal's payoff]
Let $w=(\alpha,\beta)$ with $\alpha<1$, and let $\theta_1'$ be a point with $\alpha\theta_1'>-r$ at which $V_P(\theta_1';w)=0$. If
$$\left.\frac{\partial}{\partial\theta_1}V_P(\theta_1;w)\right|_{\theta_1=\theta_1'}\ge 0,$$
then $V_P(\theta_1;w)\ge 0$ for all $\theta_1\ge\theta_1'$.
\end{lemma}

\begin{proof}
Since $\alpha\theta_1'>-r$, the floor is slack on $[\theta_1',\infty)$, where $V_P=V_S-V_A$ gives
$$\frac{\partial}{\partial\theta_1}V_P(\theta_1;w)=p(1-\alpha)\Phi\left(\frac{\theta_1-\Delta}{s}\right)+\frac{p}{s}\left((1-\alpha)\theta_1-\beta\right)\phi\left(\frac{\theta_1-\Delta}{s}\right).$$
Denote the right-hand side, regarded as a function on $\mathbb{R}$, by $g$. Differentiating,
$$\frac{\partial g}{\partial\theta_1}=\frac{p}{s}\phi\left(\frac{\theta_1-\Delta}{s}\right)\left(2(1-\alpha)-\frac{(\theta_1-\Delta)\left((1-\alpha)\theta_1-\beta\right)}{s^2}\right),$$
whose bracket is a concave quadratic in $\theta_1$ because $\alpha<1$. Moreover, $\lim_{\theta_1\to-\infty}g=0$ and $\lim_{\theta_1\to+\infty}g=p(1-\alpha)>0$. These limits rule out $g$ being nonincreasing on $\mathbb{R}$, so the quadratic is positive somewhere and therefore has two distinct real roots. The argument of Lemma~\ref{lem:surplus-shape} then yields a unique $\theta_p$ with $g<0$ for $\theta_1<\theta_p$ and $g>0$ for $\theta_1>\theta_p$. The hypothesis gives $g(\theta_1')\ge0$, which forces $\theta_1'\ge\theta_p$, so $\partial V_P/\partial\theta_1=g\ge0$ on $[\theta_1',\infty)$; hence $V_P(\,\cdot\,;w)$ is nondecreasing there and $V_P(\theta_1;w)\ge V_P(\theta_1';w)=0$. This completes the proof.
\end{proof}

\begin{proof}[Proof of Lemma~\ref{lem:guarantee-implementation}]
For $\lambda=0$, the guarantee inequality reduces to $U_P(\theta_1;w)\geq0$ for every $\theta_1\in\mathbb R$. Since $U_P$ equals $V_P$ when the agent experiments and equals zero otherwise, this condition is equivalent to Rule~1.

Now suppose that $w$ provides a uniform guarantee $\lambda>0$. Since $\overline U_P\geq0$, the contract also provides the guarantee zero and therefore satisfies Rule~1. Whenever $V_S(\theta_1)>0$, Lemma~\ref{lem:full-information-benchmark} and the guarantee inequality give $U_P(\theta_1;w)\geq\lambda V_S(\theta_1)>0$. The agent must therefore experiment at every such alternative value.

Conversely, if the agent experiments at an alternative value with $V_S(\theta_1)<0$, then $V_A(\theta_1;w)\geq0$ and $V_P(\theta_1;w)=V_S(\theta_1)-V_A(\theta_1;w)<0$, contradicting Rule~1. Thus the agent does not experiment when total surplus is negative.

Finally, consider an alternative value with $V_S(\theta_1)=0$. By Lemma~\ref{lem:surplus-shape}, $V_S$ is strictly monotone on each side of $\underline\theta_1$, so every such value can be approached by alternative values with strictly positive total surplus. The agent experiments at each of those values, so continuity of $V_A(\,\cdot\,;w)$ implies $V_A(\theta_1;w)\geq0$. Under the maintained tie-breaking rule, the agent therefore experiments at $\theta_1$ as well.

Combining these cases, the agent experiments if and only if $V_S(\theta_1)\geq0$, which is Rule~2. This completes the proof.
\end{proof}

\subsection{Proofs for Section~\ref{subsec:pessimistic-default}}

\begin{proof}[Proof of Proposition~\ref{prop:rule1}]
Suppose first that $w$ provides a uniform guarantee $\lambda>0$. By Lemma~\ref{lem:guarantee-implementation}, the agent then experiments exactly when $V_S(\theta_1)\geq0$, that is, exactly when $\theta_1\geq\theta_s$. Thus $V_A(\theta_1;w)<0$ for $\theta_1<\theta_s$ and $V_A(\theta_s;w)\geq0$, and continuity gives $V_A(\theta_s;w)=0$, which is~\eqref{eq:pessimistic-rule2}.

Conversely, consider any contract satisfying this cutoff condition. Because the guarantee is defined through the principal's payoff $U_P$ rather than the conditional payoff $V_P$, we first determine where the agent experiments. Subtracting the agent's payoff at the cutoff gives
$$V_A(\theta_1;w)=p\left(\Phi\left(\frac{\theta_1-\Delta}{s}\right)\left(\beta+\max\{\alpha\theta_1,-r\}\right)-\Phi_s(\beta+\alpha\theta_s)\right).$$
Since $c>0$, the contract cannot have both coefficients zero. On $[\theta_s,\infty)$, the liability floor is slack and $\Phi((\theta_1-\Delta)/s)(\beta+\alpha\theta_1)$ is strictly increasing, so $V_A(\theta_1;w)>0$ for $\theta_1>\theta_s$. For $\theta_1<\theta_s$, if $\beta+\alpha\theta_s>0$, then
$$\Phi\left(\frac{\theta_1-\Delta}{s}\right)\left(\beta+\max\{\alpha\theta_1,-r\}\right)\leq\Phi\left(\frac{\theta_1-\Delta}{s}\right)(\beta+\alpha\theta_s)<\Phi_s(\beta+\alpha\theta_s),$$
so $V_A(\theta_1;w)<0$. If instead $\beta+\alpha\theta_s=0$, then $\beta=\theta_s=0$ and $\alpha>0$. In this case, for every $\theta_1<0$,
$$V_A(\theta_1;w)=p\Phi\left(\frac{\theta_1-\Delta}{s}\right)\max\{\alpha\theta_1,-r\}<0,$$
because $r>0$. Under the maintained tie-breaking rule, the agent therefore experiments if and only if $\theta_1\geq\theta_s$.

For $\theta_1>\theta_s$, the cutoff identities for total surplus and the agent's payoff imply
$$V_S(\theta_1)=p\left(\theta_1\Phi\left(\frac{\theta_1-\Delta}{s}\right)-\theta_s\Phi_s\right),\quad V_A(\theta_1;w)=\alpha V_S(\theta_1)+p\beta\left(\Phi\left(\frac{\theta_1-\Delta}{s}\right)-\Phi_s\right).$$
With $H$ as in Lemma~\ref{lem:surplus-shape}, we have
$$\frac{\partial}{\partial t}V_S(t)=\frac{p}{s}\phi\left(\frac{t-\Delta}{s}\right)H(t)\quad\text{and}\quad\frac{\partial}{\partial t}\Phi\left(\frac{t-\Delta}{s}\right)=\frac{1}{s}\phi\left(\frac{t-\Delta}{s}\right).$$
Integrating these identities over $[\theta_s,\theta_1]$ gives
$$V_S(\theta_1)=\frac{p}{s}\int_{\theta_s}^{\theta_1}\phi\left(\frac{t-\Delta}{s}\right)H(t)\,dt,\quad \Phi\left(\frac{\theta_1-\Delta}{s}\right)-\Phi_s=\frac{1}{s}\int_{\theta_s}^{\theta_1}\phi\left(\frac{t-\Delta}{s}\right)dt.$$
Consequently,
\begin{equation}\label{eq:pessimistic-ratio-proof}\frac{V_P(\theta_1;w)}{V_S(\theta_1)}=1-\alpha-\beta\,\frac{\int_{\theta_s}^{\theta_1}\phi\left(\frac{t-\Delta}{s}\right)dt}{\int_{\theta_s}^{\theta_1}\phi\left(\frac{t-\Delta}{s}\right)H(t)\,dt},\end{equation}
where the fraction is the reciprocal of a weighted average of $H$ over $[\theta_s,\theta_1]$.

Since the standard normal CDF $\Phi$ is strictly log-concave, $\Phi(z)/\phi(z)$ is strictly increasing in $z$. By Lemma~\ref{lem:surplus-shape}, $H$ is strictly increasing, and it is positive on $[\theta_s,\infty)$ because $\theta_s\geq0$. The weighted average is therefore positive and strictly increasing in $\theta_1$, and it tends to $H(\theta_s)=\theta_s+\Phi_s/\phi_s$ as $\theta_1\downarrow\theta_s$. Equation~\eqref{eq:pessimistic-ratio-proof} then shows that $V_P/V_S$ is nondecreasing on $(\theta_s,\infty)$, strictly increasing if $\beta>0$, and
$$\inf_{\theta_1>\theta_s}\frac{V_P(\theta_1;w)}{V_S(\theta_1)}=1-\alpha-\frac{\beta}{H(\theta_s)}=1-\alpha-\frac{\beta\phi_s}{\Phi_s+\theta_s\phi_s}.$$
Since $V_P(\theta_s;w)=V_S(\theta_s)=0$ and $\left.\partial V_S/\partial\theta_1\right|_{\theta_1=\theta_s}=p(\Phi_s+\theta_s\phi_s)>0$, the limit at the cutoff also equals the derivative ratio stated in the proposition.

It remains to relate this infimum to the uniform guarantee. Below $\theta_s$, the agent does not experiment and the full-information benchmark is zero. At $\theta_s$, both the principal's payoff and the benchmark are zero. Above $\theta_s$, the agent experiments and the benchmark equals $V_S$. The contract therefore provides the guarantee $\lambda$ if and only if the infimum above is at least $\lambda$, which is precisely~\eqref{eq:pessimistic-positive-guarantee}. Together with the necessity of the cutoff condition established at the outset, this proves the proposition.
\end{proof}

\begin{proof}[Proof of Theorem~\ref{thm:pessimistic-default}]
Let $L(\alpha,\beta)$ denote the left-hand side of~\eqref{eq:pessimistic-positive-guarantee}. By Proposition~\ref{prop:rule1}, a contract provides a strictly positive guarantee if and only if it satisfies~\eqref{eq:pessimistic-rule2} and $L(\alpha,\beta)<1$, in which case its largest guarantee is $1-L(\alpha,\beta)$. Once a strictly positive guarantee is shown to be attainable, every optimal contract therefore satisfies~\eqref{eq:pessimistic-rule2}, and the problem over the entire affine class reduces to minimizing $L$ over contracts satisfying that cutoff condition.

Suppose first that $\theta_s\geq\theta_0$. The pure launch bonus $(\alpha_B,\beta_B)$ satisfies~\eqref{eq:pessimistic-rule2}, and $\beta_B\left(p\Phi_s+(1-p)\Phi_0\right)=c$. The efficient-cutoff identity $p\theta_s\Phi_s+(1-p)\theta_0\Phi_0=c+\kappa$ gives $\theta_s\left(p\Phi_s+(1-p)\Phi_0\right)=c+\kappa+(1-p)\Phi_0(\theta_s-\theta_0)$. Subtracting,
$$(\theta_s-\beta_B)\left(p\Phi_s+(1-p)\Phi_0\right)=\kappa+(1-p)\Phi_0(\theta_s-\theta_0)>0.$$
Therefore,
$$L(\alpha_B,\beta_B)=\frac{\beta_B\phi_s}{\Phi_s+\theta_s\phi_s}<\frac{\theta_s\phi_s}{\Phi_s+\theta_s\phi_s}<1,$$
so this contract provides a strictly positive guarantee.

Now let $(\alpha,\beta)$ be any contract satisfying~\eqref{eq:pessimistic-rule2}. Subtracting the cutoff condition for $(\alpha_B,\beta_B)$ gives
$$0=p\Phi_s\left(\beta+\alpha\theta_s-\beta_B\right)+(1-p)\Phi_0\left(\beta+\max\{\alpha\theta_0,-r\}-\beta_B\right).$$
Since $\theta_s\geq\theta_0$, $\theta_s\geq0$, and $\alpha\geq0$, we have $\alpha\theta_s\geq\max\{\alpha\theta_0,-r\}$, so the first term in parentheses is at least the second. As the weights $p\Phi_s$ and $(1-p)\Phi_0$ are strictly positive and the weighted sum is zero, the larger term is nonnegative: $\beta+\alpha\theta_s-\beta_B\geq0$. Consequently,
$$L(\alpha,\beta)-L(\alpha_B,\beta_B)=\alpha+\frac{(\beta-\beta_B)\phi_s}{\Phi_s+\theta_s\phi_s}=\frac{\alpha\Phi_s+\phi_s\left(\beta+\alpha\theta_s-\beta_B\right)}{\Phi_s+\theta_s\phi_s}\geq0.$$
Equality requires $\alpha=0$ and $\beta+\alpha\theta_s-\beta_B=0$, and hence $(\alpha,\beta)=(\alpha_B,\beta_B)$. Thus the pure launch bonus is uniquely optimal when $\theta_s\geq\theta_0$.

Suppose next that $\theta_s<\theta_0$. Then $\theta_0>\theta_s\geq0$, so the liability floor is slack at both $\theta_s$ and $\theta_0$, and the cutoff condition becomes
$$\left(p\Phi_s+(1-p)\Phi_0\right)\beta+\left(p\theta_s\Phi_s+(1-p)\theta_0\Phi_0\right)\alpha=c.$$
Replacing the coefficient on $\alpha$ by $c+\kappa$ (the efficient-cutoff identity) and dividing by $c$ gives $\alpha/\alpha_V+\beta/\beta_B=1$, which, together with $\alpha,\beta\geq0$, describes exactly the line segment joining $(\alpha_B,\beta_B)$ and $(\alpha_V,\beta_V)$. In particular, proportional sharing satisfies the cutoff condition, and $L(\alpha_V,\beta_V)=\alpha_V=c/(c+\kappa)<1$, so a strictly positive guarantee is again attainable.

The objective $L$ is affine on the segment, and at its endpoints
$$L(\alpha_B,\beta_B)=\frac{M_B}{p(\Phi_s+\theta_s\phi_s)},\quad L(\alpha_V,\beta_V)=\frac{M_V}{p(\Phi_s+\theta_s\phi_s)}.$$
Therefore, the pure launch bonus is the unique minimizer when $M_B<M_V$, proportional sharing is the unique minimizer when $M_B>M_V$, and every point on the segment is a minimizer when $M_B=M_V$.

In both cases, the minimizing contracts are independent of $r$. Finally, Proposition~\ref{prop:rule1} gives the corresponding optimal guarantees. When the pure launch bonus is optimal, $\lambda^*=1-\beta_B\phi_s/\left(\Phi_s+\theta_s\phi_s\right)$, whereas, when proportional sharing is optimal, $\lambda^*=\kappa/(c+\kappa)$. When $M_B=M_V$, these two expressions coincide. This completes the proof.
\end{proof}

\begin{proposition}[Selection in the equality case]
\label{prop:pessimistic-equality-selection}
Suppose $\theta_0\leq\overline\theta_0$, $\theta_s<\theta_0$, and $M_B=M_V$, and let $w_B=(\alpha_B,\beta_B)$ denote the pure launch bonus. For every other optimal contract $w$, $U_P(\theta_1;w_B)>U_P(\theta_1;w)$ for every $\theta_1>\theta_s$, and $U_P(\theta_1;w_B)=U_P(\theta_1;w)=0$ for every $\theta_1\leq\theta_s$. Consequently, $w_B$ is the unique undominated optimal contract.
\end{proposition}

\begin{proof}
Let $w_V=(\alpha_V,\beta_V)$ denote proportional sharing. By Theorem~\ref{thm:pessimistic-default}, the optimal contracts are exactly $w_t=(1-t)w_B+tw_V$ for $t\in[0,1]$. Each provides a strictly positive guarantee, so it is implementing by Lemma~\ref{lem:guarantee-implementation} and satisfies~\eqref{eq:pessimistic-rule2} by Proposition~\ref{prop:rule1}. The agent therefore experiments exactly when $\theta_1\geq\theta_s$, and $U_P(\theta_s;w_t)=V_S(\theta_s)-V_A(\theta_s;w_t)=0$, so $U_P(\theta_1;w_t)=0$ for every $\theta_1\leq\theta_s$ and every $t$.

Now fix $\theta_1>\theta_s$. Every optimal contract induces experimentation, so $U_P=V_P$. Because $\theta_1>\theta_s\geq0$ and $\theta_0>\theta_s\geq0$, the liability floor is slack in both states, so $V_A(\theta_1;w)$ is affine in the coefficients of $w$ and
$$V_P(\theta_1;w_t)=(1-t)V_P(\theta_1;w_B)+tV_P(\theta_1;w_V).$$
Under proportional sharing, $V_A(\theta_1;w_V)=\alpha_V\left(V_S(\theta_1)+c+\kappa\right)-c=\alpha_VV_S(\theta_1)$, so $V_P(\theta_1;w_V)=(1-\alpha_V)V_S(\theta_1)$. Under the pure launch bonus, Proposition~\ref{prop:rule1} shows that $V_P(\theta_1;w_B)/V_S(\theta_1)$ is strictly increasing on $(\theta_s,\infty)$ with infimum $1-\beta_B\phi_s/(\Phi_s+\theta_s\phi_s)$. By the definitions of $M_B$ and $M_V$, the equality $M_B=M_V$ is equivalent to $\beta_B\phi_s/(\Phi_s+\theta_s\phi_s)=\alpha_V$, so this infimum equals $1-\alpha_V$ and
$$V_P(\theta_1;w_B)>(1-\alpha_V)V_S(\theta_1)=V_P(\theta_1;w_V).$$
Hence $U_P(\theta_1;w_B)-U_P(\theta_1;w_t)=t\left(V_P(\theta_1;w_B)-V_P(\theta_1;w_V)\right)>0$ whenever $t>0$, that is, whenever $w_t\neq w_B$. Together with the equality at every $\theta_1\leq\theta_s$, this shows that $w_B$ dominates every other optimal contract.

It remains to verify that $w_B$ is itself undominated. Suppose that some contract $\tilde w$ dominates $w_B$. Since $w_B$ provides the optimal guarantee, pointwise dominance implies that $\tilde w$ provides at least the same guarantee and is therefore also optimal. But the comparison above shows that no other optimal contract can dominate $w_B$. Hence $w_B$ is undominated, while every other optimal contract is dominated by $w_B$. This proves that $w_B$ is the unique undominated optimal contract.
\end{proof}

\subsection{Proofs for Section~\ref{subsec:optimistic-default}}
\label{app:optimistic-default}

\begin{proof}[Proof of Theorem~\ref{thm:optimistic-default}]
Any implementing contract must make the agent indifferent at both efficient cutoffs. Indeed, the agent experiments at each cutoff but does not experiment at nearby values in $(\theta_{s1},\theta_{s2})$, so continuity of $V_A$ implies $V_A(\theta_{s1};w)=V_A(\theta_{s2};w)=0$. These equalities are precisely~\eqref{eq:optimistic-lower-cutoff} and~\eqref{eq:optimistic-upper-cutoff}. We first solve the two cutoff equations and then determine which solutions implement efficient experimentation.

Because $\theta_0>\overline\theta_0>0$, the liability floor does not bind at the default value. Since $\theta_{s1}<\theta_{s2}<0$ and $\alpha\geq0$, there are only three possible binding patterns at the two cutoffs.

Suppose first that the floor binds at both cutoffs, that is, $\alpha^*\theta_{s2}<-r$, so that also $\alpha^*\theta_{s1}<-r$. The cutoff equations become
$$p\Phi_{si}(\beta^*-r)+(1-p)\Phi_0(\beta^*+\alpha^*\theta_0)=c,\quad i=1,2.$$
Subtracting them and using $\Phi_{s1}\neq\Phi_{s2}$ gives $\beta^*=r$. Substitution then gives
$$\alpha^*=\frac{1}{\theta_0}\left(\frac{c}{(1-p)\Phi_0}-r\right).$$
The floor binds at the upper cutoff if and only if $\alpha^*\theta_{s2}<-r$, or equivalently
$$r<-\frac{c\theta_{s2}}{(1-p)(\theta_0-\theta_{s2})\Phi_0}=\underline r.$$
Moreover, $\alpha^*>0$ throughout this range because $\underline r<c/\left((1-p)\Phi_0\right)$. Thus this binding pattern gives a unique admissible solution precisely when $r<\underline r$.

Suppose next that the floor binds at neither cutoff, that is, $\alpha^*\theta_{s1}\geq-r$, so that also $\alpha^*\theta_{s2}\geq-r$. For $i=1,2$, the cutoff equation is
$$\left(p\Phi_{si}+(1-p)\Phi_0\right)\beta^*+\left(p\theta_{si}\Phi_{si}+(1-p)\theta_0\Phi_0\right)\alpha^*=c.$$
Since $V_S(\theta_{si})=0$, we have $p\theta_{si}\Phi_{si}+(1-p)\theta_0\Phi_0=c+\kappa$, so both equations reduce to
$$\left(p\Phi_{si}+(1-p)\Phi_0\right)\beta^*+(c+\kappa)\alpha^*=c.$$
Subtracting the two equations gives $\beta^*=0$, and therefore $\alpha^*=c/(c+\kappa)=\alpha_V$. The floor does not bind at the lower cutoff if and only if $\alpha_V\theta_{s1}\geq-r$, or equivalently $r\geq-c\theta_{s1}/(c+\kappa)=\overline r$. Thus this binding pattern gives the unique solution $w^*=(\alpha_V,0)$ precisely when $r\geq\overline r$.

The thresholds are ordered as stated. Both are positive because $\theta_{s1}<\theta_{s2}<0$. Moreover, $\underline r<\overline r$ is equivalent to
$$-\theta_{s2}(c+\kappa)<-\theta_{s1}(1-p)(\theta_0-\theta_{s2})\Phi_0.$$
Using $c+\kappa=p\theta_{s2}\Phi_{s2}+(1-p)\theta_0\Phi_0$, the difference between the right-hand and left-hand sides is
$$(1-p)\theta_0\Phi_0(\theta_{s2}-\theta_{s1})+(1-p)\theta_{s1}\theta_{s2}\Phi_0+p\theta_{s2}^2\Phi_{s2},$$
which is strictly positive.

In the remaining binding pattern, the floor binds only at the lower cutoff, that is, $\alpha^*\theta_{s1}<-r\leq\alpha^*\theta_{s2}$. The cutoff equations reduce to
\begin{equation}\left\{\begin{aligned}&\left(p\Phi_{s1}+(1-p)\Phi_0\right)\beta^*+(1-p)\theta_0\Phi_0\,\alpha^*=c+p\Phi_{s1}r,\\&\left(p\Phi_{s2}+(1-p)\Phi_0\right)\beta^*+\left(p\theta_{s2}\Phi_{s2}+(1-p)\theta_0\Phi_0\right)\alpha^*=c.\end{aligned}\right.\label{eq:optimistic-intermediate}\end{equation}
Using $p\theta_{s2}\Phi_{s2}+(1-p)\theta_0\Phi_0=c+\kappa$, the determinant of this system is
\begin{align*}
&\left(p\Phi_{s1}+(1-p)\Phi_0\right)(c+\kappa)-\left(p\Phi_{s2}+(1-p)\Phi_0\right)(1-p)\theta_0\Phi_0\\
=\,&p\left(\theta_{s2}\Phi_{s2}\left(p\Phi_{s1}+(1-p)\Phi_0\right)+(1-p)\theta_0\Phi_0(\Phi_{s1}-\Phi_{s2})\right)<0,
\end{align*}
so the system has a unique solution, which is affine in $r$. At $r=\underline r$, the solution of the equations of the first pattern has $\alpha^*\theta_{s2}=-r$, so it also solves~\eqref{eq:optimistic-intermediate}; hence the solution there is $(\alpha^*,\beta^*)=(-\underline r/\theta_{s2},\,\underline r)$. Likewise, at $r=\overline r$, proportional sharing has $\alpha_V\theta_{s1}=-r$ and also solves~\eqref{eq:optimistic-intermediate}, so the solution there is $(\alpha_V,0)$. Writing $r=(1-t)\underline r+t\overline r$ with $t\in\mathbb R$, affinity in $r$ gives
$$\alpha^*=-(1-t)\frac{\underline r}{\theta_{s2}}+t\alpha_V,\quad\beta^*=(1-t)\underline r,$$
and therefore, using $\alpha_V\theta_{s1}=-\overline r$,
$$r-\beta^*=t\overline r,\quad\alpha^*\theta_{s2}+r=t\alpha_V(\theta_{s2}-\theta_{s1}),\quad\alpha^*\theta_{s1}+r=-(1-t)\,\underline r\,\frac{\theta_{s2}-\theta_{s1}}{-\theta_{s2}}.$$
Because $\theta_{s1}<\theta_{s2}<0$, the factors multiplying $t$ and $1-t$ in these expressions are positive. Hence $-r\leq\alpha^*\theta_{s2}$ if and only if $t\geq0$, and $\alpha^*\theta_{s1}<-r$ if and only if $t<1$. This binding pattern is therefore consistent precisely when $0\leq t<1$, that is, when $\underline r\leq r<\overline r$. On this range, $\alpha^*>0$ and $\beta^*>0$, and $\beta^*<r$ except at $r=\underline r$, where $\beta^*=r$. Since the three ranges $r<\underline r$, $\underline r\leq r<\overline r$, and $r\geq\overline r$ partition $(0,\infty)$, the cutoff equations have exactly one admissible solution $w^*$ for every $r>0$.

We next determine whether $w^*$ implements efficient experimentation. In the intermediate and loose regimes, $\alpha^*>0$ and $\beta^*<r$. Lemma~\ref{lem:agent-shape}, together with the two cutoff equalities, implies
$$V_A(\theta_1;w^*)>0\text{ for }\theta_1<\theta_{s1}\text{ or }\theta_1>\theta_{s2},\quad V_A(\theta_1;w^*)<0\text{ for }\theta_1\in(\theta_{s1},\theta_{s2}).$$
The maintained tie-breaking rule gives experimentation at the two cutoffs, so $w^*$ implements efficient experimentation.

In the tight regime, $r\leq\underline r$, the solution has $\beta^*=r$ and $\alpha^*\theta_{s2}\leq-r$; it comes from the first pattern when $r<\underline r$ and from the third pattern at $r=\underline r$. The alternative-state payment $\beta^*+\max\{\alpha^*\theta_1,-r\}$ is therefore zero on $[\theta_{s1},\theta_{s2}]$, and the cutoff equations imply $V_A(\theta_1;w^*)=0$ for every $\theta_1\in[\theta_{s1},\theta_{s2}]$. The agent experiments throughout the inefficient interval under the tie-breaking rule, so $w^*$ does not implement efficient experimentation. Since every implementing contract must satisfy the two cutoff equations and their admissible solution is unique, no contract implements efficient experimentation when $r\leq\underline r$.

By Lemma~\ref{lem:guarantee-implementation}, a strictly positive uniform guarantee requires implementation. Hence, in the tight-cap regime no contract provides a strictly positive guarantee. The null contract is protected and provides the guarantee zero, so $\lambda^*=0$.

Now suppose $r>\underline r$, so that $w^*$ is the unique implementing contract. By Lemma~\ref{lem:guarantee-implementation}, every contract providing a strictly positive guarantee is implementing and therefore equals $w^*$. Suppose
$$\sup_{\theta_1:\,V_S(\theta_1)>0}\frac{V_A(\theta_1;w^*)}{V_S(\theta_1)}<1.$$
Then $V_P(\theta_1;w^*)>0$ wherever $V_S(\theta_1)>0$, and $V_P(\theta_{si};w^*)=0$ at the two cutoffs; since $w^*$ is implementing, it is protected. By the expression for $\lambda(w)$ following Lemma~\ref{lem:guarantee-implementation}, $\lambda(w^*)$ equals one minus this supremum and is strictly positive, so $w^*$ is uniquely optimal. Otherwise, $w^*$ provides no strictly positive guarantee, and neither does any other contract, so $\lambda^*=0$. In both cases,
$$\lambda^*=\max\left\{0,\,1-\sup_{\theta_1:\,V_S(\theta_1)>0}\frac{V_A(\theta_1;w^*)}{V_S(\theta_1)}\right\}.$$
When $\lambda^*=0$, a contract is optimal if and only if it provides the guarantee zero, which by Lemma~\ref{lem:guarantee-implementation} is equivalent to being protected. This proves the theorem.
\end{proof}

\begin{proposition}[Explicit guarantee conditions under an optimistic default]
\label{prop:optimistic-explicit-conditions}
Suppose $\theta_0>\overline\theta_0$, and let $w^*=(\alpha^*,\beta^*)$ be the contract of Theorem~\ref{thm:optimistic-default}.
\begin{enumerate}
\item Under an intermediate cap, the agent's share $V_A(\theta_1;w^*)/V_S(\theta_1)$ is strictly increasing on $(-\infty,\theta_{s1})$ and strictly decreasing on $(\theta_{s2},\infty)$, and its suprema over these two regions are the two terms in~\eqref{eq:optimistic-intermediate-guarantee}. Hence $\lambda^*>0$ if and only if~\eqref{eq:optimistic-intermediate-guarantee} holds.
\item Under a loose cap, $\lambda^*>0$ if and only if~\eqref{eq:optimistic-loose-guarantee} holds. The left-hand side of~\eqref{eq:optimistic-loose-guarantee} is nonincreasing in $r$ and converges to zero as $r\to\infty$. Hence the condition holds for all sufficiently large caps, and $\lambda^*$ converges to $\kappa/(c+\kappa)$ as $r\to\infty$.
\end{enumerate}
\end{proposition}

\begin{proof}
By Theorem~\ref{thm:optimistic-default}, $\lambda^*>0$ if and only if the agent's share has supremum strictly below one over the efficient region $(-\infty,\theta_{s1})\cup(\theta_{s2},\infty)$, where $V_S>0$. Let $H$ be as in Lemma~\ref{lem:surplus-shape}. Since $\theta_{s1}<\underline\theta_1<\theta_{s2}$, the sign of the marginal total surplus gives $H(\theta_{s1})<0<H(\theta_{s2})$. Moreover, for any $a<b$, the factorization in Lemma~\ref{lem:surplus-shape} and the fundamental theorem of calculus give
\begin{align*}
b\Phi\left(\frac{b-\Delta}{s}\right)-a\Phi\left(\frac{a-\Delta}{s}\right)&=\frac{1}{s}\int_a^b\phi\left(\frac{t-\Delta}{s}\right)H(t)\,dt,\\
\Phi\left(\frac{b-\Delta}{s}\right)-\Phi\left(\frac{a-\Delta}{s}\right)&=\frac{1}{s}\int_a^b\phi\left(\frac{t-\Delta}{s}\right)dt.
\end{align*}

Suppose the cap is intermediate. For $\theta_1<\theta_{s1}$, the floor binds in the alternative state, because $\alpha^*\theta_1<\alpha^*\theta_{s1}<-r$. Subtracting the cutoff equalities at $\theta_{s1}$ gives
\begin{align*}
V_A(\theta_1;w^*)&=p(r-\beta^*)\left(\Phi_{s1}-\Phi\left(\frac{\theta_1-\Delta}{s}\right)\right),\\
V_S(\theta_1)&=p\left(\theta_1\Phi\left(\frac{\theta_1-\Delta}{s}\right)-\theta_{s1}\Phi_{s1}\right),
\end{align*}
so, by the integral identities,
$$\frac{V_A(\theta_1;w^*)}{V_S(\theta_1)}=(r-\beta^*)\,\frac{\int_{\theta_1}^{\theta_{s1}}\phi\left(\frac{t-\Delta}{s}\right)dt}{\int_{\theta_1}^{\theta_{s1}}\phi\left(\frac{t-\Delta}{s}\right)\left(-H(t)\right)dt}.$$
The fraction is the reciprocal of a weighted average of $-H$ over $[\theta_1,\theta_{s1}]$. Since $-H$ is strictly decreasing and $-H(\theta_{s1})>0$, this average is positive, strictly decreasing in $\theta_1$, and converges to $-H(\theta_{s1})$ as $\theta_1\uparrow\theta_{s1}$. As $r-\beta^*>0$, the share is strictly increasing on $(-\infty,\theta_{s1})$, with supremum
$$\frac{r-\beta^*}{-H(\theta_{s1})}=\frac{r-\beta^*}{-\theta_{s1}-{\Phi_{s1}}/{\phi_{s1}}}.$$
For $\theta_1>\theta_{s2}$, the floor is slack, because $\alpha^*\theta_1>\alpha^*\theta_{s2}>-r$. Subtracting the cutoff equalities at $\theta_{s2}$ gives $V_S(\theta_1)=p\left(\theta_1\Phi((\theta_1-\Delta)/s)-\theta_{s2}\Phi_{s2}\right)$ and
$$V_A(\theta_1;w^*)=\alpha^*V_S(\theta_1)+p\beta^*\left(\Phi\left(\frac{\theta_1-\Delta}{s}\right)-\Phi_{s2}\right),$$
so
$$\frac{V_A(\theta_1;w^*)}{V_S(\theta_1)}=\alpha^*+\beta^*\,\frac{\int_{\theta_{s2}}^{\theta_1}\phi\left(\frac{t-\Delta}{s}\right)dt}{\int_{\theta_{s2}}^{\theta_1}\phi\left(\frac{t-\Delta}{s}\right)H(t)\,dt}.$$
Here the fraction is the reciprocal of a weighted average of $H$ over $[\theta_{s2},\theta_1]$, which is positive, strictly increasing in $\theta_1$, and converges to $H(\theta_{s2})>0$ as $\theta_1\downarrow\theta_{s2}$. As $\beta^*>0$, the share is strictly decreasing on $(\theta_{s2},\infty)$, with supremum
$$\alpha^*+\frac{\beta^*}{H(\theta_{s2})}=\alpha^*+\frac{\beta^*}{\theta_{s2}+{\Phi_{s2}}/{\phi_{s2}}}.$$
The supremum over the whole efficient region is the larger of the two, which proves part~1.

Suppose next that the cap is loose, so $w^*=(\alpha_V,0)$. Wherever the floor is slack, $V_A(\theta_1;w^*)=\alpha_V\left(V_S(\theta_1)+c+\kappa\right)-c=\alpha_VV_S(\theta_1)$. The floor binds in the alternative state exactly when $\theta_1<-r/\alpha_V$, and it then raises the agent's payoff by $p\Phi((\theta_1-\Delta)/s)(-r-\alpha_V\theta_1)>0$. Since $r\geq\overline r=-\alpha_V\theta_{s1}$, every such alternative value lies in the lower efficient region. Hence
$$\sup_{\theta_1:\,V_S(\theta_1)>0}\frac{V_A(\theta_1;w^*)}{V_S(\theta_1)}=\alpha_V+\sup_{\theta_1<-r/\alpha_V}\frac{p\Phi\left(\frac{\theta_1-\Delta}{s}\right)(-r-\alpha_V\theta_1)}{V_S(\theta_1)}.$$
By Theorem~\ref{thm:optimistic-default}, $\lambda^*>0$ if and only if the left-hand side is strictly less than one. Since $1-\alpha_V=\kappa/(c+\kappa)$, this holds if and only if the supremum on the right is strictly less than $\kappa/(c+\kappa)$, which is~\eqref{eq:optimistic-loose-guarantee}.

The supremum in~\eqref{eq:optimistic-loose-guarantee} is nonincreasing in $r$. Raising $r$ shrinks the region $\theta_1<-r/\alpha_V$ and lowers $-r-\alpha_V\theta_1$ at every point of it.

To see that the supremum converges to zero, note that $r>0$ gives $-r-\alpha_V\theta_1<-\alpha_V\theta_1$. Hence, on the region $\theta_1<-r/\alpha_V$,
$$0<\frac{p\Phi\left(\frac{\theta_1-\Delta}{s}\right)(-r-\alpha_V\theta_1)}{V_S(\theta_1)}<\frac{-p\alpha_V\theta_1\Phi\left(\frac{\theta_1-\Delta}{s}\right)}{V_S(\theta_1)}.$$
As $\theta_1\to-\infty$, the numerator on the right converges to zero and the denominator converges to $(1-p)\theta_0\Phi_0-c-\kappa>0$. The right-hand side therefore converges to zero. Since the region recedes to $-\infty$ as $r\to\infty$, the supremum converges to zero as well.

Consequently, \eqref{eq:optimistic-loose-guarantee} holds for all sufficiently large $r$. For such $r$, $\lambda^*$ equals $\kappa/(c+\kappa)$ minus the supremum, so $\lambda^*$ converges to $\kappa/(c+\kappa)$. This completes the proof.
\end{proof}

\begin{proof}[Proof of Theorem~\ref{thm:small-agent-cost}]
Holding $\Delta$, $p$, $\theta_0$, $s$, and $c+\kappa$ fixed, total surplus and its two zeros $\theta_{s1}<\theta_{s2}$ do not depend on $c$. Let $H$ be as in Lemma~\ref{lem:surplus-shape}. Since $\theta_{s1}<\underline\theta_1<\theta_{s2}$, that lemma gives $H(\theta_{s1})<0<H(\theta_{s2})$. Define
$$M=\max\left\{\frac{\theta_{s1}}{H(\theta_{s1})},\,1-\frac{\theta_{s1}}{H(\theta_{s2})}\right\},$$
which depends only on $\Delta$, $p$, $\theta_0$, $s$, and $c+\kappa$. Both terms exceed one, because $0<-H(\theta_{s1})=-\theta_{s1}-\Phi_{s1}/\phi_{s1}<-\theta_{s1}$ and $\theta_{s1}<0<H(\theta_{s2})$. We show that, for every $r>\underline r$,
$$\sup_{\theta_1:\,V_S(\theta_1)>0}\frac{V_A(\theta_1;w^*)}{V_S(\theta_1)}\leq\alpha_VM.$$

Suppose first that the cap is intermediate. By Proposition~\ref{prop:optimistic-explicit-conditions}, the left-hand side is the larger of the two terms in~\eqref{eq:optimistic-intermediate-guarantee}, which in terms of $H$ are $(r-\beta^*)/(-H(\theta_{s1}))$ and $\alpha^*+\beta^*/H(\theta_{s2})$. In this regime, $0<\beta^*<r<\overline r=-\alpha_V\theta_{s1}$. Because the floor is slack at the upper cutoff, that cutoff equation reads $\left(p\Phi_{s2}+(1-p)\Phi_0\right)\beta^*+(c+\kappa)\alpha^*=c$, so $\alpha^*<\alpha_V$. Hence
$$\frac{r-\beta^*}{-H(\theta_{s1})}<\frac{\overline r}{-H(\theta_{s1})}=\alpha_V\frac{\theta_{s1}}{H(\theta_{s1})},\quad\alpha^*+\frac{\beta^*}{H(\theta_{s2})}<\alpha_V+\frac{\overline r}{H(\theta_{s2})}=\alpha_V\left(1-\frac{\theta_{s1}}{H(\theta_{s2})}\right),$$
and both bounds are at most $\alpha_VM$.

Suppose next that the cap is loose, so $w^*=(\alpha_V,0)$ and $r\geq\overline r=-\alpha_V\theta_{s1}$. Wherever the floor is slack, $V_A(\theta_1;w^*)=\alpha_VV_S(\theta_1)$, so the agent's share is $\alpha_V<\alpha_VM$. The floor binds only for $\theta_1<-r/\alpha_V\leq\theta_{s1}$, where it adds $p\Phi((\theta_1-\Delta)/s)(-r-\alpha_V\theta_1)$ to the agent's payoff. Since $-r\leq\alpha_V\theta_{s1}$, this addition is at most $\alpha_Vp\Phi((\theta_1-\Delta)/s)(\theta_{s1}-\theta_1)$. Because $\Phi$ is increasing and, by the definition of $H$, $\Phi((t-\Delta)/s)=\phi((t-\Delta)/s)(H(t)-t)/s$,
$$\Phi\left(\frac{\theta_1-\Delta}{s}\right)(\theta_{s1}-\theta_1)\leq\int_{\theta_1}^{\theta_{s1}}\Phi\left(\frac{t-\Delta}{s}\right)dt=\frac{1}{s}\int_{\theta_1}^{\theta_{s1}}\phi\left(\frac{t-\Delta}{s}\right)\left(H(t)-t\right)dt.$$
Integrating the factorization in Lemma~\ref{lem:surplus-shape} over $[\theta_1,\theta_{s1}]$ and using $V_S(\theta_{s1})=0$ gives
$$V_S(\theta_1)=\frac{p}{s}\int_{\theta_1}^{\theta_{s1}}\phi\left(\frac{t-\Delta}{s}\right)\left(-H(t)\right)dt.$$
On $[\theta_1,\theta_{s1}]$, $H(t)-t=s\Phi((t-\Delta)/s)/\phi((t-\Delta)/s)\leq\Phi_{s1}/\phi_{s1}$, because the standard normal CDF is log-concave, and $-H(t)\geq-H(\theta_{s1})>0$, because $H$ is increasing. The addition is therefore at most $\alpha_VV_S(\theta_1)(\Phi_{s1}/\phi_{s1})/(-H(\theta_{s1}))$, and the agent's share is at most
$$\alpha_V\left(1+\frac{\Phi_{s1}/\phi_{s1}}{-H(\theta_{s1})}\right)=\alpha_V\frac{\theta_{s1}}{H(\theta_{s1})}\leq\alpha_VM.$$

Hence, for every $r>\underline r$, the agent's largest share is at most $\alpha_VM=cM/(c+\kappa)$. Fix $\lambda\in(0,1)$ and set $\overline c_\lambda=(1-\lambda)(c+\kappa)/M$. This threshold depends only on $\Delta$, $p$, $\theta_0$, $s$, and $c+\kappa$, and it is less than $c+\kappa$ because $M>1$. Whenever $0<c<\overline c_\lambda$ and $r>\underline r$, the agent's largest share is strictly less than $1-\lambda$. By Theorem~\ref{thm:optimistic-default}, $\lambda^*$ then exceeds $\lambda$ and $w^*$ is uniquely optimal; in particular, $w^*$ provides the guarantee $\lambda$. This completes the proof.
\end{proof}

\begin{proof}[Proof of Proposition~\ref{prop:zero-guarantee-bonus-family}]
Fix $\theta_a\geq\theta_0$. Under $w_a$, the agent's payoff is $V_A(\theta_1;w_a)=\beta_a\left(p\Phi\left(\frac{\theta_1-\Delta}{s}\right)+(1-p)\Phi_0\right)-c$, which is strictly increasing in $\theta_1$ and equals zero at $\theta_a$ by the definition of $\beta_a$. Under the maintained tie-breaking rule, the agent therefore experiments if and only if $\theta_1\geq\theta_a$.

We first show that $w_a$ is protected. Since $\theta_0>\overline\theta_0$, we have $(1-p)\theta_0\Phi_0>c+\kappa>c$, so $\beta_a\leq c/\left((1-p)\Phi_0\right)<\theta_0\leq\theta_a$. For $\theta_1\geq\theta_a$,
$$V_P(\theta_1;w_a)=p\Phi\left(\frac{\theta_1-\Delta}{s}\right)(\theta_1-\beta_a)+(1-p)\Phi_0(\theta_0-\beta_a)-\kappa,$$
whose derivative $p\Phi((\theta_1-\Delta)/s)+(p/s)\phi((\theta_1-\Delta)/s)(\theta_1-\beta_a)$ is positive. At $\theta_a$, the agent is indifferent, so $V_P(\theta_a;w_a)=V_S(\theta_a)>0$, because $\theta_a>0>\theta_{s2}$. Hence $U_P(\theta_1;w_a)>0$ for every $\theta_1\geq\theta_a$, while $U_P(\theta_1;w_a)=0$ for $\theta_1<\theta_a$, so $w_a$ is protected. Since the null contract gives the principal zero at every alternative value, $w_a$ also dominates the null contract.

Next, $\lambda(w_a)=0$: at every $\theta_1<\theta_{s1}$, total surplus is positive, but $\theta_1<\theta_a$, so $U_P(\theta_1;w_a)=0$.

It remains to prove that $w_a$ is undominated. Suppose that a contract $\tilde w=(\tilde\alpha,\tilde\beta)$ dominates $w_a$. At $\theta_a$, the contract $w_a$ gives the principal $V_S(\theta_a)$, which by Lemma~\ref{lem:full-information-benchmark} is the most any contract can give there. Dominance therefore requires $U_P(\theta_a;\tilde w)=V_S(\theta_a)$, so $\tilde w$ induces experimentation at $\theta_a$ and $V_A(\theta_a;\tilde w)=0$. Because $\theta_a\geq\theta_0>0$, the liability floor binds in neither state at $\theta_a$, and subtracting $V_A(\theta_a;w_a)=0$ gives
$$0=p\Phi\left(\frac{\theta_a-\Delta}{s}\right)\left(\tilde\beta+\tilde\alpha\theta_a-\beta_a\right)+(1-p)\Phi_0\left(\tilde\beta+\tilde\alpha\theta_0-\beta_a\right).$$
Since $\theta_0\leq\theta_a$, the first term in parentheses is at least the second. As the weights are positive and the weighted sum is zero, the first term is nonnegative: $\tilde\beta+\tilde\alpha\theta_a-\beta_a\geq0$.

Now let $\theta_1>\theta_a$. Both contracts give the agent zero payoff at $\theta_a$, and the liability floor binds under neither contract at $\theta_1$. Subtracting each contract's payoff at $\theta_a$ from its payoff at $\theta_1$ and rearranging gives
$$V_A(\theta_1;\tilde w)-V_A(\theta_1;w_a)=p\left(\Phi\left(\frac{\theta_1-\Delta}{s}\right)-\Phi\left(\frac{\theta_a-\Delta}{s}\right)\right)\left(\tilde\beta+\tilde\alpha\theta_a-\beta_a\right)+p\tilde\alpha\Phi\left(\frac{\theta_1-\Delta}{s}\right)(\theta_1-\theta_a).$$
The first term is nonnegative. If $\tilde\alpha>0$, the second is positive, so the agent earns strictly more under $\tilde w$ than under $w_a$. Because $w_a$ gives the principal positive profit at $\theta_1$, dominance requires $\tilde w$ to induce experimentation there as well, and since total surplus does not depend on the contract, $U_P(\theta_1;\tilde w)<U_P(\theta_1;w_a)$, contradicting dominance. Hence $\tilde\alpha=0$, and the zero-rent condition at $\theta_a$ gives $\tilde\beta=\beta_a$. Thus $\tilde w=w_a$, contradicting the requirement that a dominating contract yield a strict improvement at some alternative value. Therefore, $w_a$ is undominated.

Finally, let $\theta_0\leq\theta_a<\theta_b$, so that $\beta_a>\beta_b$. Both contracts give the principal zero below $\theta_a$, and $w_a$ gives strictly more on $[\theta_a,\theta_b)$, where only $w_a$ induces experimentation. On $[\theta_b,\infty)$, both induce experimentation and
$$U_P(\theta_1;w_b)-U_P(\theta_1;w_a)=(\beta_a-\beta_b)\left(p\Phi\left(\frac{\theta_1-\Delta}{s}\right)+(1-p)\Phi_0\right)>0.$$
Distinct members of the family are therefore incomparable, and the family is a continuum of distinct undominated contracts. When $\lambda^*=0$, each of them is optimal, because it is protected and provides the guarantee zero.

This completes the proof.
\end{proof}

\subsection{Proofs for Section~\ref{subsec:profitability-threshold}}
\label{app:profitability-threshold}
We now prove the results of Section~\ref{subsec:profitability-threshold} and give the exact classification and coefficients of the optimal contract throughout the upper-region case. Throughout, we maintain the optimistic-default setting of Section~\ref{subsec:optimistic-default}, including the assumption that $V_S$ has a strictly negative minimum. We first establish three facts that the proofs below use repeatedly.
\begin{lemma}[Preliminaries for threshold guarantees]\label{lem:threshold-preliminaries}\ 
\begin{enumerate}
\item As $\theta_1\to-\infty$, $V_S(\theta_1)\to V_S(0)>0$, and $V_S<V_S(0)$ on $(-\infty,0)$. Consequently,
$$\Theta_\eta=\begin{cases}[\theta_\eta,\infty)&\text{if }\eta\geq V_S(0),\\(-\infty,\theta_\eta^-]\cup[\theta_\eta^+,\infty)&\text{if }0<\eta<V_S(0),\end{cases}$$
where $\theta_\eta$, and $\theta_\eta^-<\theta_\eta^+$, are the solutions of $V_S(\theta_1)=\eta$, with $\theta_\eta\geq0$ and $\theta_\eta^-<\theta_{s1}<\theta_{s2}<\theta_\eta^+<0$.
\item If a protected contract $w$ induces experimentation throughout $\Theta_\eta$, then
$$\lambda_\eta(w)=1-\sup_{\theta_1\in\Theta_\eta}\frac{V_A(\theta_1;w)}{V_S(\theta_1)};$$
otherwise, $\lambda_\eta(w)=0$.
\item If the liability floor does not bind in the alternative state at $\theta_1$ under $w=(\alpha,\beta)$, that is, if $\alpha\theta_1\geq-r$, then
$$V_A(\theta_1;w)=\beta\left(p\Phi\left(\frac{\theta_1-\Delta}{s}\right)+(1-p)\Phi_0\right)+\alpha\left(V_S(\theta_1)+c+\kappa\right)-c.$$
\end{enumerate}
\end{lemma}

\begin{proof}
For part~1, $\theta_1\Phi((\theta_1-\Delta)/s)\to0$ as $\theta_1\to-\infty$, so
$$\lim_{\theta_1\to-\infty}V_S(\theta_1)=(1-p)\theta_0\Phi_0-c-\kappa=V_S(0)>0,$$
where the inequality follows from $\theta_0>\overline\theta_0$. Moreover,
$$V_S(\theta_1)-V_S(0)=p\theta_1\Phi\left(\frac{\theta_1-\Delta}{s}\right),$$
so $V_S(\theta_1)<V_S(0)$ for $\theta_1<0$, $V_S(\theta_1)>V_S(0)$ for $\theta_1>0$, and $V_S$ is unbounded above. By Lemma~\ref{lem:surplus-shape}, $V_S$ is strictly decreasing up to $\underline\theta_1<0$ and strictly increasing thereafter. If $\eta\geq V_S(0)$, no negative alternative value belongs to $\Theta_\eta$, and strict monotonicity on $[0,\infty)$ gives a unique $\theta_\eta\geq0$ with $V_S(\theta_\eta)=\eta$, so that $\Theta_\eta=[\theta_\eta,\infty)$; in particular, $\theta_\eta=0$ when $\eta=V_S(0)$. If $0<\eta<V_S(0)$, then $\lim_{\theta_1\to-\infty}V_S(\theta_1)>\eta$ and $V_S(\theta_{s1})=0<\eta$, so strict decrease on the lower branch gives a unique $\theta_\eta^-<\theta_{s1}$ with $V_S(\theta_\eta^-)=\eta$. Similarly, $V_S(\theta_{s2})=0<\eta<V_S(0)$, and strict increase on $[\theta_{s2},0]$ gives a unique $\theta_\eta^+\in(\theta_{s2},0)$ with $V_S(\theta_\eta^+)=\eta$. Thus $\Theta_\eta=(-\infty,\theta_\eta^-]\cup[\theta_\eta^+,\infty)$ with $\theta_\eta^-<\theta_{s1}<\theta_{s2}<\theta_\eta^+<0$.

For part~2, note that $V_S(\theta_1)\geq\eta>0$ on $\Theta_\eta$. If $w$ induces experimentation throughout $\Theta_\eta$, then $U_P(\theta_1;w)=V_S(\theta_1)-V_A(\theta_1;w)$ on this set, and therefore
$$\lambda_\eta(w)=\inf_{\theta_1\in\Theta_\eta}\left(1-\frac{V_A(\theta_1;w)}{V_S(\theta_1)}\right)=1-\sup_{\theta_1\in\Theta_\eta}\frac{V_A(\theta_1;w)}{V_S(\theta_1)}.$$
If instead $w$ fails to induce experimentation at some $\theta_1\in\Theta_\eta$, then $U_P(\theta_1;w)=0$. Protection gives $U_P\geq0$ at every alternative value, so the infimum defining $\lambda_\eta(w)$ equals zero.

For part~3, the liability floor never binds in the default state, because $\theta_0>0$ and $\alpha\geq0$. If it also does not bind in the alternative state, then
$$V_A(\theta_1;w)=\beta\left(p\Phi\left(\frac{\theta_1-\Delta}{s}\right)+(1-p)\Phi_0\right)+\alpha\left(p\theta_1\Phi\left(\frac{\theta_1-\Delta}{s}\right)+(1-p)\theta_0\Phi_0\right)-c.$$
The expression multiplying $\alpha$ is $V_S(\theta_1)+c+\kappa$, which gives the stated formula.

This completes the proof.
\end{proof}

\begin{proof}[Proof of Proposition~\ref{prop:eta-strict-bonus}]
First let $w=(\alpha,0)$ be protected and induce experimentation throughout $\Theta_\eta$. By Lemma~\ref{lem:threshold-preliminaries}, the agent's share under $w$ has supremum $1-\lambda_\eta(w)$ on $\Theta_\eta$, and this supremum is at least $\alpha>0$, because the share tends to $\alpha$ as $\theta_1\to\infty$. The floor does not bind under $w$ at the upper boundary of $\Theta_\eta$: it cannot bind at a nonnegative boundary, and binding at $\theta_\eta^+<0$ would make the agent's payoff strictly decrease from $\theta_{s2}$, where protection gives $V_A\leq0$, to $\theta_\eta^+$, contradicting participation there. Hence $\alpha(c+\kappa+\eta)\geq c$.

If $\alpha(c+\kappa+\eta)>c$, let $\tilde w=\left((1-\varepsilon)\alpha,\varepsilon c/2\right)$. Since $\max\{(1-\varepsilon)\alpha\theta,-r\}\leq(1-\varepsilon)\max\{\alpha\theta,-r\}$ and the launch probability is at most one,
$$V_A(\theta_1;\tilde w)\leq(1-\varepsilon)V_A(\theta_1;w)-\frac{\varepsilon c}{2},$$
so protection is preserved. On $\Theta_\eta$, the floor and the bonus only raise payments and $V_S\geq\eta$, so for small $\varepsilon>0$, $V_A(\theta_1;\tilde w)\geq(1-\varepsilon)\alpha(c+\kappa+\eta)-c>0$. Thus $\tilde w$ induces experimentation throughout $\Theta_\eta$, and the supremum of the agent's share falls to at most $(1-\varepsilon)\left(1-\lambda_\eta(w)\right)<1-\lambda_\eta(w)$.

Now suppose $\alpha(c+\kappa+\eta)=c$, so the agent is indifferent at the upper boundary. Let $t_\eta$ be the smallest boundary of $\Theta_\eta$ at which the agent is indifferent. The floor does not bind at any such boundary, since a binding floor would raise the payoff above $\alpha(c+\kappa+\eta)-c=0$. Let
$$\tilde\alpha=\alpha-\varepsilon,\qquad\tilde\beta=\varepsilon\,\frac{c+\kappa+\eta}{p\Phi\left(\frac{t_\eta-\Delta}{s}\right)+(1-p)\Phi_0},$$
with $\varepsilon>0$ small enough that $\tilde\alpha>0$, $\tilde\beta<r$, $p\tilde\beta\leq\tilde\alpha\eta$, and positive boundary payoffs under $w$ remain positive under $\tilde w$. The agent remains indifferent at $t_\eta$, and higher indifferent boundaries acquire positive payoffs. Where the floor does not bind under $\tilde w$,
$$V_A(\theta_1;\tilde w)=\tilde\alpha\left(V_S(\theta_1)-\eta\right)+p\tilde\beta\left(\Phi\left(\frac{\theta_1-\Delta}{s}\right)-\Phi\left(\frac{t_\eta-\Delta}{s}\right)\right)\leq\tilde\alpha V_S(\theta_1).$$
Where it binds, it also binds under $w$ and $\theta_1<t_\eta$, so
$$V_A(\theta_1;\tilde w)-V_A(\theta_1;w)=\tilde\beta\left(p\Phi\left(\frac{\theta_1-\Delta}{s}\right)+(1-p)\Phi_0\right)-\varepsilon(1-p)\theta_0\Phi_0\leq-\varepsilon\left(V_S(0)-\eta\right).$$
If $\eta<V_S(0)$, the agent therefore earns strictly less under $\tilde w$ than under $w$ wherever the floor binds, and protection there is inherited from $w$. If $\eta\geq V_S(0)$, the agent's payoff wherever the floor binds is below its limit as $\theta_1\to-\infty$, namely $\tilde\alpha\left(V_S(0)-\eta\right)-p\tilde\beta\Phi((t_\eta-\Delta)/s)<0$, so the agent does not experiment there. Together with $0<\tilde\alpha<1$, these bounds show that $\tilde w$ is protected. Participation at the boundaries, nonparticipation on $(\theta_{s1},\theta_{s2})$, and Lemma~\ref{lem:agent-shape} then give experimentation throughout $\Theta_\eta$.

Finally, compare shares on $\Theta_\eta$. When $\eta\geq V_S(0)$, the floor does not bind on $\Theta_\eta$, so the supremum under $\tilde w$ is at most $\tilde\alpha$. When $0<\eta<V_S(0)$, every value in $\Theta_\eta$ where the floor binds satisfies $0<V_S(\theta_1)<V_S(0)$, so the supremum is at most $\max\{\tilde\alpha,\,1-\lambda_\eta(w)-\varepsilon(V_S(0)-\eta)/V_S(0)\}$. In both cases it is strictly below $1-\lambda_\eta(w)$, because $\tilde\alpha<\alpha\leq1-\lambda_\eta(w)$. Thus every protected pure-sharing contract that induces experimentation throughout $\Theta_\eta$ is strictly improved.

A protected pure-sharing contract that fails to induce experimentation throughout $\Theta_\eta$ has guarantee zero. If $\eta\geq V_S(0)$, choose positive coefficients with $\tilde\alpha(c+\kappa+\eta)+\tilde\beta\left(p\Phi((\theta_\eta-\Delta)/s)+(1-p)\Phi_0\right)=c$ and $\tilde\beta$ small; the argument above for $\eta\geq V_S(0)$ gives a protected mixed contract with guarantee at least $1-\tilde\alpha>0$. If $0<\eta<V_S(0)$, the assumption $\lambda_\eta^*>0$ supplies a protected contract with a positive guarantee. It cannot be a pure launch bonus, whose increasing payoff would extend experimentation from the lower part of $\Theta_\eta$ into the negative-surplus interval $(\theta_{s1},\theta_{s2})$. It is therefore mixed already or a covering pure-sharing contract, which the argument above strictly improves to a mixed contract.

Hence every protected pure-sharing contract is strictly improved by a protected mixed contract, and every contract attaining $\lambda_\eta^*$ has a strictly positive launch bonus. This completes the proof.
\end{proof}

\begin{proof}[Proof of Proposition~\ref{prop:threshold-guarantee-frontier}]
Because $r>\underline r$, protection is equivalent to
\begin{equation}\label{eq:protection-characterization}
V_A(\theta_1;w)\leq\max\{V_S(\theta_1),0\}\quad\text{for every }\theta_1.
\end{equation}
If~\eqref{eq:protection-characterization} fails at some $\theta_1$, the agent experiments there and $V_P(\theta_1;w)=V_S(\theta_1)-V_A(\theta_1;w)<0$, so $w$ is not protected. Conversely, under~\eqref{eq:protection-characterization}, protection can fail only at a point of $(\theta_{s1},\theta_{s2})$ where the agent's payoff is zero, and~\eqref{eq:protection-characterization} makes this payoff nonpositive throughout the interval. If $\alpha=0$, the payoff is strictly increasing or identically $-c$, which rules this out. If $\alpha>0$, Lemma~\ref{lem:agent-shape} implies that such a zero requires $\beta=r$ and $V_A(\theta_1;w)=0$ for every $\theta_1\leq\theta_{s2}$, so $w$ solves both cutoff equations. Theorem~\ref{thm:optimistic-default} then gives $w=w^*$, which is impossible, since $w^*$ is implementing for $r>\underline r$ and therefore $V_A(\theta_1;w^*)<0$ on $(\theta_{s1},\theta_{s2})$.

Since $V_A(\theta_1;w)$ is continuous in $w$, \eqref{eq:protection-characterization} shows that $\mathcal W_P$ is closed. Since $V_S(\theta_{s2})=0$ and the liability floor can only raise compensation, \eqref{eq:protection-characterization} at $\theta_{s2}$ implies $\alpha(c+\kappa)+\beta\left(p\Phi_{s2}+(1-p)\Phi_0\right)\leq c$. Hence $\mathcal W_P$ is compact, and it is nonempty because it contains the null contract. For each $\theta_1$ with $V_S(\theta_1)>0$, $U_P(\theta_1;w)$ is continuous in $w$ wherever the agent is not indifferent at $\theta_1$, and equals its largest possible value $V_S(\theta_1)$ where he is. It is therefore upper semicontinuous in $w$. The same holds for $\lambda_\eta$, an infimum of such payoff ratios, so $\lambda_\eta^*$ is attained.

Since $\lambda_\eta^*$ is nondecreasing in $\eta$ and $\lambda_\eta^*\geq\lambda^*$ for every $\eta$, the limit $\hat\lambda=\lim_{\eta\downarrow0}\lambda_\eta^*$ exists and satisfies $\hat\lambda\geq\lambda^*$. Because $\lambda_\eta^*\geq\hat\lambda$ and $\lambda_\eta$ is upper semicontinuous, the sets $\left\{w\in\mathcal W_P:\lambda_\eta(w)\geq\hat\lambda\right\}$ are nonempty and compact, and they are nested as $\eta$ decreases. Their intersection is therefore nonempty. Since $\bigcup_{\eta>0}\Theta_\eta=\{\theta_1:V_S(\theta_1)>0\}$, any contract in this intersection provides the unrestricted guarantee $\hat\lambda$, so $\hat\lambda=\lambda^*$.

If $\lambda_\eta^*>0$ for every $\eta>0$, the protected contracts inducing experimentation throughout $\Theta_\eta$ likewise form nonempty nested compact sets. The same argument gives a protected contract inducing experimentation wherever $V_S>0$. By continuity, the agent's payoff under this contract is nonnegative at both efficient cutoffs, and by~\eqref{eq:protection-characterization} it is nonpositive there. The contract therefore solves both cutoff equations, so it is $w^*$ by Theorem~\ref{thm:optimistic-default}, and $w^*$ is protected.

Conversely, suppose $w^*$ is protected and fix $\eta>0$. By Theorem~\ref{thm:optimistic-default}, $\alpha^*>0$ and $\beta^*<r$, so by Lemma~\ref{lem:agent-shape} the agent's payoff under $w^*$ has no zeros other than the efficient cutoffs. It is therefore strictly positive at every boundary point of $\Theta_\eta$. For $q\in(0,1)$ sufficiently close to one, the same holds under $qw^*$. Since $\max\{q\alpha^*\theta,-r\}\leq q\max\{\alpha^*\theta,-r\}$ for every $\theta$, $V_A(\theta_1;qw^*)\leq qV_A(\theta_1;w^*)-(1-q)c$, so $qw^*$ satisfies~\eqref{eq:protection-characterization} and is protected. This bound also makes the agent's payoff under $qw^*$ negative on $(\theta_{s1},\theta_{s2})$, so Lemma~\ref{lem:agent-shape}, applied to $qw^*$, gives experimentation throughout $\Theta_\eta$, where $U_P(\theta_1;qw^*)\geq V_S(\theta_1)-qV_A(\theta_1;w^*)+(1-q)c\geq(1-q)\left(V_S(\theta_1)+c\right)$. Hence $\lambda_\eta^*\geq1-q>0$. If $w^*$ is unprotected, the equivalence just proved and monotonicity imply $\lambda_\eta^*=0$ for all sufficiently small $\eta>0$. 

This completes the proof.
\end{proof}

\begin{proof}[Proof of Proposition~\ref{prop:bonus-threshold-guarantee}]
The contracts leaving the agent indifferent at $\theta_\eta$ form the segment $$\alpha(c+\kappa+\eta)+\beta\left(p\Phi\left(\frac{\theta_\eta-\Delta}{s}\right)+(1-p)\Phi_0\right)=c,\quad\alpha,\beta\geq0.$$ The cutoff comparison in the proof of Proposition~\ref{prop:rule1} applies with $\theta_\eta$ in place of $\theta_s$: since $\theta_\eta\geq0$, every segment contract induces experimentation exactly on $\Theta_\eta$. Any protected contract $w$ with a positive guarantee must induce experimentation at $\theta_\eta$. If $V_A(\theta_\eta;w)>0$, scaling both coefficients by the $q\in(0,1)$ that makes the agent indifferent at $\theta_\eta$ preserves protection and strictly improves the guarantee, by $$V_A(\theta_1;qw)\leq qV_A(\theta_1;w)-(1-q)c.$$ Thus no positive-guarantee optimum lies outside the segment.

On $\Theta_\eta$ the floor does not bind, so by Lemma~\ref{lem:threshold-preliminaries} the agent's share $V_A(\theta_1;w)/V_S(\theta_1)$ is affine in $w$, with coefficients bounded uniformly over $\Theta_\eta$ because $V_S\geq\eta>0$ there. Its supremum over $\Theta_\eta$ is therefore convex and continuous in $w$ and attains a minimum on the segment. At the pure-sharing end of the segment, $\beta=0$ and $\alpha=c/(c+\kappa+\eta)$, so the share is $\alpha\left(1-\eta/V_S(\theta_1)\right)$, with supremum $\alpha<1$. Every minimizer is therefore protected, because experimentation occurs exactly on $\Theta_\eta$. Hence $\lambda_\eta^*$ is one minus the minimum and is positive, and the optimal contracts are exactly the minimizers.

Suppose $w_1=(\alpha_1,\beta_1)$ and $w_2=(\alpha_2,\beta_2)$ are distinct minimizers with $\alpha_1<\alpha_2$, so that $\beta_1>\beta_2$. By convexity, their midpoint is also a minimizer. The agent's share under the midpoint is zero at $\theta_\eta$ and tends to $(\alpha_1+\alpha_2)/2$ as $\theta_1\to\infty$, whereas its supremum equals that under $w_2$, which is at least $\alpha_2$. This supremum is therefore attained at some $x>\theta_\eta$, and since the share is affine in $w$, the shares under $w_1$ and $w_2$ attain their suprema at $x$ as well. Hence $V_A(\theta_1;w_2)-V_A(\theta_1;w_1)$, which is zero at $\theta_\eta$, has value and derivative zero at $x$. Since the floor does not bind on $\Theta_\eta$, the derivative of this difference has the sign of $(\alpha_2-\alpha_1)H(\theta_1)-(\beta_1-\beta_2)$, which is strictly increasing in $\theta_1$ by Lemma~\ref{lem:surplus-shape}. The difference is therefore strictly decreasing on $[\theta_\eta,x]$, contradicting its being zero at both $\theta_\eta$ and $x$. The optimal contract is thus unique. Neither the segment nor the agent's share on $\Theta_\eta$ depends on $r$, which proves the claimed independence.

Finally, suppose $\theta_\eta\geq\theta_0$. The comparison in the proof of Proposition~\ref{prop:zero-guarantee-bonus-family}, with $\theta_\eta$ in place of $\theta_a$, shows that every segment contract with $\alpha>0$ gives the agent strictly more than the pure launch bonus on the segment at every $\theta_1>\theta_\eta$. The pure launch bonus therefore minimizes the supremum of the agent's share, and by uniqueness it is the optimal contract $w_\eta^*$. This completes the proof.
\end{proof}

We now give the exact classification and coefficients of the optimal contract in the upper-region case.

\begin{proposition}[Exact solution in the upper region]
\label{prop:upper-region-classification}
Suppose $\Theta_\eta=[\theta_\eta,\infty)$ with $\theta_\eta\geq0$, consider the pure launch bonus and the pure-sharing contract that leave the agent indifferent at $\theta_\eta$, and let $H$ be as in Lemma~\ref{lem:surplus-shape}. Under the pure launch bonus, the agent's share of total surplus attains its maximum over $\Theta_\eta$ at a unique point $x$.
\begin{enumerate}
\item If, at $x$, the agent's payoff under the pure-sharing contract is at least his payoff under the pure launch bonus, the pure launch bonus is optimal and $$\lambda_\eta^*=1-\frac{c}{\left(p\Phi\left(\frac{\theta_\eta-\Delta}{s}\right)+(1-p)\Phi_0\right)H(x)}.$$
\item Otherwise, the two contracts give the agent the same payoff at a unique $y>\theta_\eta$, and the optimal contract is the mixed contract $(\alpha,\beta)$ solving $$\alpha(c+\kappa+\eta)+\beta\left(p\Phi\left(\frac{\theta_\eta-\Delta}{s}\right)+(1-p)\Phi_0\right)=c$$ and $$\alpha+\frac{\beta}{H(y)}=\frac{c}{c+\kappa+\eta}\left(1-\frac{\eta}{V_S(y)}\right).$$ Its guarantee is $\lambda_\eta^*=1-\left(c/(c+\kappa+\eta)\right)\left(1-\eta/V_S(y)\right)$.
\end{enumerate}
\end{proposition}

\begin{proof}
By the proof of Proposition~\ref{prop:bonus-threshold-guarantee}, the optimal contract minimizes the supremum of the agent's share over $\Theta_\eta$ among the contracts $w=(\alpha,\beta)$ satisfying the first equation in part~2, whose two ends are the contracts in the statement. For such $w$ and $t\geq\theta_\eta$, Lemma~\ref{lem:threshold-preliminaries} gives $$V_A(t;w)=\alpha\left(V_S(t)-\eta\right)+p\beta\left(\Phi\left(\frac{t-\Delta}{s}\right)-\Phi\left(\frac{\theta_\eta-\Delta}{s}\right)\right).$$ This payoff is affine in $w$, and by Lemma~\ref{lem:surplus-shape} its derivative in $t$ is $\alpha+\beta/H(t)$ times the derivative of $V_S$, which is positive. The share $V_A(t;w)/V_S(t)$ therefore increases where it is below $\alpha+\beta/H(t)$ and decreases where it is above, and $\alpha+\beta/H(t)$ minus the share equals $1/V_S(t)$ times $$\alpha\eta+\beta\left(\frac{V_S(t)}{H(t)}-p\left(\Phi\left(\frac{t-\Delta}{s}\right)-\Phi\left(\frac{\theta_\eta-\Delta}{s}\right)\right)\right).$$ The coefficient of $\beta$ is $\eta/H(\theta_\eta)>0$ at $\theta_\eta$, tends to $-p\left(1-\Phi\left(\frac{\theta_\eta-\Delta}{s}\right)\right)<0$, and is strictly decreasing, with derivative $-V_S(t)\left(\partial H(t)/\partial t\right)/H(t)^2$. Under the pure launch bonus, the share therefore has a unique maximizer $x$, the zero of this coefficient, and its maximum is $\beta/H(x)$, which gives the guarantee in part~1.

On $\Theta_\eta$, the agent's payoff under each segment contract lies between his payoffs under the two ends. If, at $x$, pure sharing pays him at least as much as the pure launch bonus, every segment contract does too, so its supremum share is at least the maximum under the pure launch bonus. The pure launch bonus is then optimal, and uniqueness follows from Proposition~\ref{prop:bonus-threshold-guarantee}.

Otherwise, his payoff under pure sharing minus that under the pure launch bonus is zero at $\theta_\eta$, negative at $x$, and tends to infinity, and its derivative has the sign of $cH(t)/(c+\kappa+\eta)-\beta$, with $\beta$ the coefficient of the pure launch bonus; this is strictly increasing in $t$. This difference therefore has a unique zero $y>\theta_\eta$, and $y>x$. At $y$, every segment contract gives the agent the same payoff, so its share there equals $\left(c/(c+\kappa+\eta)\right)\left(1-\eta/V_S(y)\right)$, the share under pure sharing, and this value bounds its supremum share from below. For a segment contract, the second equation in part~2 thus says that the share at $y$ equals $\alpha+\beta/H(y)$, so that the displayed expression is zero at $t=y$. Since $y>x$, the coefficient of $\beta$ is negative at $y$, so exactly one segment contract satisfies this condition, and it has $\alpha,\beta>0$. For this contract the displayed expression is strictly decreasing in $t$, so its share is maximized at $y$ and attains the lower bound. It is therefore optimal and has the stated guarantee, and uniqueness again follows from Proposition~\ref{prop:bonus-threshold-guarantee}. 

This completes the proof.
\end{proof}

\subsection{Proofs for Section~\ref{sec:distributional-uncertainty}}
The proofs below also establish the restricted claims in the footnotes to Proposition~\ref{prop:distributional-unrestricted-impossibility} and Theorem~\ref{thm:distributional-mixed-guarantees}.

\begin{proof}[Proof of Proposition~\ref{prop:distributional-unrestricted-impossibility}]
It suffices to prove both claims on the subfamily $$\mu=(1-p)\delta_{\theta_0}+p\delta_{\theta_1},\quad\theta_0\geq0,\quad\theta_1\leq0.$$ Within it, total surplus is strictly increasing in $\theta_0$, because $\theta\Phi((\theta-\Delta)/s)$ increases strictly from zero to infinity on $[0,\infty)$. For each $\theta_1\leq0$, total surplus is therefore zero at a unique $\theta_0>0$, which equals $\overline\theta_0$ when $\theta_1=0$ and tends to $\overline\theta_0$ as $\theta_1\to-\infty$.

Suppose a contract $w=(\alpha,\beta)$ either implements efficient experimentation or provides a strictly positive uniform guarantee on this subfamily. Either property requires experimentation wherever $V_S>0$ and no experimentation wherever $V_S<0$. Raising or lowering $\theta_0$ slightly around a break-even distribution therefore gives, by continuity, $V_A(\mu;w)=0$ whenever $V_S(\mu)=0$. When $\theta_1=0$, the floor does not bind, so this indifference reads $$\alpha(c+\kappa)+\beta\left((1-p)\Phi\left(\frac{\overline\theta_0-\Delta}{s}\right)+p\Phi\left(-\frac{\Delta}{s}\right)\right)=c.$$ As $\theta_1\to-\infty$ along break-even distributions, the expected payment in the alternative state tends to zero, because its launch probability tends to zero and its conditional payment lies between $\beta-r$ and $\beta$. Taking limits in the indifference condition gives $$\alpha(c+\kappa)+\beta(1-p)\Phi\left(\frac{\overline\theta_0-\Delta}{s}\right)=c.$$ Subtracting yields $\beta p\Phi(-\Delta/s)=0$, so $\beta=0$ and $\alpha=c/(c+\kappa)=\alpha_V$.

Without the floor, proportional sharing would give the agent $\alpha_VV_S(\mu)$, which is zero at every break-even distribution. Take the break-even distribution with $\theta_1<-r/\alpha_V$. The floor binds in its alternative state, so $$V_A(\mu;w)=p\Phi\left(\frac{\theta_1-\Delta}{s}\right)\left(-r-\alpha_V\theta_1\right)>0,$$ contradicting indifference. Hence no contract implements efficient experimentation or provides a strictly positive guarantee, even on this subfamily. The null contract provides the guarantee zero throughout $\mathcal M$, so $\lambda^*=0$. This completes the proof.
\end{proof}

\begin{proof}[Proof of Theorem~\ref{thm:distributional-mixed-guarantees}]
Write $$\Phi_\mu=\int_{\mathbb R}\Phi\left(\frac{\theta-\Delta}{s}\right)\,\mu(d\theta)$$ for the launch probability under $\mu$, conditional on experimentation. 

Let $w=(0,\beta)$ be protected. Under a pure launch bonus, the agent's payoff $\beta\Phi_\mu-c$ depends on $\mu$ only through $\Phi_\mu$. Total surplus at the zero-quality distribution $\delta_0$ is $-c-\kappa<0$, so protection requires $\beta\Phi(-\Delta/s)<c$. The agent therefore does not experiment at any $\mu$ with $\Phi_\mu\leq\Phi(-\Delta/s)$, the launch probability under $\delta_0$. Some of these distributions have arbitrarily large total surplus. Fix $q\in(0,\Phi(-\Delta/s))$, choose $x>0$ with $(1-q)\Phi((-x-\Delta)/s)+q<\Phi(-\Delta/s)$, and let $\mu_y=(1-p)\delta_0+p(1-q)\delta_{-x}+pq\delta_y$ for $y>0$. Each $\mu_y$ lies in $\mathcal M$ and has launch probability below $\Phi(-\Delta/s)$, so $w$ induces no experimentation at $\mu_y$. Since $V_S(\mu_y)\to\infty$ as $y\to\infty$, every $\mathcal M_\eta$ contains some $\mu_y$, and hence $\lambda_\eta(w)=0$ for every $\eta>0$. A positive guarantee therefore requires a positive value share.

For part~2, we first bound how much the floor can raise the agent's payoff. Let $$\overline\eta=p\sup_{\theta<0}\left(-\theta\Phi\left(\frac{\theta-\Delta}{s}\right)\right),$$ which is positive and finite because $-\theta\Phi((\theta-\Delta)/s)$ is positive on $(-\infty,0)$ and tends to zero as $\theta\to-\infty$ and as $\theta\uparrow0$. Under $w=(\alpha,\beta)$, the floor adds at most $-\alpha\theta$ to the payment after an outcome $\theta<0$ and nothing after $\theta\geq0$. Since negative outcomes have probability at most $p$ and $(1-p)\Phi(-\Delta/s)\leq\Phi_\mu\leq1$, every $\mu\in\mathcal M$ satisfies $$\beta(1-p)\Phi\left(-\frac{\Delta}{s}\right)\leq V_A(\mu;w)+c-\alpha\left(V_S(\mu)+c+\kappa\right)\leq\beta+\alpha\overline\eta.$$

Now fix $\eta>\overline\eta$ and add a small launch bonus to value sharing. For $\beta\in(0,c)$, let $$\alpha=\frac{c-\beta(1-p)\Phi(-\Delta/s)}{c+\kappa+\eta},$$ which lies in $(0,1)$, and let $w=(\alpha,\beta)$. This choice makes $\alpha(c+\kappa+\eta)+\beta(1-p)\Phi(-\Delta/s)=c$, so the lower bound becomes $V_A(\mu;w)\geq\alpha\left(V_S(\mu)-\eta\right)$. The agent therefore experiments at every $\mu\in\mathcal M_\eta$.

It remains to choose $\beta$ so that $w$ is protected. As $\beta\downarrow0$, $\alpha$ tends to $c/(c+\kappa+\eta)$, so $\alpha(c+\kappa+\overline\eta)+\beta$ tends to $c(c+\kappa+\overline\eta)/(c+\kappa+\eta)$, which is below $c$ because $\eta>\overline\eta$. Fix $\beta$ small enough that $\alpha(c+\kappa+\overline\eta)+\beta<c$. The upper bound then gives $V_A(\mu;w)\leq\alpha V_S(\mu)+\alpha(c+\kappa+\overline\eta)+\beta-c<\alpha V_S(\mu)$ for every $\mu\in\mathcal M$. Hence the agent experiments only where $V_S(\mu)>0$, and there the principal receives $V_S(\mu)-V_A(\mu;w)>(1-\alpha)V_S(\mu)>0$. So $w$ is protected, and on $\mathcal M_\eta$ the agent's share is below $\alpha$, which gives $\lambda_\eta(w)\geq1-\alpha>0$.

Pure sharing does worse. A protected pure-sharing contract with a positive guarantee induces experimentation throughout $\mathcal M_\eta$. At a nonnegative-support $\mu$ with $V_S(\mu)=\eta$, the floor does not bind, so its value share is at least $c/(c+\kappa+\eta)$, which exceeds $\alpha$. As $V_S(\mu)\to\infty$ along nonnegative-support distributions, the agent's share under it tends to its value share, so its guarantee is at most $1-c/(c+\kappa+\eta)<1-\alpha\leq\lambda_\eta(w)$. Combined with the first paragraph, $w$ beats every protected single-instrument contract.

It remains to show that $\lambda_\eta^*$ is attained. At a nonnegative-support $\mu$ with $V_S(\mu)=0$, every protected contract $(\alpha,\beta)$ satisfies $\alpha(c+\kappa)+\beta\Phi_\mu\leq c$, so $\mathcal W_P$ is bounded. To see that it is closed, let protected contracts converge to $w=(\alpha,\beta)$. Since $V_A(\mu;\cdot)$ is continuous, $V_A(\mu;w)\leq\max\{V_S(\mu),0\}$ for every $\mu\in\mathcal M$, so $w$ can fail to be protected only if $V_A(\mu;w)=0$ at some $\mu$ with $V_S(\mu)<0$. Suppose this happens. Then $\alpha>0$ or $\beta>c$: if $\alpha=0$ and $\beta\leq c$, then $V_A(\mu;w)=\beta\Phi_\mu-c<0$ because $\Phi_\mu<1$. Under a point mass at a nonnegative quality, the agent's payoff tends to infinity as the quality grows if $\alpha>0$, and to $\beta-c>0$ if $\alpha=0$ and $\beta>c$. In either case, it is positive at a point mass on a sufficiently high quality. Mix a small weight on this point mass into $\mu$. The mixture lies in $\mathcal M$, and since total surplus and the agent's payoff are affine under mixing, its total surplus is still negative while the agent's payoff becomes positive. This contradicts $V_A(\mu;w)\leq\max\{V_S(\mu),0\}$ for every $\mu\in\mathcal M$, so $w$ is protected and $\mathcal W_P$ is compact. For each $\mu\in\mathcal M_\eta$, the ratio $U_P(\mu;w)/V_S(\mu)$ is upper semicontinuous in $w$. It is continuous wherever the agent is not indifferent at $\mu$; where he is indifferent, the tie-breaking rule makes him experiment, so $U_P(\mu;w)=V_S(\mu)$ and the ratio equals one, the largest value it can take. Hence $\lambda_\eta$, the infimum of these ratios, is upper semicontinuous and attains its maximum on the compact set $\mathcal W_P$. For $\eta>\overline\eta$, this maximum is at least the guarantee of the mixed contract above, which is positive and exceeds that of every single-instrument contract, so every optimal contract has $\alpha>0$ and $\beta>0$. 

This completes the proof.
\end{proof}
The same argument proves the claim in the footnote to Theorem~\ref{thm:distributional-mixed-guarantees}. There, improvements occur in the alternative state with a fixed conditional probability, and ``sufficiently rare'' means that this probability is below $\Phi(-\Delta/s)$. Every distribution the proof constructs can be chosen from the three-outcome family.

\section{Efficient Implementation Without Principal Protection}
\label{app:rule1-counterexample}

\setcounter{lemma}{0}
\setcounter{proposition}{0}
\setcounter{equation}{0}
\setcounter{figure}{0}
\setcounter{table}{0}

In this appendix, we show that the unique implementing contract $w^*$ in Theorem~\ref{thm:optimistic-default} need not be protected. The two examples below differ only in the liability cap. In both, $w^*$ implements efficient experimentation but is unprotected, so $\lambda^*=0$.

\begin{figure}[!htbp]\centering
\begin{subfigure}[t]{0.49\textwidth}\centering
\begin{tikzpicture}
\tikzset{every path/.append style={-}}
\begin{axis}[payoffaxis, xmin=-1.0, xmax=0.0, ymin=-0.0003, ymax=0.002, xtick distance=0.2, ytick distance=0.0005, yticklabel style={/pgf/number format/precision=4},
   legend style={font=\footnotesize,at={(0.03,0.97)},anchor=north west,draw=none,fill=white,row sep=-1.5pt},
   legend cell align=left]
  \fill[cVP!12] (axis cs:-0.2337,-0.0003) rectangle (axis cs:-0.0248,0.002);
  \draw[gray,thin] (axis cs:-1.0,0)--(axis cs:0.0,0);
  \draw[cutline] (axis cs:-0.4404,-0.0003)--(axis cs:-0.4404,0.002);
  \draw[cutline] (axis cs:-0.2337,-0.0003)--(axis cs:-0.2337,0.002);
  \node[vlabel] at (axis cs:-0.4404,0.00197){$\theta_{s1}$};
  \node[vlabel] at (axis cs:-0.2337,0.00197){$\theta_{s2}$};
  \addplot[vsline] table[x=theta,y=VS]{intermediate.dat}; \addlegendentry{$V_S(\theta_1)$}
  \addplot[valine] table[x=theta,y=VA]{intermediate.dat}; \addlegendentry{$V_A(\theta_1;w^*)$}
  \addplot[vpline] table[x=theta,y=VP]{intermediate.dat}; \addlegendentry{$V_P(\theta_1;w^*)$}
\end{axis}
\end{tikzpicture}
\caption{Full range.}
\end{subfigure}\hfill
\begin{subfigure}[t]{0.49\textwidth}\centering
\begin{tikzpicture}
\tikzset{every path/.append style={-}}
\begin{axis}[payoffaxis, xmin=-0.50, xmax=0.0, ymin=-0.00025, ymax=0.0005,
   xtick distance=0.1, ytick distance=0.0002, yticklabel style={/pgf/number format/precision=4}]
  \fill[cVP!12] (axis cs:-0.2337,-0.00025) rectangle (axis cs:-0.0248,0.0005);
  \draw[gray,thin] (axis cs:-0.50,0)--(axis cs:0.0,0);
  \draw[cutline] (axis cs:-0.4404,-0.00025)--(axis cs:-0.4404,0.0005);
  \draw[cutline] (axis cs:-0.2337,-0.00025)--(axis cs:-0.2337,0.0005);
  \node[vlabel] at (axis cs:-0.4404,0.000493){$\theta_{s1}$};
  \node[vlabel] at (axis cs:-0.2337,0.000493){$\theta_{s2}$};
  \addplot[vsline] table[x=theta,y=VS]{intermediate.dat};
  \addplot[valine] table[x=theta,y=VA]{intermediate.dat};
  \addplot[vpline] table[x=theta,y=VP]{intermediate.dat};
\end{axis}
\end{tikzpicture}
\caption{Near $\theta_{s2}$.}
\end{subfigure}
\caption{Intermediate cap ($r=1/5$). On the shaded interval, the agent experiments and $U_P=V_P<0$.}
\label{fig:rule1-intermediate}
\end{figure}
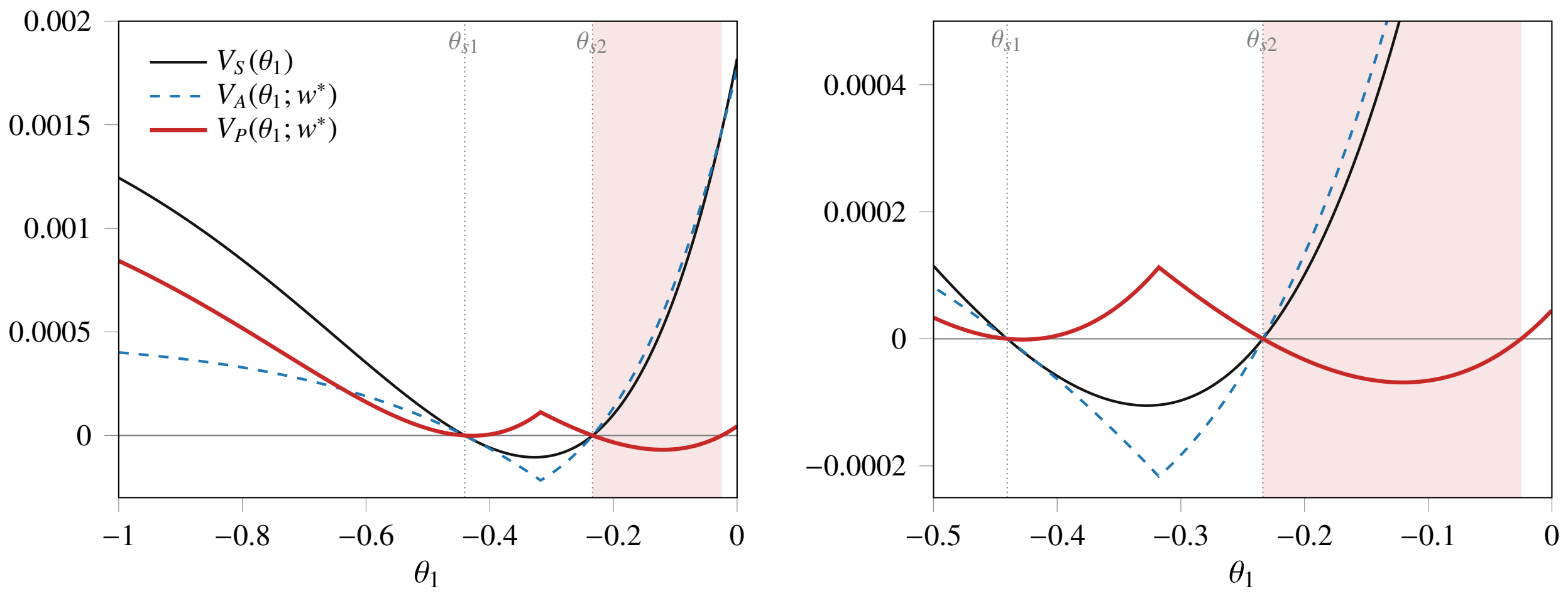

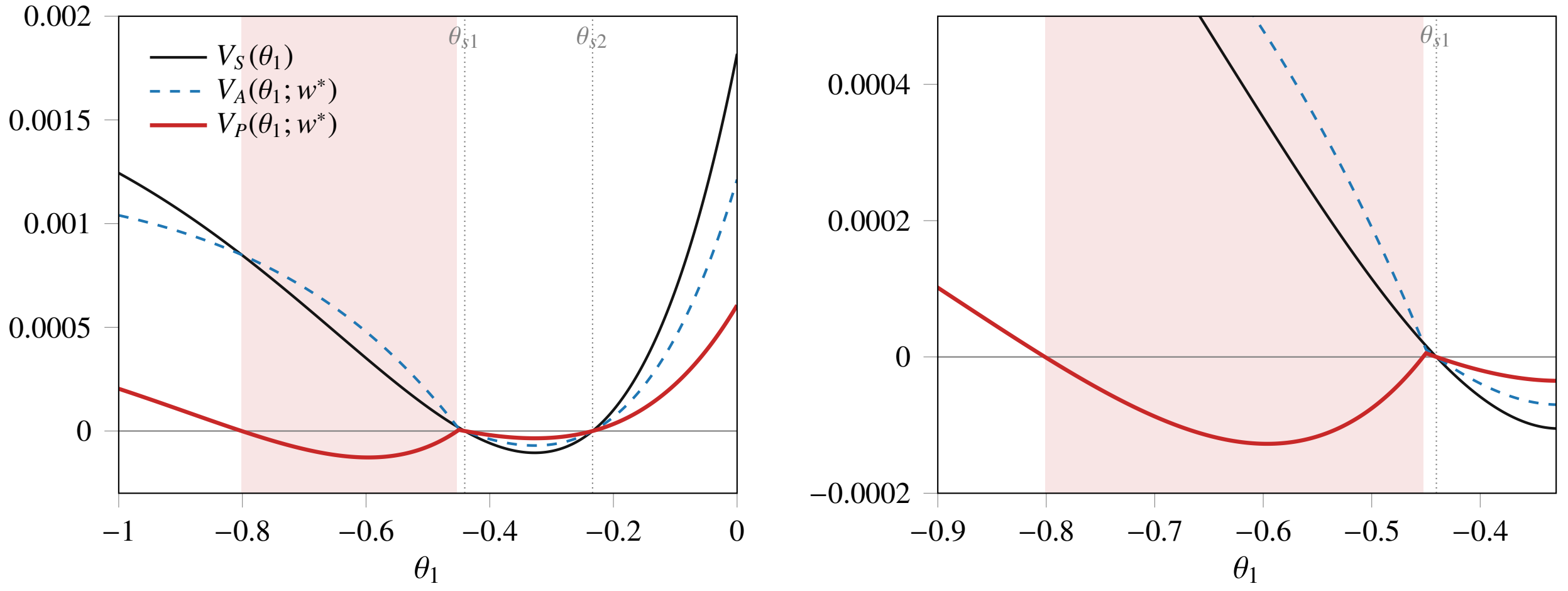
\begin{figure}[!htbp]\centering
\begin{subfigure}[t]{0.49\textwidth}\centering
\begin{tikzpicture}
\tikzset{every path/.append style={-}}
\begin{axis}[payoffaxis, xmin=-1.0, xmax=0.0, ymin=-0.0003, ymax=0.002, xtick distance=0.2, ytick distance=0.0005, yticklabel style={/pgf/number format/precision=4},
   legend style={font=\footnotesize,at={(0.03,0.97)},anchor=north west,draw=none,fill=none,row sep=-1.5pt},
   legend cell align=left]
  \fill[cVP!12] (axis cs:-0.8012,-0.0003) rectangle (axis cs:-0.4527,0.002);
  \draw[gray,thin] (axis cs:-1.0,0)--(axis cs:0.0,0);
  \draw[cutline] (axis cs:-0.4404,-0.0003)--(axis cs:-0.4404,0.002);
  \draw[cutline] (axis cs:-0.2337,-0.0003)--(axis cs:-0.2337,0.002);
  \node[vlabel] at (axis cs:-0.4404,0.00197){$\theta_{s1}$};
  \node[vlabel] at (axis cs:-0.2337,0.00197){$\theta_{s2}$};
  \addplot[vsline] table[x=theta,y=VS]{loose.dat}; \addlegendentry{$V_S(\theta_1)$}
  \addplot[valine] table[x=theta,y=VA]{loose.dat}; \addlegendentry{$V_A(\theta_1;w^*)$}
  \addplot[vpline] table[x=theta,y=VP]{loose.dat}; \addlegendentry{$V_P(\theta_1;w^*)$}
\end{axis}
\end{tikzpicture}
\caption{Full range.}
\end{subfigure}\hfill
\begin{subfigure}[t]{0.49\textwidth}\centering
\begin{tikzpicture}
\tikzset{every path/.append style={-}}
\begin{axis}[payoffaxis, xmin=-0.90, xmax=-0.33, ymin=-0.0002, ymax=0.0005,
   xtick distance=0.1, ytick distance=0.0002, yticklabel style={/pgf/number format/precision=4}]
  \fill[cVP!12] (axis cs:-0.8012,-0.0002) rectangle (axis cs:-0.4527,0.0005);
  \draw[gray,thin] (axis cs:-0.90,0)--(axis cs:-0.33,0);
  \draw[cutline] (axis cs:-0.4404,-0.0002)--(axis cs:-0.4404,0.0005);
  \node[vlabel] at (axis cs:-0.4404,0.000493){$\theta_{s1}$};
  \addplot[vsline] table[x=theta,y=VS]{loose.dat};
  \addplot[valine] table[x=theta,y=VA]{loose.dat};
  \addplot[vpline] table[x=theta,y=VP]{loose.dat};
\end{axis}
\end{tikzpicture}
\caption{Near $\theta_{s1}$.}
\end{subfigure}
\caption{Loose cap ($r=3/10$). On the shaded interval, the agent experiments and $U_P=V_P<0$.}
\label{fig:rule1-loose}
\end{figure}

Fix $p=3/5$, $s=7/8$, $\Delta=1.96s$, $c=1/2$, $\kappa=1/4$, and $\theta_0=2.4$. With this threshold, a feature is launched if and only if the experiment reports a statistically significant positive effect in a two-sided test at the 5\% level ($\hat\theta/s>1.96$). The default is optimistic, so experimentation is efficient exactly when $\theta_1\le\theta_{s1}$ or $\theta_1\ge\theta_{s2}$. Figures~\ref{fig:rule1-intermediate} and~\ref{fig:rule1-loose} plot the payoffs $V_S$, $V_A$, and $V_P$ conditional on experimentation. The principal's payoff $U_P$ equals $V_P$ where the agent experiments ($V_A\geq0$) and is zero elsewhere.

\paragraph{Intermediate cap.} Suppose that $r=1/5\in(\underline r,\overline r)$. The unique implementing contract is the mixed contract $w^{*}=(\alpha^{*},\beta^{*})$, under which the floor binds at the lower cutoff $\theta_{s1}$ but not at the upper cutoff $\theta_{s2}$. As is shown in Figure~\ref{fig:rule1-intermediate}, on part of the upper efficient region the agent experiments and $U_P(\theta_1;w^*)=V_P(\theta_1;w^*)<0$. Since $\alpha^{*}\theta_1>-r$ on this part of the region, the floor does not bind, and the loss is caused by the launch bonus. The principal pays $\beta^{*}$ on every launch, including launches of the harmful alternative. Near the upper cutoff, total surplus is small, and the agent's expected payoff exceeds it.

\paragraph{Loose cap.} Suppose that $r=3/10>\overline r$. The unique implementing contract is proportional sharing, $w^*=(\alpha_V,0)$, under which the floor binds at neither cutoff. As is shown in Figure~\ref{fig:rule1-loose}, on part of the lower efficient region the agent experiments and $U_P(\theta_1;w^*)=V_P(\theta_1;w^*)<0$. Since $\alpha_V\theta_1<-r$ on this part of the region, the floor binds, and the loss is caused by the liability floor. After a launch of the harmful alternative, the floor limits the agent's loss to $r$, below his share of the loss, and the principal bears the remainder. Total surplus is small here, and the agent's expected payoff exceeds it.

\section{Comparison Between Affine and Unrestricted Monotone Contracts}
\label{app:affine-vs-unrestricted}

\setcounter{lemma}{0}
\setcounter{proposition}{0}
\setcounter{equation}{0}
\setcounter{figure}{0}
\setcounter{table}{0}

In this appendix, we numerically compare optimal affine contracts with optimal monotone contracts under the uniform profit guarantee. A monotone contract pays the agent $w(\theta)\geq-r$ after a launch with realized quality $\theta$, where $w$ is nondecreasing. The rest of the baseline model is unchanged.

\begin{figure}[!htbp]\centering
\begin{subfigure}[t]{0.49\textwidth}\centering
\includegraphics[width=\linewidth]{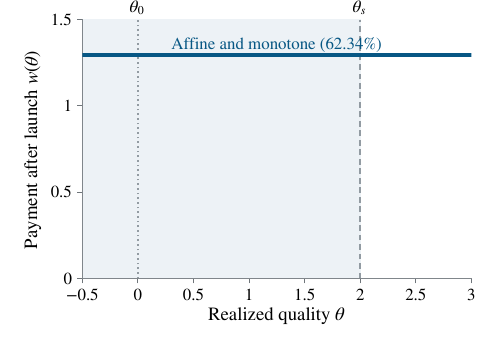}
\caption{Default value $\theta_0=0$.}\label{fig:affine-monotone-a}
\end{subfigure}\hfill
\begin{subfigure}[t]{0.49\textwidth}\centering
\includegraphics[width=\linewidth]{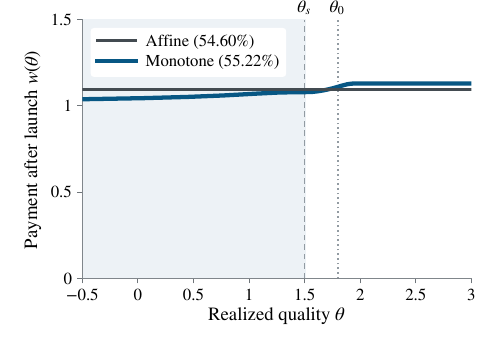}
\caption{Default value $\theta_0=1.8$.}\label{fig:affine-monotone-b}
\end{subfigure}
\caption{Payments after launch under optimal affine and monotone contracts.}
\label{fig:monotone-payment-schedules}
\end{figure}

Fix $p=3/5$, $s=7/8$, $\Delta=1.96s$, $c=1/2$, and $\kappa=1/4$, as in Section~\ref{sec:numerical-illustration}, and consider two pessimistic default values, $\theta_0=0$ and $\theta_0=1.8$. Figure~\ref{fig:monotone-payment-schedules} plots the payment $w(\theta)$ from the principal to the agent after launch as a function of realized quality $\theta$ under the optimal affine and monotone contracts, with the shaded region marking qualities below the efficient cutoff $\theta_s$.

When $\theta_0=0$, shown in panel~\subref{fig:affine-monotone-a}, the same pure launch bonus is optimal in both classes, so the two curves coincide and restricting attention to affine contracts is without loss. When $\theta_0=1.8$, shown in panel~\subref{fig:affine-monotone-b}, the optimal affine contract is again a pure launch bonus. Its uniform profit guarantee is $54.60\%$, compared with $55.22\%$ for the optimal monotone contract. To raise the principal's profit ratio near the cutoff, the monotone contract pays slightly less than the affine contract after launches just above the cutoff and more after launches of quality $\theta_0$ or higher. The principal's expected payoff is therefore higher under the affine contract at every $\theta_1\geq\theta_0$.

\end{appendices}

\end{document}